\documentclass[12pt,twocolumn,numberedappendix,twocolappendix,tighten,letterpaper]{openjournal}

\usepackage{xcolor}
\usepackage{textgreek}
\usepackage{amsfonts,amsmath,amssymb,bm}
\allowdisplaybreaks
\usepackage[utf8]{inputenc}
\usepackage[T1]{fontenc}
\usepackage[english]{babel}
\usepackage{hyperref}
\hypersetup{
    colorlinks=true,
    linkcolor=blue,
    filecolor=blue,      
    urlcolor=red,
    citecolor=blue,
}
\usepackage{color,colortbl}
\usepackage{tensind}
\usepackage[export]{adjustbox}
\tensordelimiter{?}
\DeclareGraphicsExtensions{.bmp,.png,.jpg,.pdf}
\usepackage{verbatim}
\usepackage[normalem]{ulem}
\usepackage{orcidlink}
\usepackage{soul}
\usepackage[]{cleveref}
\crefname{section}{\S\!}{\S\S\!}
\crefname{appendix}{\S\!}{\S\S\!}
\crefname{equation}{Eq.}{Eqs.}
\Crefname{equation}{Equation}{Equations}
\crefname{figure}{Fig.}{Figs.}
\Crefname{figure}{Figure}{Figures}

\newcommand{\ffold}{}

\providecommand{\dodoi}[1]{\href{https://doi.org/#1}{#1}}

\newcommand{\mean}[1]{\overline{#1}}
\newcommand{\meanz}[1]{\langle {#1}\rangle}
\newcommand{\va}{v_{\rm A}}
\newcommand{\hpaper}{H+24}

\newcommand{\revchng}[1]{{#1}}

\begin{document}
\title{Rapid, strongly magnetized accretion in the zero-net-vertical-flux shearing box\vspace{-20pt}}
\author{Jonathan Squire\orcidlink{0000-0001-8479-962X}}
\email{jonathan.squire@otago.ac.nz}
\affiliation{Physics Department, University of Otago, Dunedin 9010, New Zealand}
\author{Eliot Quataert}
\affiliation{Department of Astrophysical Sciences, Princeton University, Princeton, NJ 08544, USA}
\author{Philip F.~Hopkins}
\affiliation{TAPIR, Mailcode 350-17, California Institute of Technology, Pasadena, CA 91125, USA}

\begin{abstract}

We show that there exist two qualitatively distinct turbulent states of the zero-net-vertical-flux shearing box.  The first, which has been studied in detail previously, is characterized by a weakly magnetized ($\beta\sim50$) midplane with slow periodic reversals of the mean azimuthal field (dynamo cycles). The second, the `low-$\beta$ state,' which is the main subject of this paper, is characterized by a strongly magnetized $\beta\sim 1$ midplane dominated by a coherent azimuthal field with much stronger turbulence and much larger accretion stress ($\alpha \sim 1$). The low-$\beta$ state emerges in simulations initialized with sufficiently strong azimuthal magnetic fields. The mean azimuthal field in the low-$\beta$ state is quasi steady (no cycles) and is sustained by a dynamo mechanism that compensates for the continued loss of magnetic flux through the vertical boundaries; we attribute the dynamo to the combination of differential rotation and the Parker instability, although many of its details remain unclear. Vertical force balance in the low-$\beta$ state is dominated by the mean magnetic pressure except at the midplane, where thermal pressure support is always important (this holds true even when simulations are initialized at $\beta \ll 1$, provided the thermal scale height of the disk is well resolved). The efficient angular momentum transport in the low-$\beta$ state may resolve long-standing tension between predictions of magnetorotational turbulence (at high $\beta$) and observations; likewise, the bifurcation in accretion states we identify may be important for understanding the state transitions observed in dwarf novae, X-ray binaries, and changing-look AGN. We discuss directions for future work, including the implications of our results for global accretion disk models and simulations.

\end{abstract}


\maketitle

\section{Introduction}\label{sec:intro}

Accretion disks power a variety of the most luminous and interesting sources in the Universe,
forming the crucial intermediary that allows matter to lose its angular momentum and 
fall into a central object. Hotter systems, such as  quasars, X-ray binaries, and dwarf novae 
are mostly  well ionized, with the required outwards angular momentum  transport thought to be dominated by turbulence 
induced by the magnetorotational instability (MRI; \citealp{Balbus1991}). 
Significant work has thus focused on understanding the nonlinear turbulent saturation of the MRI, particularly 
the so-called \cite{Shakura1973} $\alpha$ parameter, which quantifies the rate of 
angular momentum transport. Local and global simulation studies
have found $\alpha$ values between ${\simeq}0.001$ and ${\simeq}1$ depending on the physical parameters 
and resolution \citep{Hawley1995,Stone1996,Fromang2010,Bai2013a,Ryan2017}, as well as various more complex pictures
involving e.g., transport via winds and coherent fields \citep[e.g.,][]{Lesur2013,Zhu2018}. Difficulties in 
the agreement between theory and observations remain \citep[e.g.,][]{King2007,Tetarenko2018} and it 
is fair to say that our understanding of accretion remains incomplete.

\revchng{An intriguing possibility for addressing related observational and theoretical challenges is that disks may host magnetic fields with pressures comparable to, or even exceeding, the gas pressure.
By supporting the gas against vertical collapse, such superthermal fields enhance the disk's thermal, viscous, and  gravitational stability compared to standard models, potentially allowing much higher accretion rates \citep[e.g.,][]{Pariev2003,Begelman2007,Oda2009,Sadowski2016}. 
A net vertical flux (NVF) threading the disk appears  to aid in the growth and sustenance of such fields via the MRI \citep[e.g.,][]{Miller2000,Suzuki2009,Zhu2018,Mishra2022}, but presumably requires the coherent vertical field to pre-exist as a property of the central object.
However, even without a clear source of NVF, various global simulation studies have reported the formation of (apparently) stable, 
rapidly accreting disks  with strong toroidal magnetic fields \citep{Machida2006,Gaburov2012,Sadowski2016,Guo2024}, suggesting that such states may be a common outcome of accretion. A striking recent example is the ``hyper-refined'' simulation of 
\cite{Hopkins2024b} and \citet{Hopkins2024} (hereafter \hpaper), which traced gas inflow  from cosmological initial conditions on galactic scales into a sub-pc steady state
accretion disk. The disk's toroidal magnetic field becomes highly superthermal,
with a plasma $\beta$ (the ratio of thermal 
to magnetic pressure) of ${\sim}10^{-4}$. This allows it to avoid fragmentation across a wide range of radii (${\gtrsim}1{\rm pc}$) while sustaining very high accretion rates $\alpha\gg1$.
The magnetized system's enhanced stability arises because magnetic support and strong turbulence produce a thicker disk, enabling higher inflow rates for similar surface densities, or equivalently, lower densities for a given mass supply rate\footnote{Indeed,
the simple numerical experiment of removing the magnetic field in \hpaper\ yields a vastly different solution, which 
fragments on small scales as expected and may not even accrete (see \hpaper\  figures 28-30).} \citep{Hopkins2024a}.}

The global results discussed above  fundamentally rely on the disk being 
\emph{able} to sustain  $\beta\lesssim 1$ (or $\beta\ll1$) magnetic fields against destruction by turbulence or vertical escape from the disk. 
Some previous works have argued that a $\beta\lesssim 1$ midplane cannot 
be maintained without NVF, because strong  toroidal fields rise rapidly (in several orbits) 
out of the disk midplane \citep{Salvesen2016,Fragile2017}.
While \hpaper\ and \citet{Gaburov2012} argue that the toroidal field in their simulations is replenished primarily via simple advection
 from further out in the disk,  if  loss processes acted more 
 rapidly than the accretion of more flux, the system would lose its field, undermining all other aspects of the strongly magnetized solution. 
 \revchng{As mentioned above, while simulations with NVF do maintain a $\beta\lesssim1$ midplane in both local \citep{Suzuki2009,Bai2013a} and global \citep{Zhu2018,Mishra2020} setups, the global solutions seem to posses a morphology that is rather different from \hpaper\ and other $\beta\lesssim1$ global simulations without external NVF. In addition, as far as we are aware, NVF simulations have never been observed to sustain   $\beta\ll1$ magnetic fields, as seen 
 in \hpaper\ and \citet{Guo2024}.}

\revchng{These considerations motivate a re-examination of whether strong magnetic fields can be sustained against vertical escape without NVF 
in the simplest viable model: the local stratified shearing box. 
Early results from  \citet{Johansen2008} suggested that zero-net-vertical-flux (ZNVF) boxes \emph{could} sustain a toroidal field with  $\beta \sim 1 $ against escape. However, this was challenged by  \citet{Salvesen2016}, who showed that  \citet{Johansen2008}'s results 
were affected by their vertical boundary conditions. When open boundaries were used, \citet{Salvesen2016} argued that $\beta\sim1 $ purely toroidal fields 
always escape, leaving behind a midplane with $\beta\sim 50$.
Here, using a setup similar to \cite{Salvesen2016}'s but with stronger initial toroidal fields, we argue that a ZNVF $\beta\sim1 $ state  \emph{can}, in fact, self sustain in the local shearing box and --- as a consequence --- that vertical flux is not a requirement for
 large accretion stresses.}
The simulations resemble those of \citet{Johansen2008} but (crucially) differ in that  flux continuously 
escapes from the system's top and bottom boundaries while being regenerated at the midplane. Our solutions are also 
similar to those of \hpaper, \citet{Gaburov2012}, and \citet{Guo2024} in various relevant respects, \revchng{although they do not sustain 
a $\beta\ll1$ midplane in steady state unless numerically limited (we provide some evidence and commentary on additional 
global effects that could act to decrease $\beta$ to be  $\ll1$).}
  Compared to global simulations, our results  thus provide 
a much simpler way to study the main physical mechanisms that could enable  
strongly magnetized accretion.

An interesting outcome of our findings is the existence of a sudden and sharp 
transition between two radically different forms of shearing-box turbulence. 
The first --- termed the ``high-$\beta$ state'' here, and explored in many previous works \citep[e.g.,][]{Miller2000,Guan2009,Simon2012} --- 
involves a thermal-pressure dominated ($\beta\simeq50$) midplane, with super-Alfv\'enic turbulence, cyclic reversals of
the weak mean toroidal field over ${\simeq}10$-orbit timescales (often termed `MRI dynamo cycles'), and weak transport (angular-momentum stress $\alpha\simeq 0.02 $).
The second ---  termed the ``low-$\beta$ state'' and the subject of this work --- involves  a continuous (non-cyclic) 
strong toroidal field with $\beta\lesssim1$, modestly sub-Alfv\'enic turbulence, and much stronger angular momentum transport ($\alpha\simeq 0.5 $).
So long as the box has sufficient vertical extent (which is required to realize the low-$\beta$ state), the transition between 
the two is determined by the initial total flux available to the system: 
if the system starts out sufficiently strongly magnetized, it can sustain this low-$\beta$ state; if the
magnetic field is too weak, all of the flux inevitably escapes, transitioning into the high-$\beta$ state.
This bifurcation may have interesting implications for understanding state transitions observed in 
X-ray transients  and dwarf novae \cite[e.g.,][]{Smak1984,Hameury2020}, suggesting 
that additional magnetic flux in almost any configuration (not just a
coherent vertical field in particular) could trigger high accretion rates.  The same abrupt changes in angular momentum transport efficiency could be important for understanding the emerging class of changing-look active galactic nuclei \citep{Ricci2023}. 

Our work here is organized around a wide set of stratified shearing-box simulations starting 
from initial conditions with strong ($\beta\lesssim1$) toroidal fields. 
In order to convincingly characterize the low-$\beta$ state and  transition, a range of different simulations are needed, which also 
leads to a proliferation of complications about the effects of numerical options, initial conditions etc. To provide the reader with a ``road map'' before starting, 
in \S\ref{sec: key points} we outline, without providing any detailed interpretation, various outcomes of our numerical experiments that inform our results and interpretation.
We then describe the numerical  setup in \S\ref{sec:methods}, with the various
different numerical options, boundary conditions, and initial conditions described in \S\ref{subsec: bcs} and \S\ref{subsec: ics}. These 
are mostly chosen to be  standard, matching previous works.
For reference in the simulation analysis, we also provide an overview of strongly magnetized vertical equilibria (\S\ref{subsec: equilibria}), as
well as a recapitulation of the  
 MRI and Parker instabilities (\S\ref{sec: linear behavior}), which we argue are key to the  
 field-sustenance mechanism, consistent with past studies \citep{Johansen2008,Kudoh2020,Held2024}.
Our main results on the properties of the low-$\beta$ state are detailed in  \S\ref{sec: results}, while those 
related to the field-sustenance mechanism (dynamo) are presented in  \S\ref{sec: dynamo}.
Finally, we summarize and conclude in \S\ref{sec: conclusion} with a detailed 
discussion of how our results apply to previous global results and $\beta\ll1$ fields in \S\ref{subsec: global apps}.
The appendices explore the effect of \revchng{numerical options (box aspect ratio, resolution, 
and an artificial density floor) and present various subsidiary results.}

\section{Summary of key behaviors}\label{sec: key points}

In \S\ref{sec: results} and \S\ref{sec: dynamo}, we explore results from a range of different setups
in order to build a picture of the local behavior of the strongly accreting, low-$\beta$ state.
However, various factors add complexity to the story (e.g., initial
conditions, boundary conditions, resolution); so, for reference, we 
provide a list of the key behaviors that  will be discussed. The following  describes various {observations}  from the set of numerical experiments; a more detailed {interpretation} of the important physics will be given in the results (\S\ref{sec: results}-\S\ref{sec: dynamo}) and discussion (\S\ref{sec: conclusion}). 
\begin{enumerate}
\item There exist (at least) two qualitatively different turbulent states of ZNVF shearing-box turbulence. 
The first, which  has been regularly studied previously, is characterized by a weakly magnetized ($\beta\sim50$) midplane with slow
periodic reversals of the mean azimuthal field (dynamo cycles). The second, which is the main subject of this article, 
is characterized by a strongly magnetized $\beta\sim 1 $ midplane dominated by a coherent azimuthal field in quasi steady state (no cycles), with much stronger
turbulence and much larger accretion stress $\alpha$.
\item In the absence of a vertical field, the transition between these two states is primarily controlled by the magnetization of the system: 
if the system has sufficient magnetic flux available  such that it retains $\beta\lesssim1$ in the midplane
once it reorganizes from its initial state, it can maintain this steady state; if $\beta$ in the midplane rises above $\simeq1$, 
it inevitably loses nearly all of its flux and transitions into the high-$\beta$ weakly accreting state.
\item Turbulence in the low-$\beta$ state is trans-Alfv\'enic with $\delta u/\va\simeq 0.3$ across its full vertical profile, becoming
highly supersonic in  lower-$\beta$ regions above/below the midplane. Its profile is well fit by a simple model based on the MRI and Parker instability.
 \item By virtue of point 3, the vertical equilibrium of the low-$\beta$ state is dominated by the mean magnetic pressure, except
within ${\lesssim} H_{\rm th}$ of the midplane, where the density collapses down to approximately its (Gaussian) thermal profile. Above the
midplane, both azimuthal field and density are approximately  power laws.
\item The low-$\beta$ magnetized state \revchng{remains robust regardless of the chosen vertical boundary conditions, as long as material and magnetic flux can escape the domain. It is also insensitive to the vertical domain size, provided the domain is sufficiently large }compared to $H_{\rm th}$. However, when starting from 
initial conditions that lie near those at which the system changes between the low- and high-$\beta$ state (with constant-$\beta$ Gaussian initial conditions, 
this occurs when $\beta\approx0.1$ initially), the transition can depend on resolution and boundary conditions, presumably because they affect the efficiency of the flux transport through the vertical domain boundaries.  
 \item A sufficiently large  domain in the azimuthal direction is necessary to sustain the low-$\beta$ state. The radial domain size does not appear to have 
 a significant influence. 
\item With progressively more vertical magnetic flux,  the ZNVF low-$\beta$ state  changes 
into the stratified low-$\beta $ NVF state described in previous works \citep{Suzuki2009,Bai2013a,Salvesen2016a}. This 
transition happens for $\beta_{z}\lesssim 1000$, \emph{viz.,} a vertical field with  $\beta_{z}\geq 1000$  makes little difference to $\alpha$ or the morphology of the low-$\beta$ state.
Consistent with past works, stronger vertical fields cause
a more strongly magnetized midplane ($\beta<1$), with a larger $\alpha$  that is dominated by the contribution from 
the mean magnetic fields. 
\item The  net azimuthal field in the low-$\beta$ state is actively sustained by a dynamo mechanism. Strong turbulent
diffusion and vertical outflows continuously remove the flux out of the top and bottom boundaries of the simulation. This azimuthal flux loss is balanced by shearing of the radial field, which is replenished from the azimuthal field by turbulence to maintain steady state. 
\item Both rotation and the mean shear flow are needed to maintain the low-$\beta$ strongly accreting state.
\item Systems that are initially even more strongly magnetized, with $\beta\ll1$, remain in quasi-equilibrium while relaxing back down towards $\beta\simeq 1$ in the midplane. As this occurs, the density profile collapses towards the thermal scale height.  This relaxation process is relatively slow, operating on a timescale of $5$-$10$ orbits.
\item If the numerical resolution is sufficiently \revchng{coarse} that the density scale height near the midplane becomes limited by grid resolution before  thermal pressure, the system reaches a $\beta\ll1$ steady state with supersonic (though still sub-Alfv\'enic) turbulence in the midplane. The mean fields  are sustained via a similar dynamo 
  mechanism to that which operates when the scale-height near the midplane is well resolved. This suggests that if the collapse of the density to a thermally supported region is halted by some mechanism (e.g., another source of turbulence), the system may self-sustain a $\beta\ll1$ state.
    \item The mechanism that closes the low-$\beta$ ZNVF dynamo loop, generating a mean radial field from the mean azimuthal field, remains unclear. Although 
  the turbulence has a net helicity away from the midplane, suggesting that a dynamo ``alpha effect'' could operate, the sign of the net helicity 
  is $\beta$ dependent and not always that needed to sustain the mean field  \citep[see also][]{Tharakkal2023}. 
  \end{enumerate}
  In addition to these observations relating to the low-$\beta$ ZNVF  state, our numerical experiments have revealed some subsidiary points relevant to accretion physics
  in other contexts:
  \begin{enumerate}
    \setcounter{enumi}{13}
  \item In the high-$\beta$ state, dynamo cycles  (temporal reversals in the sign of the mean azimuthal field) persist  in the 
absence of rotation so long as  mean shear is present,  with or without  NVF.  This shows that, despite being traditionally associated with the MRI, differential rotation and the MRI are sufficient but not necessary for producing  dynamo cycles.
  \item With NVF, the  strong mean azimuthal fields are 
effectively sustained as a single MRI channel mode that  stretches across the full vertical extent of the domain. 
While the symmetry of this state is likely inconsistent with global constraints \citep{Bai2013a}, 
global simulations \citep{Zhu2018,Mishra2020} appear to exhibit a similar global channel mode but with opposite parity,  with the ordered radial magnetic field maintained by an equatorial outflow and inflows away from the midplane.
  \end{enumerate}

\section{Theory \& Methods}\label{sec:methods}

Our numerical methods and setup are designed to be as standard as possible
in order to understand the relationship of our results to previous works. We use the 
\textsc{Athena++} code \citep{Stone2020} with the shearing box method, thus 
modelling a local patch of the disk in Cartesian coordinates \citep{Hawley1995}.
We  define the Cartesian coordinates in the standard way: $x\equiv (R-R_{0})/L_{0}$, $y\equiv R_{0}\phi/L_{0}$, $ z\equiv z/L_{0}$,
where $R$, $\phi$, and $z$ refer to the usual cylindrical coordinate system 
for a disk in the $z=0$ plane. The parameter $L_{0}$ allows the local coordinate system to 
be rescaled to a convenient length scale, which we will generally take to be that of a magnetized 
equilibrium for reasons discussed below. 

We  solve the isothermal shearing-box MHD equations:
\vspace{-0.4cm}
\begin{subequations}
\begin{align}
\frac{\partial \rho}{\partial t}= & -\nabla \cdot(\rho \bm{U}),\label{eq:MHD rho}\\
 \frac{\partial(\rho \bm{U})}{\partial t}= & -\nabla \cdot\left[\rho \bm{U} \bm{U}-\frac{\bm{B} \bm{B}}{4\pi}+\left(c_{s}^{2}\rho+\frac{B^2}{8\pi}\right) \bm{1}\right] \nonumber\\ & +2 q \rho \Omega^2 x \hat{\bm{x}}-2 \Omega \hat{\bm{z}} \times(\rho \bm{U})-\rho \Omega^2 z \hat{\bm{z}},\label{eq:MHD u}\\
\frac{\partial \bm{B}}{\partial t}=& \nabla \times(\bm{U} \times \bm{B}).\label{eq:MHD B}
\end{align}
\end{subequations}
\vskip 0.2cm
\noindent Here $\bm{B}$ is the magnetic field, $\bm{U}=-q\Omega x\bm{\hat{y}}+\bm{u}$ is the total flow velocity (including the background shear $-q\Omega x\bm{\hat{y}}$), $\rho$ is the mass density, $\Omega$ is the local rotation rate, which is the natural time unit for the system, $q\equiv -d\ln \Omega/d\ln R=3/2$ for a Keplerian disk, and $c_{s}$ is the 
(constant) sound speed.

Although idealized, many studies have shown that the turbulent dynamics induced 
by \crefrange{eq:MHD rho}{eq:MHD B} are widely varied and dynamically complex, involving the natural generation of 
long-time coherent large-scale fields and flows, and other complex features.
The system also has various undesirable properties such as solutions that depend on the vertical boundary conditions, resolution,
and box size (e.g., \citealt{Simon2012,Shi2016,Ryan2017}), as well as  the obvious
neglect of potentially important global effects (e.g., field-line curvature).
For our purposes, it retains  two important advantages compared to global simulations: 
first, it is  cheaper computationally than global simulations for similar disk resolutions; second,
it provides a much simpler and more controllable testbed for understanding 
the key physics at play in the system, particularly in steady state. We will see that
vertical stratification is absolutely crucial to the physics we describe, meaning that
 the yet simpler unstratified version of the shearing box would be unsuitable.

\subsection{Static vertical equilibria in the local shearing box}\label{subsec: equilibria}

The form of the magneto-static solutions to \cref{eq:MHD u} will  play an 
important role in our narrative and numerical setup, so are helpful to outline in detail here. 
Assuming $\bm{u}=0$ and that all variables are time independent and depend only on $z$, 
\cref{eq:MHD rho} is trivial, \cref{eq:MHD B} just implies $B_{x}=0$, while \cref{eq:MHD u} becomes
\vskip 2pt
\begin{equation}
\beta \frac{d\ln P_{\rm th}}{dz} + \frac{d\ln P_{B}}{dz} + \beta \frac{2z}{H_{\rm th}^{2}}=0.\label{eq: equilibrium}
\end{equation}
\vskip -0.1cm
\noindent Here $P_{\rm th}\equiv c_{s}^{2}\rho$ is the thermal pressure, $P_{B}\equiv B^{2}/8\pi$ is the magnetic 
pressure,  $\beta(z)\equiv P_{\rm th}/P_{B}$, and $H_{\rm th}^{2}\equiv 2 c_{s}^{2}/\Omega^{2}$ is the thermal scale height.   
We will find that the azimuthal field  dominates strongly over the
other components. This implies that so long as the turbulence is modestly subsonic where $\beta\gtrsim 1$
and modestly sub-Alfv\'enic where $\beta\lesssim 1 $, then  \cref{eq: equilibrium} establishes the relationship between $\rho$ and $B_{y} \approx \sqrt{8\pi P_{B}}$.
We will also see that the mean vertical outflows that develop do not provide significant contributions to the overall pressure balance,
further justifying the magneto-static assumption.

The simplest class of solutions to \cref{eq: equilibrium} comes from assuming $\beta$ remains constant with height. 
Then $P_{B}= P_{\rm th}/\beta$ and \cref{eq: equilibrium} becomes $(1+\beta^{-1})d\ln P_{\rm th}/dz = -2z/H_{\rm th}^{2}$ with solution 
\begin{equation}
\rho \propto P_{\rm th} \propto B_{y}^{2} \propto \exp (-z^{2}/H_{\beta}^{2}), \label{eq: Gaussian equilibrium}
\end{equation}
where $H_{\beta}^{2}\equiv H_{\rm th}^{2} (1+\beta^{-1})$. As expected, this reduces to the thermal profile, $\rho\propto \exp(-z^{2}/H_{\rm th}^{2})$, as $P_{B}\rightarrow 0$ ($\beta\rightarrow \infty$), but the profile widens with decreasing $\beta$ as 
the magnetic field provides more pressure support.

\begin{figure}
\begin{center}
\includegraphics[width=0.9\columnwidth]{\ffold 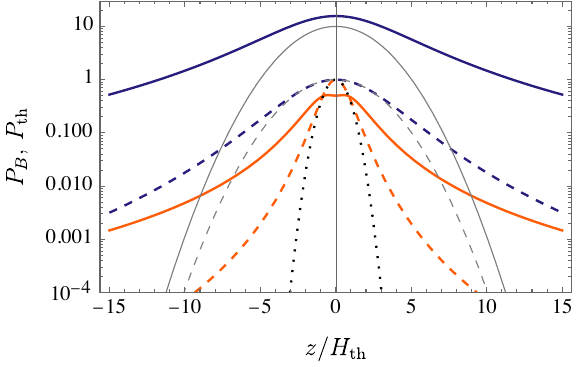}
\caption{\revchng{Various magneto-static equilibrium solutions to \cref{eq: equilibrium}, showing $P_{B}$ (solid lines) and $P_{\rm th}$ (dashed lines). Colored line pairs show the Lorentzian-like equilibrium \eqref{eq: lorenztian equilibrium} with $a=3$; the blue curves show $H/H_{\rm th}=5$, such that 
$\beta(z=0)\approx 0.1$,  and the orange curves show $H/H_{\rm th}=1.5$, which is just below where $P_{B}$ develops a midplane dip and $\beta(z=0)>1$ (see text).
The dotted black line shows the thermal density profile $\exp(-z^{2}/H_{\rm th}^{2})$ and the thin gray curves show the constant-$\beta$ equilibrium \eqref{eq: Gaussian equilibrium} with $\beta=0.1$ ($H_{\beta}\approx 3.3 H_{\rm th}$), as used to initialize most simulations. All profiles are normalized with $P_{\rm th}(z=0)=1$ for ease of comparison.} }
\label{fig: analytic}
\end{center}
\end{figure}

However, although we use \cref{eq: equilibrium}  for most initial conditions, it does 
not even slightly resemble the profiles that arise dynamically in our simulations, which 
become increasingly magnetically dominated away from the midplane.  Given that if $\beta\ll1$,
the system has no way to know about the sound speed and is thus ignorant of $H_{\rm th}$, which is the only length scale in \cref{eq: equilibrium},  the $\beta\ll1$ limit must be described by    
scale-free power-law solutions. Inserting the ansatz $P_{B}\propto z^{-a}$ and $P_{{\rm th}}\propto z^{-b}$, we see that the 
first term in \cref{eq: equilibrium} must become subdominant to the others at large $z$, leading
to $\beta\propto z^{-2}$ and 
\begin{equation}
P_{B}\propto z^{-a},\quad P_{\rm th}\propto z^{-a-2},\quad \text{for}\quad \frac{z^{2}}{H_{\rm th}^{2}}\gg \frac{a}{2}+1\label{eq: pure power law equilibria}
\end{equation}
(i.e., $b=a+2$). Note that this also implies $\va^{2} = 2\Omega^{2}z^{2}/a$ in order to satisfy \cref{eq: equilibrium}. In such an
equilibrium, the magnetic field provides all of the pressure support while 
the gas just provides the mass for the gravitational field. The idea can 
be generalized to the full range of $z$ by regularizing  the $P_{\rm th}$ profile at $z=0$ as in a Lorentzian function. We thus replace 
$z^{2}$ with $z^{2}+H^{2}$, where $H$ is the density profile's width near $z=0$ , 
solving \cref{eq: equilibrium} 
to yield,\vspace{-5pt}
\begin{gather}
P_{\rm th}=\mathcal{N}_{a}\frac{\rho_{0}c_{s}^{2}}{\sqrt{\pi}}\frac{H^{a+2}}{(z^{2}+H^{2})^{(a+2)/2}},\nonumber\\ 
P_{\rm B} = \mathcal{N}_{a}\frac{\rho_{0}\Omega^{2}}{\sqrt{\pi}a}\frac{ H^{a+2}(z^{2}+H^{2}-aH_{\rm th}^{2}/2 )}{   (z^{2}+H^{2})^{(a+2)/2}}.\label{eq: lorenztian equilibrium}
\end{gather}
This solution is valid for all $z$, asymptoting to \cref{eq: pure power law equilibria} at large $z$, and the normalization factor  $\mathcal{N}_{a}=\Gamma(a/2+1)/\Gamma(a/2+1/2)$ ensures $\int dz\,\rho =\rho_{0}H$. \Cref{eq: lorenztian equilibrium}  is useful for understanding the profiles observed in simulations, as well as providing a different equilibrium initial condition to test the robustness of 
our findings.
From the expression for $P_{B}$ it is clear that the profile is physical only for $H>\sqrt{a/2}H_{\rm th}$, since 
otherwise $P_{B}$ becomes negative as $z$ approaches $0$. This is a consequence of the fact that if $\rho$ becomes \emph{steeper} than the thermal profile,  $P_{B}$ must  provide an inwards force instead of outwards support, causing 
a dip in $P_{B}$ near the midplane when $H^{2}\leq (a+2)H_{\rm th}^{2}$/2 (otherwise $P_{B}$ increases monotonically to $z=0$). Also useful are expressions for the density scale height (defined such that $H_{\rho}=H_{\rm th}$ for a thermal profile),
$H_{\rho}\equiv\sqrt{\pi}(\rho_{0}H)^{-1}\int dz\,|z| \rho = 2  \mathcal{N}_{a}H /a\approx H$, and  the  midplane $\beta$, $\beta(z=0) =  a/(2 H^{2}/H_{\rm th}^{2}-a)$. These show that at large $H/H_{\rm th}$ the solution has a wide density profile, $H_{\rho}\gg H_{\rm th}$, and satisfies $\beta\ll1$ everywhere, including in the midplane. \revchng{These properties are highlighted in \cref{fig: analytic}, which compares the solutions
\eqref{eq: lorenztian equilibrium} to Gaussian equilibria.}


\subsection{Boundary conditions}\label{subsec: bcs}

In all simulations we use periodic boundary conditions in $y$, and ``shear-periodic'' boundary conditions in $x$, whereby 
quantities are periodic in the frame that remains stationary on the background shear flow $-q\Omega x \bm{\hat{y}}$.
However, previous works have shown significant
dependence of results on the vertical boundary conditions \citep[e.g.,][]{Johansen2008,Gressel2013,Salvesen2016,Coleman2017},
so we use a number of different choices through this work.  All boundary conditions
set $u_{x}$, $u_{y}$, and $B_{z}$ to have zero gradient at the vertical boundary, and set the vertical flow $u_{z}$ to have zero gradient unless there is an inflow ($u_{z} z <0$), in which case it is set to $u_{z}=0$ \citep{Stone1996}. The conditions on $\rho$, $B_{x}$, and $B_{y}$ are as follows, 
with the naming scheme used to label simulations in \cref{tab:sims}: 
\begin{description}
\item[Outflow] Zero-gradient boundary conditions for $B_{x}$ and $B_{y}$, while  $\rho$ is extrapolated 
into the ghost cells assuming the initial hydrostatic equilibrium as $\rho(z) = \rho_{e} \exp(-z^{2}/H_{\beta}^{2})/\exp(-z_{e}^{2}/H_{\beta}^{2})$ (where the $z_{e}$ and $\rho_{e}$ here refer to the values at the boundary).  Following \cite{Simon2011}, these boundary conditions have been widely used and allow the total toroidal and radial magnetic flux within the system to vary.
\item[Simon13] Following \cite{Simon2013} and \cite{Salvesen2016}, these are the same as Outflow, except that 
they also extrapolate $B_{x}$ and $B_{y}$ similarly to the density, with $B_{i}(z) = B_{i,e} \exp(-z^{2}/H_{\beta}^{2})/\exp(-z_{e}^{2}/H_{\beta}^{2})$. This enhances the removal of magnetic flux from the domain. 
\item[Lesur13] Following \cite{Lesur2013}, these set $B_{x}=0$ at the boundaries and use zero-gradient conditions for $\rho$ and $B_{y}$. These boundary conditions are designed to enable
winds to form more efficiently, better approximating global simulations. 
\item[Power-law] Motivated by the  low-$\beta$ equilibria discussed above (\cref{eq: pure power law equilibria}), 
$B_{x}$ and $B_{y}$ are extrapolated into the ghost cells as $B_{i}(z)=B_{i,e} |z/z_{e}|^{-a/2}$ and $\rho$ is extrapolated 
as $\rho(z) = \rho_{e}|z/z_{e}|^{-a-2}$. We set $a=3$, which is approximately the index observed at large $|z|$ in our simulations. 
\end{description}
Outflow, Simon13, and Power-law effectively differ only in the gradient of $\bm{B}_{\perp}$ and $\rho$ assigned at the boundary, as set by the various extrapolation techniques. For context, the Simon13 hydrostatic scheme (Gaussian extrapolation) is most extreme, fixing quantities to larger negative gradients; the power-law scheme is more moderate, while outflow conditions (with zero gradient) are the least extreme. From a close examination of the 
shape  of $\rho$ near  the boundaries with Outflow or Simon13, it is clear that the latter forces the  
profile to be steeper than is natural. This was the primary motivation for the new Power-law scheme --- to disturb 
the solution as little as possible from the profiles that are naturally set up in coronal regions above the midplane.   

Overall, we will see that our results are mostly independent of these vertical boundary condition choices, aside from minor changes to time-averaged vertical profiles and $\alpha$. 
The solution can be strongly impacted by other choices --- for instance, conditions that do not allow the escape of magnetic flux, or those that allow inflows ($u_{z} z <0$ at the boundaries) --- 
but these mostly cause features that would be expected to be unphysical in a global setting.

\subsection{Initial conditions and numerical setup}\label{subsec: ics}

Following \citet{Salvesen2016,Salvesen2016a}, most of our simulations are initialized 
with the magnetized constant-$\beta$ equilibrium \eqref{eq: Gaussian equilibrium}.  Most simulations 
use a vertical  domain size $z\in(-L_{z}/2,L_{z}/2)$ with $H_{\beta}^{2}=2$ and $L_{z}=10$, giving 
significant space away from the midplane to form a low-$\beta$ corona (it transpires that 
this is necessary to maintain the low-$\beta$ state). Based on the results of
\citet{Simon2013} and subsequent works, we use $L_{x}=L_{z}$ and $L_{y}=2 L_{z}$ for most runs, but also test the
effect of increasing and decreasing both $L_{x}$ and $L_{y}$ by a factor of two (App.~\ref{app: box size dep}).  The equilibrium is then defined 
by its initial $\beta= \beta_{y0}=8\pi P_{\rm th}/B_{y}^{2}$, which sets the simulation's isothermal sound speed as $c_{s}^{2}= \Omega^{2} H_{\beta}^{2}(1+\beta^{-1})^{-1}$. In \cref{tab:sims}, we label simulations by this initial $\beta$, although the midplane $\beta$ invariably 
increases significantly from this initial value as the field redistributes and flux is lost through the boundaries. 

In some simulations, we additionally add a weak initial radial field of strength $B_{x}^{2}=8\pi P_{\rm th}/\beta_{x0}$. This radial field takes the same spatial form as $B_{y}$ (i.e., $B_{x}/B_{y}=\sqrt{\beta/\beta_{x}}$), and is sufficiently weak that it hardly modifies the equilibrium; however, it is sheared at early times, strengthening $B_{y}$, thus effectively adding extra azimuthal flux. 
Since, like $B_{y}$, $B_{x}$ can  escape the boundary, the details of the initial form of $B_{x}$ do not 
persist into the saturated state, other than its role in determining whether the weakly or strongly magnetized solution 
is reached. 
Two other simulations, labelled `-Lor' in \cref{tab:sims}, are initialized from the power-law equilibrium \cref{eq: lorenztian equilibrium} using different $H/H_{\rm th}$, and thus different midplane $\beta$. Finally, some simulations
also include a vertical flux via a constant $B_{z}$, which is conserved through the simulation. These 
are parameterized by $\beta_{z}\equiv 8\pi c_{s}^{2}\rho(z=0)/B_{z}^{2}$ in the standard way.

Nearly all simulations use the piecewise parabolic method with the Van-Leer time integrator and the basic 
HLL Reimman solver. Although this solver is more diffusive than the HLLD solver, we have found this low-$\beta$ turbulence  to be 
rather sensitive to numerical instabilities, also requiring very conservative choices 
for the CFL timestep criterion. The strongly magnetized simulations are also quite expensive due to the small timesteps required to capture the fast Alfv\'en speeds in the very low-$\beta$ coronae.
To help mitigate this, which is made worse by the large density variation in such regions,
 in most simulations we also apply a density floor $\rho_{\rm flr}=10^{-4}\rho_{0}$, as used in previous works\footnote{Other 
approaches include limiting the Alfv\'en speed \citep{Johansen2008} or  a hyper-resistive-like term applied to regions of large $|\nabla\times \bm{B}|/\sqrt{\rho}$ \citep{Bai2013a}. } (e.g., \citealt{Salvesen2016}; here $\rho_{0}$ is initial density maximum).   Unfortunately, this density floor was found to have 
 adversely affected some simulations  by limiting the mean density profile 
far from the midplane. The issue occurs in the high-$\beta$ states at lower $H_{\rm th}$ (see below), which are not 
our main focus in this work, because the thermal  density profile involves an extremely sharp drop off away from the midplane (unlike the low-$\beta$ states, which develop a 
much flatter power-law profile). In  App.~\ref{app: density floor dep}  we explicitly test the effect of changing the density floor on the low-$\beta$ state, observing only tiny changes to mean profiles or transport rates. 

In all simulations except those with net vertical flux, the mass loss over the duration of the simulation is minimal. In the strongest NVF case with $\beta_{z}=100$, we 
implement a mass replenishment  by multiplying the density by a factor chosen to maintain constant total mass at every time step (as done in previous work; \citealp{Bai2013a}).

  Note that, compared to previous works, we have run most simulations for a relatively short time (in units of $\Omega^{-1}$). This is motivated by a desire to redeploy computational
resources to explore {finer} resolutions (as needed for $\beta\ll1$ and small $H_{\rm th}$) and
different simulations, rather than measuring precise values of $\alpha$ via  long-time statistics. In support of this choice, we note that turbulence in stratified shearing boxes does not yet seem to be converged even at exceptionally 
fine resolution \citep{Ryan2017}, and detailed study along such lines is beyond the scope of this work.
To ensure that the system is truly converged in time, we have run one simulation of the low-$\beta$ state to beyond $100$ orbits; this 
revealed a nearly constant statistical steady state with only small 
changes over long timescales.


\subsection{Notation \& Units}\label{subsec: units}

An average over the horizontal directions ($x$ and $y$) is henceforth denoted by $\mean{f}(z) = A_{\perp}^{-1}\int dx\,dy\,f$ 
for any variable $f$ with $A_{\perp}=L_{x}L_{y}$. A full volume average is $\meanz{f}=A_{\perp}^{-1}L_{z}^{-1}\int dx\,dy\,dz\,f$, and 
an average over the thermal midplane is $\meanz{f}_{\rm th}=\tfrac12 A_{\perp}^{-1}H_{\rm th}^{-1}\int_{-H_{\rm th}}^{H_{\rm th}} dx\,dy\,dz\,f$.
These quantities are often also averaged over time after the initial transient phase in each simulation.
``Fluctuations'' are denoted with a $\delta$ and defined as $\delta f\equiv f-\mean{f}$. 
The Alfv\'en speed 
is defined from the mean azimuthal field, which dominates all other mean components, as $\va=\mean{B_{y}}/\sqrt{4\pi\mean{\rho}}$, 
and $\beta$ is defined as usual as the ratio of thermal to magnetic pressure (we average over $x$ and $y$ before taking their ratio).  The dimensionless accretion
stress is  $\alpha = (\meanz{\rho u_{x} u_{y}}-\meanz{B_{x}B_{y}})/\meanz{P_{\rm th}}$ and we also  define 
$\alpha(z) = (\mean{\rho u_{x} u_{y}}-\mean{B_{x}B_{y}})/\meanz{P_{\rm th}}_{{\rm th}}$ to show its $z$ dependence, further breaking
this into $\alpha_{K}\equiv\mean{\rho u_{x} u_{y}}/\meanz{P_{\rm th}}_{{\rm th}}$, $\alpha_{\delta B}\equiv-\mean{\delta B_{x} \delta B_{y}}/\meanz{P_{\rm th}}_{{\rm th}}$, 
and $\alpha_{\mean{B}}\equiv-\mean{ B_{x}}\,\mean{  B_{y}}/\meanz{P_{\rm th}}_{{\rm th}}$ such that $\alpha(z)=\alpha_{K}+\alpha_{\delta B}+\alpha_{\mean{B}}$.\footnote{Note that with these 
definitions, $\meanz{\alpha(z)}\neq\alpha$ because of the different normalizations ($\meanz{P_{\rm th}}$ versus $\meanz{P_{\rm th}}_{{\rm th}}$); however,
this definition of $\alpha(z)$ provides a more intuitive picture of the midplane, which dominates the stress.}

The shearing-box system is normalized to the arbitrary lengthscale $L_{0}$,
yielding freedom to define the simulation's units. To more clearly compare 
cases with strong magnetization, we do not normalize our box lengthscales to 
the thermal scale height $H_{\rm th}$, unlike most previous works. Instead, in most simulations, because  $H_{\beta}^{2}=2$ and $\beta<1$, we have 
 $H_{\rm th}<1$ in box units.
However, everything in the system could, if desired, be rescaled to $H_{\rm th}$, in which case our domain sizes 
would vary significantly between simulations and be much larger than $(L_{x},L_{y},L_{z})=(10,20,10)$ (this would 
also rescale the density and velocity). Indeed, we find that small $H_{\rm th}$ (in box units) seems to be required 
to see the low-$\beta$ state, which is equivalent to saying 
that very large domains $L_{z}\gtrsim 20$ are needed in thermal units.  All plots 
and diagnostics are normalized to be dimensionless so can be meaningfully compared to previous works, showing 
good order-of-magnitude agreement where appropriate (see \cref{tab:sims}).

\section{The Parker and magnetorotational instabilities}\label{sec: linear behavior}


Here we provide an overview of the magnetorotational and Parker instabilities \citep{Parker1958,Balbus1991}, which appear to be the primary linear instabilities driving the dynamics. While these 
instabilities and their relation to each other have been extensively explored in the literature \citep[e.g.,][]{Foglizzo1995,Balbus1998}
we focus here on the less-well-studied regime with $\beta\lesssim 1$ trans- or super-thermal magnetic fields, 
so it is helpful to collect and simplify some past results for more straightforward application to our simulations. 

\subsection{Magnetorotational instability}\label{subsec: MRI}

The MRI is driven by the energy in the background shear flow (unrelated to vertical stratification), and
can be derived from \crefrange{eq:MHD rho}{eq:MHD B} by linearizing the 
equations near the midplane, dropping all $z$ dependence. Assuming
an equilibrium with constant azimuthal and vertical fields and no flows other than the background 
shear, we take solutions of the form $\delta f(\bm{x})=\delta \hat{f} \exp(i k_{x}x+i k_{y}y+ik_{z}z-i\omega t)$
in order to derive a dispersion relation for $\omega(\bm{k}) $. Due to the
background shear, $k_{x}$ here should really be considered a function of time when $k_{y}$ is non-zero (so called shearing waves with $k_{z}= k_{x0}+q\Omega k_{y}$; \citealp{Johnson2007,Lesur2008,Squire2014})
but we ignore this effect here, effectively considering mode growth over short timescales.
The full dispersion relation is complicated, but by assuming $B_{y}\gg B_{z}$ (and $\omega\sim \Omega$) it simplifies 
to \vspace{-2pt}
\begin{equation}
\bar{\omega}^{4}-\bar{\omega}^{2} \left( 4-2q+n^{2} \frac{1+2x}{1+x}\right) + n^{2}(n^{2}-2q) \frac{x}{1+x}=0,\label{eq: mri dispersion rel}
\end{equation}
 where $\bar{\omega}=\omega/\Omega$, $n = (k_{y}v_{{\rm A}y}+k_{z}v_{{\rm A}z})/\Omega$, and $x=c_{s}^{2}/v_{{\rm A}y}^{2}\approx \beta$ following the notation of \cite{Begelman2023} 
 (expanding \eqref{eq: mri dispersion rel} in $x\ll1$ with $\bar{\omega}\sim x^{1/2}$ yields their equation (12) with $y=0$).
  The full solutions to \cref{eq: mri dispersion rel} are unhelpful, but can be expanded in various limits to yield simple 
  results.
  
For high $\beta$ ($x\gg1$), \cref{eq: mri dispersion rel} recovers standard MRI results, which are independent of $v_{{\rm A}y}$ other than through $n$  \citep{Balbus1998}: for $0\leq q<2$, $\omega$ is 
   imaginary for $|n|<\sqrt{2q}$ (a purely growing mode), with  ${\rm Im}(\bar{\omega})=n\sqrt{q/(2-q)}$ at $n\ll1$ (long wavelengths) and a peak growth rate ${\rm Im}(\bar{\omega})=q/2$ at wavelength $n^{2}=q-q^{2}/4$. For Keplerian rotation, $q=3/2$, this yields ${\rm Im}(\bar{\omega})=3/4$ at $n=\sqrt{15}/4$.
At low $\beta$ ($x\ll1$) the shape of $\bar{\omega}(n)$ is very similar and it remains 
  purely growing for $|n|<\sqrt{2q}$, but the growth rate decreases to ${\rm Im}(\bar{\omega})=n\sqrt{xq/(2-q)}$ at $n\ll1$, giving ${\rm Im}(\bar{\omega})=\sqrt{3x}n$ at $q=3/2$ \citep{Begelman2023}. Its peak growth rate shifts to ${\rm Im}(\bar{\omega}) =  \sqrt{2x}\sqrt{4-q-2\sqrt{4-2q}}$ at wavenumber 
  $n^{2}=2(q-2+\sqrt{4-2q})$. For $q=3/2$, these evaluate to ${\rm Im}(\bar{\omega}) =\sqrt{x}$ at $n=1$.
   With $q=0$ (no shear) there is no instability, while with no rotation ($q\rightarrow \infty$ with $q \Omega={\rm const.}$) 
  the growth rate peaks instead  at $n=0$ but scales as ${\rm Im}({\omega})\propto 1/q$ (meaning it disappears in the 
  non-rotating limit at both small and large $\beta$).

Overall, we see the only major change to the MRI  with strong ($\beta\ll1$) azimuthal fields is that the growth rate is suppressed by a factor $\sqrt{x}\approx \sqrt{\beta}$ compared to the $\beta>1$ instability. 
This will presumably suppress the level of turbulence compared to the background magnetic field $v_{{\rm A}y}$ at low $\beta$ compared 
to the standard instability.
The fastest growing MRI mode at $n\approx 1$ has azimuthal scale $k_{y}^{-1}\approx v_{{\rm A}y}/\Omega \sim H_{\rm th}/\sqrt{\beta}$ so should be well resolved in our low-$\beta$ simulations, which end up with $\beta\lesssim 1$ in the midplane (less so, perhaps, in the high-$\beta$ simulations, if considering the mean-field $\beta$; but for these our parameters are mostly similar to previous work,
which exhibit well-known convergence issues; \citealp{Ryan2017})

Also worth mentioning are  other related, but fundamentally global instabilities  at $\beta\ll1 $, which
were first characterized  by \citet{Pessah2005} (the relevant dispersion relation was derived but left unexplored in \citealt{Chandrasekhar1961}; see also \citealt{Knobloch1992,Gammie1994,Das2018}). 
These instabilities  are current driven, depending on the radial gradient of the (super-thermal) toroidal field, 
 \vspace{5pt}
\begin{equation}
\hat{B}\equiv \frac{d\ln B_{\phi}}{d\ln R},
\end{equation}
and reduce to the local MRI  when $\hat{B}=-1$ such that $\bm{J}\propto \nabla\times (B_{\phi}\hat{\bm{\phi}})\approx0$  (this point is 
not obvious in \citealt{Pessah2005}, who set $\hat{B}=0$). \citet{Begelman2023}
give a simplified and more complete treatment, showing that their existence is controlled 
by the parameter $y\equiv (1+\hat{B})^{2} x^{-2} H_{\rm th}^{2}/R^{2}$,  with a maximum growth rate that 
can be much faster than the standard low-$\beta$ MRI when $y>1$ and is not suppressed by $\sqrt{\beta}$ at low $\beta$. 
Of course, in the local simulations studied here, these instabilities do not exist,
and we should thus think of the standard shearing box as being relevant only to the current-free case with $\hat{B}=-1$, in which case the 
global treatment reduces to the local one discussed above.

  
\subsection{Parker instability}\label{subsec: parker}

The Parker instability \citep{Parker1958} is driven 
by the vertical stratification of the magnetic field. Most commonly, it is 
studied while ignoring the effect of the radial shear;  indeed, \citet{Shu1974} argued that
the instability cannot ever be fully stabilized by shear or rotation, because the small-scale radial
modes remain unstable (although the growth rates of longer-wavelength modes can be suppressed and shear can 
play an important role by changing the radial wavenumber in time; \citealp{Foglizzo1994,Foglizzo1995}). 
Convenient expressions are found in \citet{Goedbloed2004} Chapter 7,
who consider a gravitating slab threaded by a perpendicular magnetic field $B_{y}(z)$ (without shear or rotation).
Because the background equilibrium and gravitational acceleration $\hat{g}=-\Omega^{2}z$ are $z$ dependent, a correct treatment requires solving the boundary-value 
differential equation in $z$ \citep{Horiuchi1988a}; however, useful insight can be gained by considering local modes that 
vary rapidly over the length scale $L\sim z\sim B_{y}/B_{y}'$ on which the background varies (i.e., $\delta f\sim\exp(i\kappa z)$ with 
$\kappa L\gg1$). This yields an effective local dispersion relation for $\omega(k_{x},k_{y},\kappa)$, 
which shows that two different types of modes can dominate the instability in different regimes: (i) the ``pure interchange mode'' 
with $k_{\|}=k_{y}=0$, which does not bend the magnetic field, and (ii) the ``quasi-interchange mode,''
with $k_{x}\gg k_{y}$ but $k_{y}\neq 0$, which bends the field. We obtain simple  expressions
below by simplifying those of \citet{Goedbloed2004} assuming an isothermal equation of state, a $\beta\ll1$ azimuthally directed mean field, and linear vertical gravity $\hat{g}=-\Omega^{2}z$.\footnote{In the notation of \citet{Goedbloed2004}, the disk atmosphere of interest becomes $\Gamma=-\hat{g}\,d\ln\rho/dz$,
$\Gamma_{B} = c_{s}^{-2}\hat{g}^{2}$, $\Gamma_{0}=c_{s}^{-2}\hat{g}^{2}/(1+2/\beta)^{2}$, with $X^{2}\equiv \Gamma/\Gamma_{B}$. The 
quasi-interchange  dominates for $\Gamma_{0}\leq \Gamma < \Gamma_{B}$ and the pure interchange
dominates for $\Gamma \leq \Gamma_{0}$.}
\revchng{In a true global disk, the vertical gravitational force and rotation change with height far from the midplane; given our  neglect of rotation and local treatment, the resultant modifications to the Parker instability can be captured by rescaling $\hat{g}$ \citep{Horiuchi1988a}.}

Considering $z>0$ regions with  $d\ln \rho /dz<0$ for concreteness,
 the general instability criterion becomes simply $d P_{B}/dz<0$. 
Defining  $X^{2}\equiv |d\ln\rho/d\ln z|\,H_{\rm th}^{2}/2z^{2}$, one finds quasi-interchange modes dominate when $X> \beta/(2+\beta)$,
while interchange modes dominate when $ X <   \beta/(2+\beta)$. Noting that in low-$\beta$ regions away from the midplane, $1\gg\beta\propto z^{-2}$ while $|d\ln\rho/d\ln z|\sim {\rm const.}$, we see that  the quasi-interchange modes should dominate. 
Their maximum growth rate is,\vspace{2pt}
\begin{equation}
{\rm Im}(\omega)= \frac{\hat{g}}{\va} (1- X),\label{eq: parker growth rate}
\end{equation}
which, for the low-$\beta$ power-law profile \eqref{eq: pure power law equilibria} is nearly constant with $z$ because $\va= c_{s}/\sqrt{2\beta}\propto z$ and $X\ll1$ at $z\gg H_{\rm th}$.
Simplifying various expressions from \citet{Goedbloed2004}, we find that the maximum   growth rate occurs for parallel wavenumber
 \vspace{-7pt}
\begin{align}
k_{\gamma,{\rm max}}&=\frac{\sqrt{2}z}{H_{\rm th}} \frac{\Omega}{\va}[(1+\beta)X-(1+\beta/2)X^{2}-\beta/2]^{1/2}\nonumber\\
&
\approx \bigg( \frac{{a\sqrt{1+a/2}}}{{z H_{\rm th}}}\bigg)^{1/2}\quad \text{for}\: z\gg H_{\rm th},\,\beta\ll1,\label{eq: parker wavenumber}
\end{align}
where in the second line we used the power-law equilibrium \eqref{eq: pure power law equilibria}, which fixes $\va=\sqrt{2/a}\Omega z$.
This applies at short radial length scales (there is no explicit dependence on $k_{x}$ but shear and rotation affect small-$k_{x}$ modes more) and small 
$\kappa$ (large vertical length scales). Although the latter condition could be problematic for the local approximation,   the growth rate becomes independent of $\kappa$ for $\kappa\ll k_{x}$. This fastest-growing mode is captured by the azimuthal box size in our low-$\beta$ simulations, since its lengthscale  is 
smaller than $z$ for $z\gg H_{\rm th}$ ($k_{y}^{-1}\sim\sqrt{z H_{\rm th}}$) and $L_{y}=2L_{z}$ in most simulations. 


\begin{figure*}
\includegraphics[width=1.0\textwidth]{\ffold 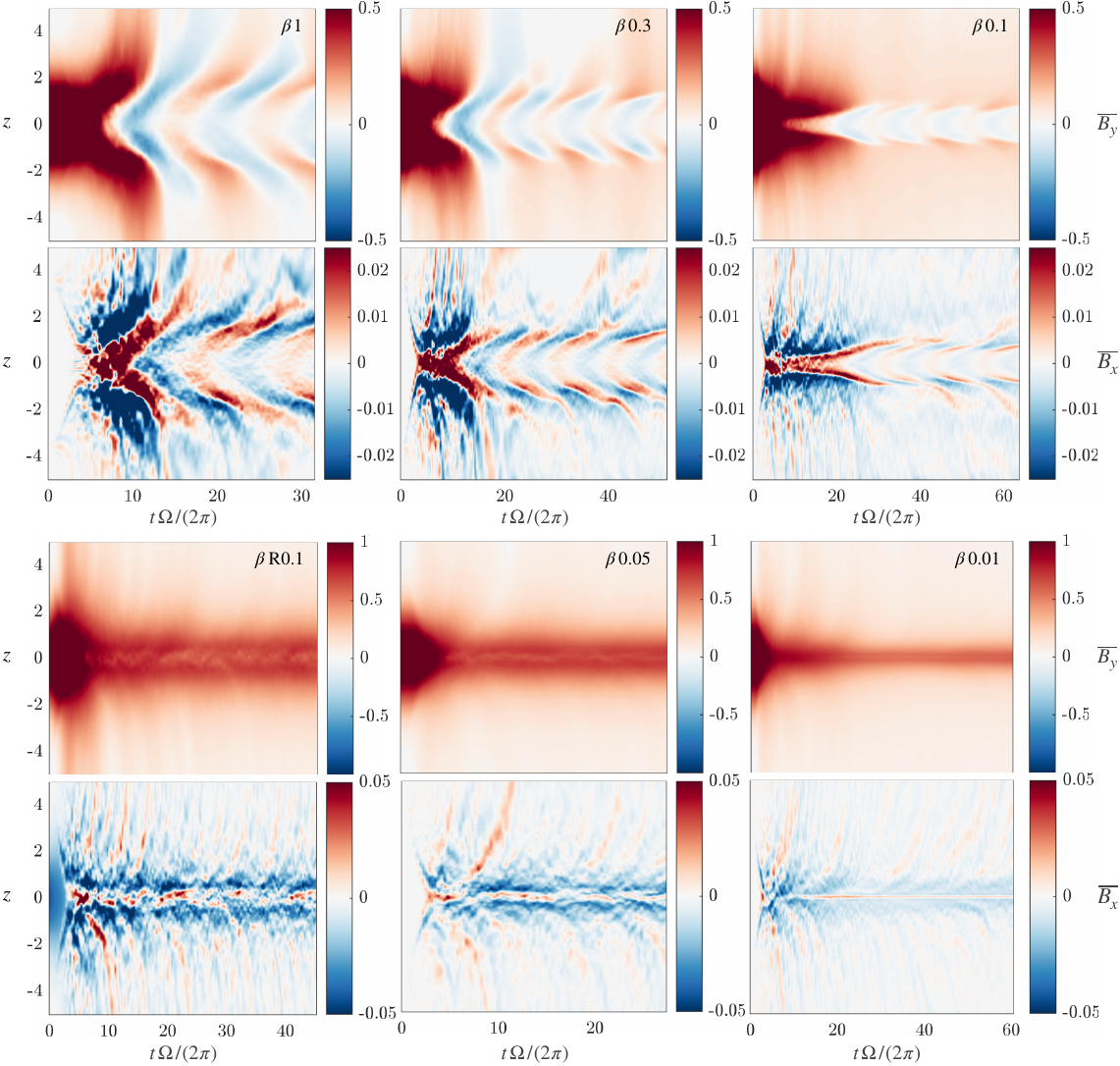}\\
\caption{Space-time `butterfly' plots for a range of ZNVF simulations with different initial conditions and sound speeds. Each case 
shows $\mean{B}_{y}(z,t)$ in the top subpanel and $\mean{B}_{x}(z,t)$ in the bottom subpanel. In the top row, we see cases that transition 
into the high-$\beta$ weakly accreting state ($\beta$1, $\beta$0.3, and  $\beta$0.1) because they do not retain sufficient magnetic flux during their initial relaxation phase; these develop  into well-studied  dynamo cycles. The bottom panels show various cases that maintain the lower-$\beta$ strongly accreting state ($\beta$R0.1, $\beta$0.05, and  $\beta$0.01); these maintain a quasi-steady state with $\beta\lesssim 1 $ in the midplane, losing significantly less flux through the boundaries during the initial phases and then maintaining a strong mean field in steady state.  }
\label{fig: butterfly}
\end{figure*}

\section{Results}\label{sec: results}

\subsection{Transition between the low-$\beta$ and high$-\beta$ states}

\label{sec: transition}

The Gaussian  initial conditions used for most simulations are in approximate vertical equilibrium but highly unstable to the  Parker instability and MRI.
The system thus rapidly reorganizes over the course of $2$-$3$ orbits, transporting significant magnetic flux to larger $z$ where it can escape out of the top and bottom boundaries, thereby organizing into a lower-$\beta$ corona if the vertical domain is sufficiently large. During this reorganization phase, around half of the initial azimuthal magnetic flux is lost from the domain if starting from 
Gaussian initial conditions (this fraction seems to be independent of $\beta$), while for the quasi-Lorentzian initial conditions
of \cref{eq: lorenztian equilibrium} the fraction lost is lower because the initial state is rather similar to the final one. 
As per points 1-2 in \S\ref{sec: key points}, we find that two very different quasi-steady states result after it settles, the transition between 
which seems to be controlled by the midplane $\beta$ following the transient reorganization phase. If too little flux remains 
after this phase, it all escapes. Alternatively, if the system remains sufficiently magnetized, the flux's escape is balanced by its regeneration and is maintained in steady state.  
For constant-$\beta$ Gaussian initial conditions, we find that the tipping point occurs for an initial $\beta\simeq 0.1$: higher-$\beta$ initial conditions 
transition into the high-$\beta$ state, lower-$\beta$ ones transition into the low-$\beta$ state, and those with $\beta=0.1$ are sensitive to the details, as described below. \revchng{Note, however, that this $\beta\simeq0.1$ threshold is specific to the choice of the constant-$\beta$ Gaussian: for the Lorentzian-like form \eqref{eq: lorenztian equilibrium}, the system transitions into the low-$\beta$ state from larger initial midplane $\beta$ (e.g., $\approx0.35$ for $\beta$0.1-LorH) because the vertical equilibrium starts closer to its steady state structure.   }


We illustrate this transition  in \cref{fig: butterfly}, which shows so-called ``butterfly diagrams'' of the time evolution 
of $\mean{\bm{B}}(z)$ (the $x,y$ mean of $\bm{B}$).  
 The top panels show cases that  transition into the
high-$\beta$ state, which is characterized by dynamo cycles that quasi-periodically  reverse the azimuthal field, in good agreement with previous work \citep{Simon2012,Salvesen2016}. 
The three cases shown, $\beta$1, $\beta$0.3, and $\beta$0.1 have the same initial $H_{\beta}$ but successively decreasing initial $\beta$, and thus
$H_{\rm th}$,  in the Gaussian initial conditions (\cref{eq: Gaussian equilibrium}). Each case loses nearly all of its initial  midplane magnetic flux to yield a thermally dominated 
midplane. The dynamo
cycles are well formed and coherent, each with a similar period of $\simeq 10$ orbits, showing that the cycle period 
is  independent of the sound speed (as expected since similar physics seems to occur in homogenous and/or incompressible models; \citealp[e.g.,][]{Lesur2008a,Shi2016}). 

The lower panels show results initialized with more flux, which instead sustain the low-$\beta$ state. The first, $\beta$R0.1 is 
identical to $\beta$0.1 except that it is also initialized with a mean radial field with $\beta_{x}=100$ (the sign is  such that $B_{x}B_{y}<0$ as  
required to strengthen the initial $B_{y}$). 
The dynamics are very different to $\beta$0.1 --- there are no cycles and the system maintains a much stronger $B_{y}$ in the midplane. 
Lower-initial-$\beta$ simulations  with smaller $H_{\rm th}$ ($\beta$0.05
and $\beta$0.01) reach a very similar state, albeit with a narrower midplane peak because this width is ultimately set by $H_{\rm th}$ (see below).
Changes to  $\mean{B_{y}}$ in time are small amplitude, slow, and  seemingly random. There also occurs times over which the midplane field  strengthens, showing that these simulations are not simply taking longer to lose their flux. 
Initial conditions with $\beta=0.1$ seem to lie on the boundary between reaching the low- and high-$\beta$ states. This is evidenced 
by the fact that $\beta$0.1 loses its flux significantly more slowly than $\beta$1 and $\beta$0.3,
and a coarser-resolution version of the same setup ($\beta0.1$-mr) reaches the low-$\beta$ state. This latter 
outcome appears to be a consequence of the expulsion of the 
flux being somewhat less efficient at coarser resolution: when initialized with an additional very weak ($\beta_{x}=1000$) radial field 
that weakens the initial $B_{y}$ (i.e., $B_{x}B_{y}>0$), the same setup instead transitions into the high-$\beta$ state (simulation $\beta$nR0.1-mr).
The simulation $\beta$0.1-mr has been run over a significantly longer time, around $100$ orbits, in order 
to test for sudden or long-term changes to the low-$\beta$ state. Its lack of significant change over this time
suggests that the system can maintain the 
low-$\beta$ state indefinitely.


\begin{table*}
\tiny
\begin{center}
\scalebox{1}{
 \begin{tabular}{c c c c c c c c c c c}  Name  & $\beta_{y0} $ & $\beta_{x0} $ &  $\beta_{z} $ &  $H_{\rm th}=\sqrt{2}\tfrac{c_{s}}{\Omega} $ & $N_{x}\times N_{y}\times N_{z}$ & $(t_{\rm av},t_{\rm fin})\tfrac{\Omega}{2\pi}$ & BCs &  Notes & Outcome & $\langle \alpha\rangle$ \\ [0.5ex] 
 \hline\hline
 $\beta$1 & 1 & $\infty$ & $\infty$ & 1 & $448^{3}$ & (20,32) & Outflow &  & High-$\beta$  &  0.039  \\
\hline
 $\beta$0.3 & 0.3 & $\infty$ & $\infty$ & 0.68 & $448^{3}$ & (20,52) & Outflow &  & High-$\beta$  &  0.029  \\
 \hline
$\beta$0.1 & 0.1 & $\infty$ & $\infty$ & 0.43 & $448^{3}$ & (30,62) & Outflow &  & High-$\beta$  &  0.017  \\
\hline
$\beta$R0.1 & 0.1 & 100 & $\infty$ & 0.43 & $448^{3}$ & (10,45) & Outflow &  & Low-$\beta$  &  0.62  \\
\hline
$\beta$0.1-mr & 0.1 & $\infty$ & $\infty$ & 0.43 & $336^{3}$ & (10,100) & Outflow &  & Low-$\beta$  &  0.49  \\
\hline
$\beta$nR0.1-mr & 0.1 & $1000$ & $\infty$ & 0.43 & $336^{3}$ & (30,64) & Outflow & $B_{x} B_{y}>0$ & High-$\beta$  & <0.02  \\
\hline
$\beta$0.05 & 0.05 & $\infty$  & $\infty$ & 0.31 & $448^{3}$ & (10,27) & Outflow &  & Low-$\beta$  &  0.57  \\
\hline
$\beta$0.01 & 0.01 & $\infty$  & $\infty$ & 0.14 & $336^{3}$ & (10,60) & Outflow & $H_{\rm th}$ marg.~resolved   & Low-$\beta$  &  0.61  \\
\hline
$\beta$0.01-hr & 0.01 & $\infty$  & $\infty$ & 0.14 & $672^{3}$ & (4,8) & Outflow & Restart from $\beta$0.01 & Low-$\beta$  &  0.46  \\
\hline
$\beta$0.001 & 0.001 & $\infty$  & $\infty$ & 0.045 & $336^{3}$ & (10,18) & Outflow & $H_{\rm th}$ unresolved & Low-$\beta$  &  15  \\
\hline\hline
$\beta$0.1-H4 & 0.012 & $\infty$ & $\infty$ & 0.43 & $336^{3}$ & (18,23) & Outflow &  & Low-$\beta$  &  0.35  \\
\hline
$\beta$0.1-LorH & $\approx 0.35$ & $\infty$ & $\infty$ & 0.43 & $336^{3}$ & (20,40) & Outflow & \cref{eq: lorenztian equilibrium}: $a=3$, $H=1$ & Low-$\beta$  &  0.47  \\
\hline
$\beta$0.1-LorL & $\approx 3$ & $\infty$ & $\infty$ & 0.43 & $336^{3}$ & (30,42) & Outflow & \cref{eq: lorenztian equilibrium}: $a=2$, $H=0.45$ & High-$\beta$  &  0.024  \\
\hline
$\beta$R0.1-S13 & 0.1 & 100  & $\infty$ & 0.43 & $336^{3}$ & (10,18) & {Simon13} &  & Low-$\beta$  &  0.81  \\
\hline
$\beta$R0.1-L13 & 0.1 & 100  & $\infty$ & 0.43 & $336^{3}$ & (10,19) & {Lesur13} &  & Low-$\beta$  &  0.64  \\
\hline
$\beta$R0.1-PL & 0.1 & 100  & $\infty$ & 0.43 & $336^{3}$ & (10,20) & Power-law &  & Low-$\beta$  &  0.68  \\
\hline
$\beta$R0.1-rec2 & 0.1 & 100  & $\infty$ & 0.43 & $336^{3}$ & (10,14) & Power-law & HLLD $2^{\rm nd}$-order & Low-$\beta$  &  0.74  \\
\hline
$\beta$0.1-tall & 0.1 & $\infty$  & $\infty$ & 0.43 & $224^{2}\times336$ & (15,27) & Outflow & $L_{z}=7.5$, $\rho_{\rm flr}=10^{-9}$ & Low-$\beta$  &  0.14  \\
\hline\hline
$\beta$R0.1-NVF3 & 0.1 & 100 & $1000$ & 0.43 & $336^{3}$ & (10,24) & Power-law &  & Low-$\beta$  &  0.75  \\
\hline
$\beta$R0.1-NVF2 & 0.1 & 100 & $100$ & 0.43 & $336^{3}$ & (10,24) & Power-law & Density replenished & Low-$\beta$  &  2.1  \\
\hline\hline
\end{tabular}
}
\end{center}\caption{A list of the main simulations analyzed in this work with relevant parameters, initial conditions, and other properties. 
The simulation Name is chosen based on the initial $\beta$ in a Gaussian equilibrium \eqref{eq: Gaussian equilibrium}  with $H_{\beta}=\sqrt{2}$, 
thus effectively providing a correspondence  with the thermal scale height in box units (column 5). The seventh column lists the final time and the ``steady state'' time $t_{\rm av}$, with
averages performed for $t\in (t_{\rm av}, t_{\rm fin})$. Boundary conditions (``BCs'') are described in \S\ref{subsec: bcs}. 
``Outcome'' lists whether the simulation reaches the low- or high-$\beta$ state, which is determined by its time-averaged profile and $\alpha$ (final column). Other simulations used for numerical tests are listed in \cref{tab:sims app}.}
\label{tab:sims}
\end{table*}

\normalsize
\begin{figure*}
\includegraphics[width=1.0\textwidth]{\ffold 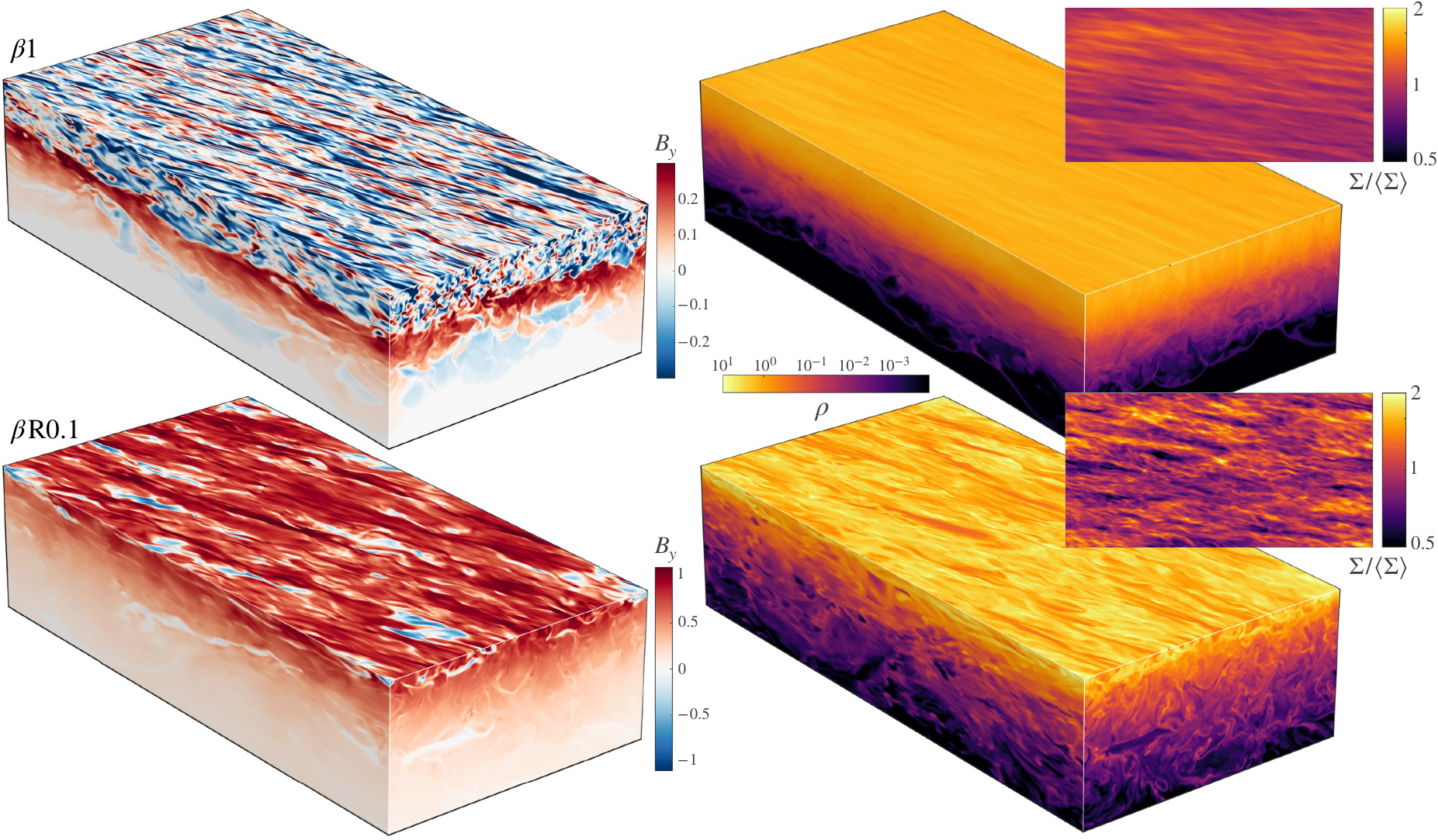}\vspace{-0.06cm}\\
\caption{Azimuthal-field (left) and density (right) snapshots from the quasi-steady state ($t\Omega/2\pi\approx 30$) of $\beta$1 in the high-$\beta$ state (top panels) and $\beta$R0.1
in the low-$\beta$ state (bottom panels). Each 3-D rendering shows just the bottom half of the domain to illustrate the midplane structure, and 
the inset on the right-hand panels shows the surface-density fluctuations ($\Sigma\equiv\int dz\rho$). The  structures are very different 
between the two states, with significantly larger fluctuations in density in the low-$\beta$ state due to its more vigorous turbulence. }
\label{fig: snapshot}
\end{figure*}

The dramatically different appearance of the low- and high-$\beta$ states is illustrated in \cref{fig: snapshot}.
The azimuthal field structure is very different: the high-$\beta$ state is characterized by a super-Alfv\'enic midplane 
with $\delta \bm{B}\gg \mean{B}_{y}$ and  small-scale fluctuations (their scale is likely resolution dependent; see \citealt{Ryan2017});  the low-$\beta$ state is mean-field 
dominated ($\delta \bm{B}< \mean{B}_{y}$) but clearly quite intermittent with small localized regions where the 
field reverses. Likewise, the density, which is much closer to perfectly thermal  (Gaussian) in 
the high-$\beta$ than the low-$\beta$ state, looks very different, with much larger fluctuations in the midplane 
at low $\beta$ because of its stronger turbulence. This also causes larger surface-density variation in the low-$\beta$ state (see insets).

\begin{figure}
\includegraphics[width=1.0\columnwidth]{\ffold 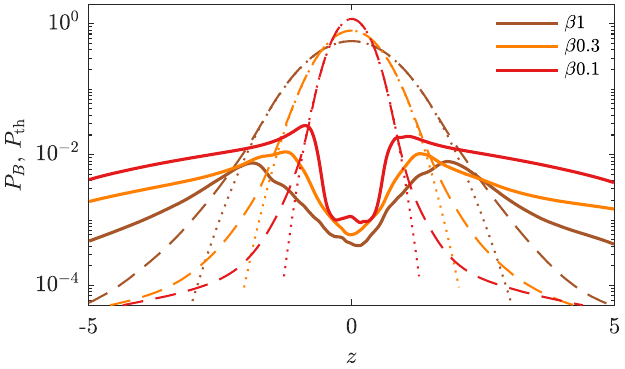}\vspace{-0.8cm}
\includegraphics[width=1.0\columnwidth]{\ffold 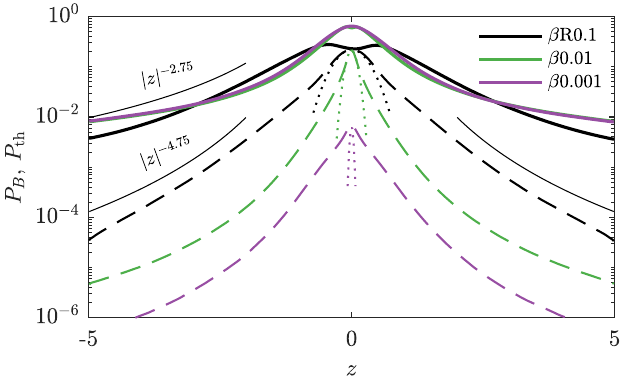}\\
\caption{In each panel,  solid lines show the average-field magnetic pressure $P_{B}=\mean{B_{y}}^{2}/8\pi$ and dashed lines
show the thermal pressure $P_{\rm th}=c_{s}^{2}\mean{\rho}$, each averaged over the quasi-steady state. In the top 
panel, we show simulations that reach the high-$\beta$ cycles (with the time average taken over these high-$\beta$ cycles); in the bottom panel we show several that reach the low-$\beta$ state.
In each case, the dotted line shows the thermal solution, $\rho\propto \exp(-z^{2}/H_{\rm th}^{2})$ and the total 
normalization is such that $\int dz\, (P_{\rm th}+P_{B})=1$.
}
\label{fig: z profiles}
\end{figure}

\begin{figure}
\includegraphics[width=0.987\columnwidth,right]{\ffold 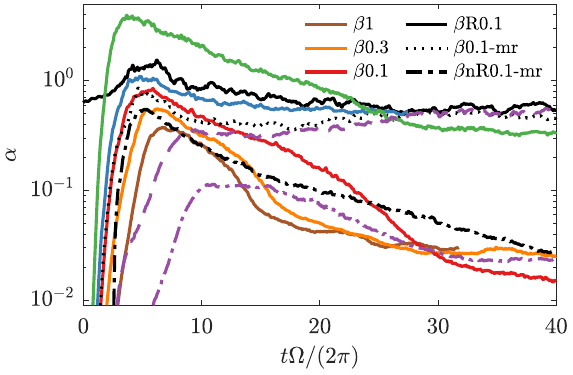}\vspace{-1.06cm}\\\includegraphics[width=0.997\columnwidth,right]{\ffold 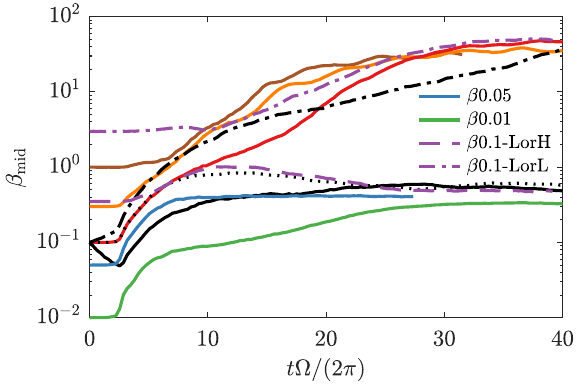}\vspace{-0.90cm}\\\includegraphics[width=1.0\columnwidth,right]{\ffold 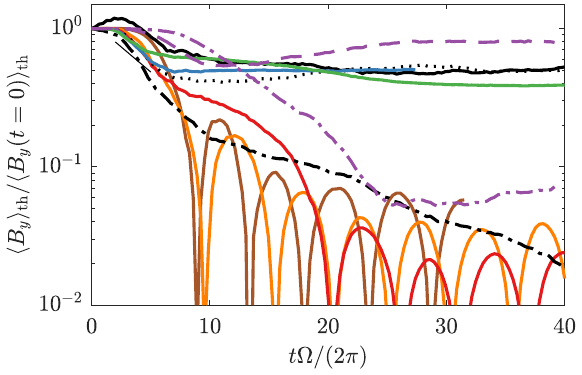}\vspace{-0.46cm}
\begin{flushright}
\includegraphics[width=0.94\columnwidth]{\ffold 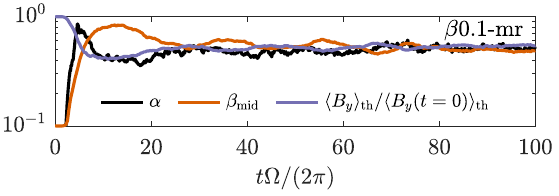}\vspace{-0.06cm}
\end{flushright}
\caption{Time evolution of various important quantities from the simulations shown in \cref{fig: butterfly}, as well as several others 
that illustrate important features. The top panel shows the total accretion stress {$\alpha$}, and the second and third panels show the midplane $\beta$ and magnetic flux $\meanz{B_{y}}_{\rm th}/\meanz{B_{y}(t=0)}_{\rm th}$, respectively, 
where the midplane is defined here as the region $|z|\leq H_{\rm th}$ (the thin black line on $\langle{B_{y}}_{\rm th}\rangle/\langle{B_{y}(t=0)}_{\rm th}\rangle$ shows $2\pi \Omega^{-1} d\ln {\meanz{B_{y}}_{\rm th}}/dt\approx  1/6$ for later discussion).
We see clearly bimodal behavior in all quantities for various initial conditions and resolutions, 
although which  state it reaches can have dependence on numerical parameters for cases near the transition (cf. $\beta$0.1 and $\beta$0.1-mr).
The  initial $\beta_{\rm mid}$ is higher for the cases starting from Lorentzian-like initial conditions  because 
this equilibrium already has a strongly magnetized corona at larger $z$. The
bottom panel shows the same quantities for $\beta$0.1-mr over its full time duration of ${\simeq}100 $ orbits. We do not observe any long-term or sudden changes to the low-$\beta$ state. 
}
\label{fig: flux beta alpha}
\end{figure}

The aforementioned behavior is illustrated more quantitatively in \cref{fig: z profiles,fig: flux beta alpha},
which show, respectively,  time-averaged density and magnetic-pressure profiles, and the time evolution of
key quantities. The former shows how high-$\beta$ profiles are thermally 
dominated out to $\simeq 2H_{\rm th}$, with a lower-$\beta$ corona above this (note that $P_{B}$ here is defined using the mean fields as $P_{B}=\mean{B_{y}}^{2}/8\pi$ averaged over time; when fluctuations 
are also included in $P_{B}$,  $\beta\approx 40$ in the midplane).\footnote{The mean density 
profile in all three cases shown here approaches the artificial density floor of $10^{-4}$ at the $z$ boundaries.
This may artificially flatten the density profile and, as a consequence,  $P_{B}$ at large $z$, thus also
reducing high-$z$ turbulent fluctuations  due to the weaker Parker instability.}
The low-$\beta$ state is magnetically dominated everywhere, though increasingly so at large $z$, where the profiles are approximate
power laws with an index $a\approx 2.75$ in \cref{eq: pure power law equilibria}. The lowest-$\beta$ initial conditions ($\beta$0.01 and $\beta$0.001) also retain 
$\beta\ll1$ midplanes in steady state, although we argue below that this is likely a consequence of not adequately resolving $H_{\rm th}$, 
which  sets the shape of $\rho$ for $z\lesssim H_{\rm th}$ (see dotted lines for $\beta$R0.1).
The small dip in $P_{B}$ in the midplane is a persistent feature when properly resolved; this is a consequence  of the pressure support being 
thermal rather than magnetic at the midplane and also implies that the Parker instability is quenched there since $z dP_{B}/dz>0$.
\revchng{By solving \cref{eq: equilibrium} for $\mean{B_{y}}$  from the measured $\mean{\rho}$, then comparing this to the measured $\mean{B_{y}}$ (or vice versa from $\mean{B_{y}}$ to $\mean{\rho}$), we find that the mean low-$\beta$ profiles in \cref{fig: z profiles} 
are almost exactly equilibria (not shown). This is expected because the turbulence remains sub-Alv\'enic, justifying directly the magneto-static approach of \S\ref{subsec: equilibria}.}

\Cref{fig: flux beta alpha} shows the time evolution of the total accretion stress $\alpha = (\meanz{\rho u_{x} u_{y}}-\meanz{B_{x}B_{y}})/\meanz{P_{\rm th}}$, the midplane $\beta$ including  fluctuations ($\beta_{\rm mid}\equiv 8\pi c_{s}^{2}\meanz{\rho}_{\rm th}/\meanz{B^{2}}_{\rm th}$), and the midplane $\mean{B_{y}}$ for the simulations shown in \cref{fig: butterfly}
as well as several other illustrative examples. 
Most importantly, we see a large ``gap'' between the low-$\beta$ and high-$\beta$ states for each quantity. As well as clearly illustrating 
the bimodality in $\beta$ between the two states, these show more concretely that (i) the low-$\beta$ state supports much higher accretion rates and (ii) the low-$\beta$ state retains much more of the 
midplane magnetic flux from the initial conditions, whereas nearly all of the flux  ($\gtrsim 95\%$) is lost in 
the evolution towards the high-$\beta$ state. It also highlights how the $\beta=0.1$, $H_{\beta}=\sqrt{2}$ initial conditions 
($\beta$0.1, $\beta$R0.1, $\beta$0.1-mr, and $\beta$nR0.1-mr)
are sensitive to slight changes in resolution and initial conditions: $\beta$0.1 transitions to the high-$\beta$ state rather slowly, while $\beta$R0.1 stays at low $\beta$ due to the extra contribution from $B_{x}$ \revchng{that initially strengthens $B_{y}$ ($B_{x}B_{y}<0$)};  $\beta$0.1-mr (at coarser resolution)
evolves back to lower $\beta$ and higher $\alpha$ after initially evolving similarly to $\beta$0.1, but adding a very weak  radial field with $B_{x}B_{y}>0$ \revchng{such that $B_{y}$ is initially made weaker} ($\beta$nR0.1-mr) causes it to transition to the high-$\beta$ state (note that this $B_{r}$ with $\beta_{x}=1000$, is weaker than that which occurs self-consistently in the low-$\beta$ saturated state).
We also show two cases that start instead from the quasi-Lorentzian initial conditions of \cref{eq: lorenztian equilibrium}. Because these profiles are already magnetically dominated at large $z$, they evolve less violently over the 
initial phases. Nonetheless, the bimodal behavior persists, with the lower-$\beta$ initial condition maintaining 
the strongly magnetized state and $\mean{B_{y}}$ increasing, while the higher-$\beta$ initial condition loses nearly 
all of its midplane flux.\footnote{At $336^{3}$ resolution and  $H_{\rm th}=0.43$, the high-$\beta$ state only sometimes  
produces well-formed cyclic behavior. Given the coarse effective resolution of the thermal midplane, and previous 
results documenting the resolution and box-size dependence of MRI turbulence \citep[e.g.,][]{Davis2010,Ryan2017}, this is not unexpected. 
}

The bottom panel of  \cref{fig: flux beta alpha} shows the evolution of the same three 
quantities for the longest-run simulation ($\beta$0.1-mr). After its initial dip in $\alpha$ (described above), 
the system remains remarkably steady, other than low-amplitude, slow (${\simeq}20$ orbit-period) oscillations in the field and $\beta_{\rm mid}$. Importantly, there is no evidence for the system transitioning out of the low-$\beta$ state once it is reached, although
we obviously cannot rule out the possibility that this could occur over longer timescales.

\begin{figure}
\includegraphics[width=1.0\columnwidth]{\ffold 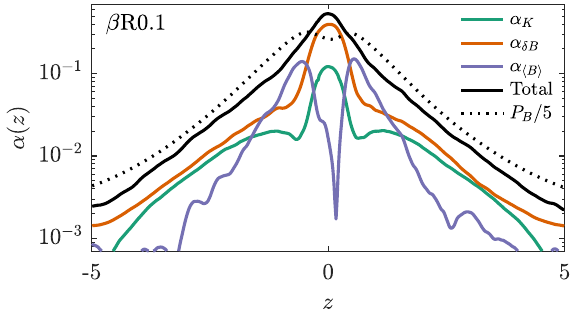}\vspace{-0.95cm}
\includegraphics[width=1.0\columnwidth]{\ffold 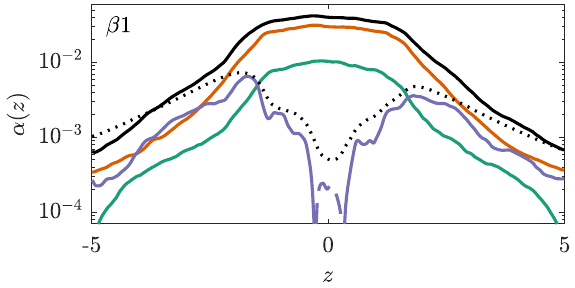}\\
\caption{Spatial profile of the various components of the accretion stress in the low-$\beta$ state  ($\beta$R0.1; top) \revchng{and the high-$\beta$ state ($\beta$1; bottom)}. We split $\alpha(z)$ into $\alpha_{K}\equiv \mean{ u_{x}u_{y}}$, $\alpha_{\delta B}= - \mean{\delta B_{x} \delta B_{y}}$, and $\alpha_{\mean{B}}=- {\mean{B_{x}}\,\mean{B_{y}}}$, each normalized by $ \meanz{\rho}_{\rm th}c_{s}^{2}$. Magnetic contributions are dominant, with the total stress following a similar profile to $P_{B}$ except right 
in the midplane. \vspace{0.1cm} }
\label{fig: alpha}
\end{figure}
\begin{figure}
\includegraphics[width=1.0\columnwidth]{\ffold 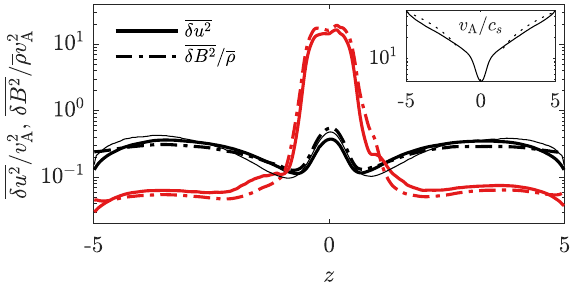}\vspace{-0.9cm}
\includegraphics[width=1.0\columnwidth]{\ffold 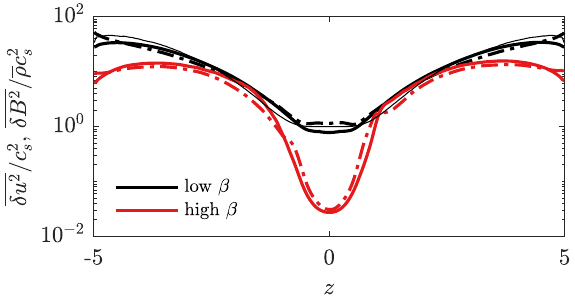}\\
\caption{Comparison of the turbulence fluctuation profiles in the low-$\beta$ state ($\beta$R0.1; black) and 
the high-$\beta$ state at otherwise identical parameters ($\beta$0.1; red). The
top panel normalizes to the local Alfv\'en speed $\va = \mean{B_{y}}/\sqrt{4\pi\mean{\rho}}$; 
the bottom panel normalizes to $c_{s}$. The thin black line shows 
a simple model of MRI \& Parker-instability driven turbulence (see text), 
and the top-panel inset shows the $\va(z)$ profile with $\propto z$ shown in dotted lines. }
\label{fig: turbulence}
\end{figure}

\subsection{Turbulence and transport at low $\beta$}

The turbulence profiles in the low-$\beta$ state are shown in \cref{fig: alpha,fig: turbulence}. \Cref{fig: alpha}
shows ${\alpha}(z)$ in $\beta$R0.1 and $\beta$1, splitting the total into its contributions from velocity fluctuations
$\alpha_{K}=\mean{\rho u_{x}u_{y}}/\meanz{P_{\rm th}}_{\rm th}$, magnetic fluctuations $\alpha_{\delta B}=-\mean{\delta B_{x}  \delta B_{y}}/\meanz{P_{\rm th}}_{\rm th}$, 
and the mean magnetic field $\alpha_{\mean{B}}=-\mean{B_{x}}\,\mean{B_{y}}/\meanz{P_{\rm th}}_{\rm th}$ (note 
that the mean shear flow has been subtracted from $u_{y}$ before computing $\alpha_{K}$).
The total ${\alpha}(z)$ profile is similar to that of the equilibrium-field  $P_{B}$ (dotted line), aside from in the midplane, 
where there is a  peak in the turbulent contributions and a drop off in the mean contribution. \revchng{We interpret this change within $ |z| \lesssim H_{\rm th} $ as a result of the MRI becoming dominant, whereas the Parker instability prevails at larger $z$. This interpretation is supported by the similar ratios between the $\alpha(z)$ components in the best-resolved high-$\beta$ simulation ($\beta$1), where the MRI remains active across a broad midplane region (cf.~lower panel). We see that both MRI- and Parker-driven} turbulence enable accretion ($\alpha>0$), but  
the MRI-driven midplane regions  have a stronger relative contribution from magnetic, compared to velocity fluctuations. 


\Cref{fig: turbulence} compares  the time-averaged fluctuation amplitudes between $\beta$R0.1 (low-$\beta$; black) and $\beta$0.1 (high-$\beta$; red), normalizing by either the local mean $\va$ (top) or $c_{s}$ (bottom). 
The low-$\beta$ state exhibits trans-Alfv\'enic turbulence across its full profile with $\mean{\delta u^{2}}\simeq \mean{\delta B^{2}}/\mean{\rho}\approx 0.3\va$, which is similar to that observed in \hpaper. This also implies the turbulence is highly supersonic at larger $z$ with $\mathcal{M}=\delta u/c_{s}\gtrsim 10$.

The thin black line in both panels illustrates a simple model of the turbulence, based on driving by a
combination of  the MRI and Parker instability (see also \citealt{Johansen2008}).  For both instabilities, 
we apply the standard reasoning that the turbulence saturates when the nonlinear turnover time $\tau_{\rm nl}\sim \ell/\delta u$ 
at the outer scale $\ell$ balances the instability's growth rate. 
In regions far above the midplane, where the profiles of $B_{y}$, $\rho$, and $\va$ are 
all power laws, it is most natural to take $\ell\propto z$, such that the turbulence remains self-similar with $z$.
Another possible choice, the lengthscale of the fastest growing mode, would scale in the same way for both the
Parker and MRI: for Parker because it grows fastest at the largest available vertical scales (the global scales, $\propto z$), and for MRI because 
it grows fastest where $n \sim \Omega^{-1} \va /\ell\sim 1$ and $\va \propto z$ in such regions (see \S\ref{subsec: equilibria} and inset of \cref{fig: turbulence}).\footnote{Similar 
arguments applied to the high-$\beta$ state would suggest that we should take $\ell \sim \va/ \Omega$ of the fastest growing mode as the outer scale, where the $\va$ is that 
of the self-induced \emph{mean} azimuthal field that can drive the MRI. This scale is very small, around $\ell \sim 4\times 10^{-2}H_{\rm th}$ for the $\beta$1 simulation 
(effectively the grid scale), in reasonable agreement with the morphology seen in \cref{fig: snapshot}. If 
$\ell$ is indeed grid limited, then since $\delta u\propto  \ell$ (assuming nearly constant growth rate), we would expect 
the turbulence amplitude to decrease with finer resolution (complicated, perhaps, by the 
fact that $\mean{\va}$ may itself depend on $\delta u$). This is consistent with the results of \citet{Ryan2017} who found
that the turbulence decreased its amplitude and scale with finer resolution in the high-$\beta$ state, even at extremely fine resolutions.  }
In contrast, near the midplane, since the shape of $\rho$ is set by $H_{\rm th}$, we require $\ell\lesssim H_{\rm th}$.
The Parker instability will not operate here because $B_{y}$ is nearly 
flat, while  the scale at which the MRI growth rate peaks ($n\approx 1$ in \cref{eq: mri dispersion rel} with $\va\sim  c_{s}$)
is $\ell\sim \va/\Omega\sim H_{\rm th}$, suggesting $\ell\sim H_{\rm th}$ is a reasonable choice (this is also plausibly consistent
with the morphology seen in \cref{fig: snapshot} for $\beta$R0.1).

Combining these estimates for $\ell$ with the growth rates, for the MRI we have $\gamma \sim \sqrt{\beta}\Omega$ at $\beta\lesssim1$, 
giving  $\delta u \sim H_{\rm th}\Omega\sim c_{s}$ in the midplane, and $\delta u\sim z \sqrt{\beta} \Omega \sim c_{s} z \Omega/\va \sim c_{s}$ for $z\gg H_{\rm th}$
(using $\va^{2}\approx 2\Omega^{2}z^{2}/a$). We thus assume that the MRI contributes appoximately 
trans-sonic turbulence $\mean{\delta u^{2}}\approx c_{s}^{2}$ at all $z$.
For the Parker instability contribution, we compute 
the maximum growth rate directly from \cref{eq: parker growth rate} based on the measured $\mean{B_{y}}$ and $\mean{\rho}$, 
then fix $\ell\propto z$ with proportionality coefficient $0.4$ chosen to fit the data. Adding these contributions 
leads to the thin black lines in \cref{fig: turbulence}, which provide an excellent match to the measured turbulence. 
The same prescription, with the same 0.4 coefficient, works well for the other low-$\beta$ ZNVF simulations.

\begin{figure}
\includegraphics[width=1.0\columnwidth]{\ffold 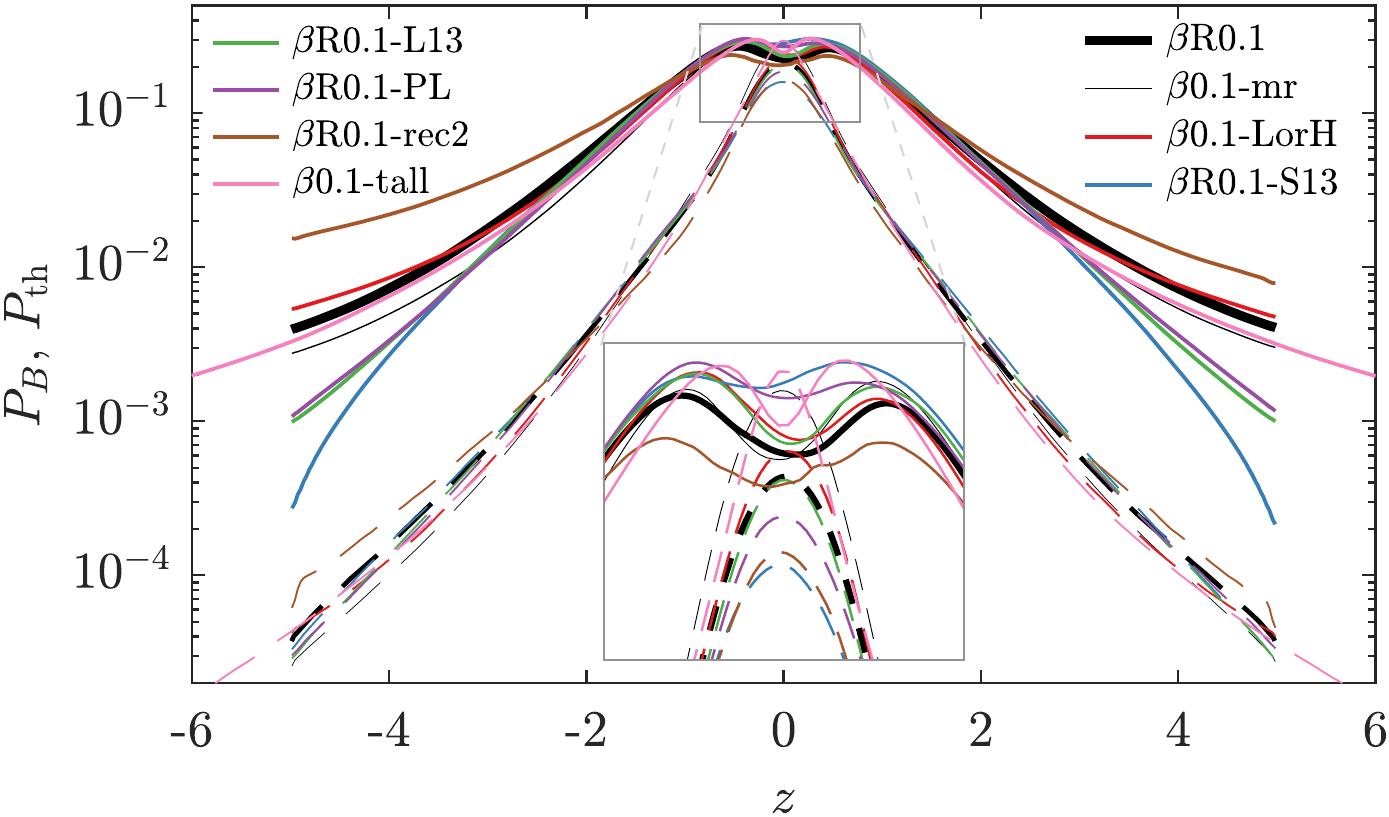}\\
\caption{Testing different vertical boundary conditions, numerical options, and initial conditions (see \cref{tab:sims}). As in \cref{fig: z profiles},  solid lines 
show $P_{B}$ and dashed lines show $P_{\rm th}$.\revchng{The inset zooms into the
midplane region.}\vspace{1pt}}
\label{fig: z profiles numerics}
\end{figure}

\subsection{The effect of boundary conditions and numerical options}

In \cref{fig: z profiles numerics} we illustrate the effect of changing 
the vertical boundary conditions and other numerical options on the low-$\beta$ state.
As listed in \cref{tab:sims},  these simulations explore 4 different vertical 
boundary conditions based on previous literature (see \S\ref{subsec: bcs}), different initial conditions (\cref{eq: lorenztian equilibrium}), 
a taller box with $L_{z}=15$, and second-order reconstruction with an HLLD Reimman solver, 
but otherwise have the same parameters (e.g., $c_{s}$). They 
all reach a quasi-steady low-$\beta$ state and we plot  the steady-state time average of
 $P_{B}=\mean{B_{y}}^{2}$ and $P_{\rm th}=c_{s}^{2}\mean{\rho}$. Overall, we see that the general 
 form is maintained for all cases, although there are certainly differences. Unsurprisingly, given that 
 Simon13 boundary conditions fix the gradients of $B_{x}$ and $B_{y}$ to be large and negative at the boundary, these 
 lead to the steepest $P_{B}$ profile at larger $z$, while the Outflow boundaries, which set $B_{x}'=B_{y}'=0$, do the opposite
 (the Lesur13 and Power-Law cases lie between these limits). Less obvious, however, is that these steeper $P_{B}$ profiles lead to states 
 that are more strongly magnetized (a lower-$\beta$ midplane) with  larger accretion rates $\alpha$ (see \cref{tab:sims}), 
 even though naively one might expect that carrying the  flux out of the domain more efficiently would lead to lower midplane magnetic fields. 
Using $2^{\rm nd}$-order reconstruction leads to the flattest profiles in $P_{B}$ and $\rho$, which 
 appears to be a consequence of it supporting modestly  less vigorous turbulence at large $z$, and 
 thus transporting the flux out of the midplane less efficiently.
 Finally, $\beta$0.1-tall --- with an extended vertical domain (the full extent is not shown), fewer grid cells per $H_{\rm th}$, and a much lower density floor --- produces similar vertical profiles, albeit with  lower midplane turbulence 
 levels, which lead to lower $\alpha$ and larger $\beta_{\rm mid}$.  This difference is at least
 partially a consequence of its coarser resolution per $H_{\rm th}$ (see App.~\ref{app: resolution dep}),
  perhaps also influenced by its different domain aspect ratio  (see App.~\ref{app: box size dep}; other than the initial conditions, this case is equivalent to rescaling the vertical box to a lower $H_{\rm th}$, 
 as for $\beta$0.05 or $\beta$0.01).

Various other numerical tests are presented in the Appendices, which describe the dependence on
domain aspect ratio (App.~\ref{app: box size dep}), resolution (App.~\ref{app: resolution dep}), and the density floor (App.~\ref{app: density floor dep}). We briefly summarize these results here. 
We find that the azimuthal length of the 
box can have an important influence, with   $\beta=0.1$ initial conditions that transition to the low-$\beta$ state in a longer 
box instead transitioning to the high-$\beta$ state in a shorter box. We interpret this behavior 
as relating to the  Parker instability, which seems to be
needed to regenerate the field against vertical escape. The fastest growing Parker-instability modes in this
regime are of the  ``quasi-interchange'' variety,  which have a specific 
azimuthal scale 
\eqref{eq: parker wavenumber} that approaches the box scale with shorter boxes, thus potentially
interfering with the turbulence.
The radial box size does not appear to have an important influence.
The dependence on resolution is interestingly non-monotonic, with the transport decreasing with coarsening resolution 
until $1$-$2$ zones per $H_{\rm th}$, then suddenly increasing dramatically at yet coarser resolutions. 
This behavior arises because the turbulence in the midplane is suppressed with coarser resolution, 
thus reducing the transport for similar profiles; but, once the resolution drops so low that it acts to support 
the collapse of density to the thermal profile in the midplane, the system no longer loses its magnetic flux and supports a 
$\beta\ll1$ midplane with higher transport (see \S\ref{sec: low beta} and e.g., $\beta$0.001 in \cref{fig: z profiles}).
Several direct tests show effectively no dependence of the low-$\beta$ state on the choice of density floor. 
\revchng{Finally, in App.~\ref{app: shear and rotation} we demonstrate that both shear and rotation are required to maintain the low-$\beta$ state. Interestingly, in the absence of rotation the system still exhibits high-$\beta$ dynamo cycles; since 
the MRI does not exist in this case (see \S\ref{subsec: MRI}), this
suggests that the cyclic dynamo is not associated with the MRI, as usually assumed. }

A final numerical issue worth discussing is that  the low-$\beta$ state requires 
a large separation between the vertical domain size and 
$H_{\rm th}$. \revchng{This conclusion is based on test simulations at the same $H_{\rm th}$ as $\beta$1 or $\beta$0.3, but with either a stronger initial magnetic field with $\beta<0.1$ (and thus a larger $H_{\beta}$) or with  initial radial flux that is sheared into an azimuthal field in the simulation's early stages (as in e.g., $\beta$R0.1). These tests} lose their flux and transition into the high-$\beta$ state. The same behavior occurs at resolutions with fewer zones per $H_{\rm th}$ than some low-$\beta$ simulations (e.g.,  $\beta$R0.1), \revchng{showing that it is not an indirect consequence of changing the resolution per $H_{\rm th}$}.  This suggests that
 the low-$\beta$ state can be maintained only when there exists an extended strongly magnetized corona, perhaps because the Parker-driven dynamo requires an atmosphere of sufficient vertical extent 
and/or due to direct effects from the boundary conditions.

Overall, we see that while  numerical options do have an important impact on the results, as is also
the case for the high-$\beta$ shearing-box turbulence \citep[e.g.,][]{Guan2009,Gressel2013,Ryan2017},  
the general features of the low-$\beta$ state seem robust.

\subsection{The effect of vertical flux}\label{sub: NVF}

\begin{figure}
\includegraphics[width=0.9\columnwidth]{\ffold 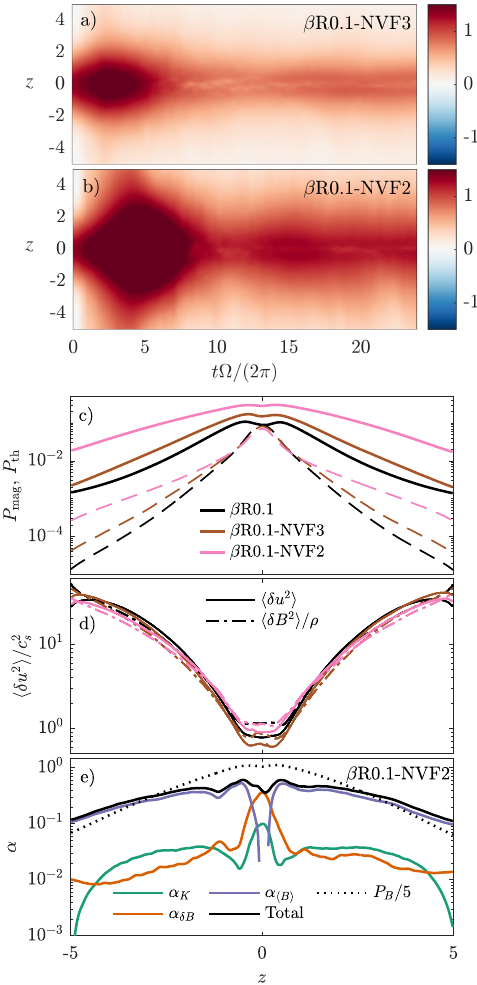}\vspace{-00.06cm}\\
\caption{ The effect of a net vertical flux on the low-$\beta$ state. Panels a and b show butterfly diagrams of $\mean{B_{y}}$ (cf. \cref{fig: butterfly}) starting from 
the same initial conditions as $\beta$R0.1, but with a net vertical flux of $\beta_{z}=1000$ or $\beta_{z}=100$. Diagnostics of the vertical equilibrium  (cf. \cref{fig: z profiles})
and  turbulence (cf. \cref{fig: turbulence})  are shown in panels c and d, respectively, while panel e shows the contributions to $\alpha(z)$ 
 for the $\beta_{z}=100$ case (cf. \cref{fig: alpha}). The $\beta_{z}=1000$ field makes little noticeable difference to the morphology, raising $\alpha$ slightly, 
while $\beta_{z}=100$ causes a stronger mean field, lowering the midplane $\beta$, in line with previous results \citep{Suzuki2009, Bai2013a,Salvesen2016a}. }
\label{fig: with bz}
\end{figure}

Previous works have found that strongly magnetized ($\beta<1$) profiles are sustained in the presence
of a net vertical flux threading the box \citep[e.g.,][]{Suzuki2009,Bai2013a}. We explore here 
how the ZNVF low-$\beta$ state changes with $\beta_{z}$, finding that it smoothly transitions into 
this low-$\beta$ NVF state as $\beta_{z}$ decreases below $\beta_{z}\simeq10^{3}$. 
The general shape of the profiles remains rather similar as this occurs. 
In other words, while for $\beta_{z}\gtrsim 10^{3}$ there exist
two qualitatively different self-sustaining accretion states depending on the initial toroidal flux threading the system, for $\beta_{z}\lesssim 10^{3}$ the system 
always reaches a strongly accreting state characterized by a $\beta\lesssim 1$ midplane, wider density profiles, and a lack of dynamo cycles
(much longer-time-duration simulations are shown in \citealt{Bai2013a,Salvesen2016a} with \citealt{Salvesen2016a}'s exhibiting widely spaced, aperiodic flips in polarity at $\beta_{z}=100$).
However,  as argued in \S\ref{sec: dynamo} below, despite the similar appearance of the ZNVF and NVF states, the mechanism that sustains the strong azimuthal field with NVF is very different: NVF
drives a single large-scale MRI channel mode, while the ZNVF state is supported by the turbulent dynamo.

Our results are collected in \cref{fig: with bz}. In the top panel we show butterfly diagrams for the same setup as $\beta$R0.1 but
with $\beta_{z}=10^{3}$ (panel a) and $\beta_{z}=100$ (panel b). 
The profiles and time evolution appear similar to  \citet{Bai2013a,Salvesen2016a}, as well as those with ZNVF (cf. \cref{fig: butterfly}), although with $\beta_{z}=100$ the field
becomes significantly stronger and less peaked  in the midplane after its initial phase (see also the quasi-steady-state profiles
 in \cref{fig: with bz}c). The accretion rate  is also  higher at $\beta_{z}=100$ (${\alpha}\approx 2.1$ at $\beta_{z}=100$, 
 versus ${\alpha}\approx 0.7$ and ${\alpha}\approx 0.6$ for $\beta_{z}=10^{3}$ and $\beta_{z}=\infty$, respectively).\footnote{\citet{Salvesen2016a} and \citet{Bai2013a} report ${\alpha}\simeq 1.0$
 at $\beta_{z}=100$  when measured at the midplane or ${\alpha}\simeq 1.5$ when measured over the full profile. The larger ${\alpha}$ we report here
 is due to the larger $L_{z}/H_{\rm th}$ in our simulation, since ${\alpha}$ is normalized by $\meanz{\rho}$ and $\mean{\rho}$ decreases more rapidly with $z$ than $\mean{B_{x}}\,\mean{B_{y}}$ 
 (see \cref{fig: with bz}e). } 
 This larger ${\alpha}$ at lower $\beta_{z}$ is not a consequence 
 of the turbulent fluctuations, which are hardly stronger  at $\beta_{z}=100$ than at $\beta_{z}=\infty$ (see 
 panel d); 
 instead its accretion stress is nearly flat with $z$ and almost entirely due to the mean field $-\mean{B_{x}}\,\mean{B_{y}}$ (see panel e). This agrees with the conclusions of  \citealt{Bai2013a} (see their figure 6).
 This behavior results from $\mean{B_{x}}$ being generated from $\mean{B_{y}}$ via direct stretching of the net vertical field (as opposed to via the 
 turbulence in ZNVF simulations): this more efficient feedback, \revchng{which is effectively the aforementioned domain-scale  MRI channel 
 mode}, increases $\alpha$  
 and lowers $\beta$ due to the stronger fields (see \S\ref{sec: dynamo}).
 As in previous works, we also observe much stronger outflows at $\beta_{z}\lesssim 10^{3}$, which  cause significant mass loss over time if the 
 density is not continuously  replenished via the artificial source term. Despite the lower $\beta$ at $\beta_{z}=100$, the density still collapses towards the thermal 
 scale height within $|z|\lesssim H_{\rm th}$, with a  similar midplane density structure as the ZNVF runs.

\subsection{Collapse towards the $\beta\simeq 1$ midplane}\label{sec: low beta}

\begin{figure}
\includegraphics[width=0.85\columnwidth]{\ffold 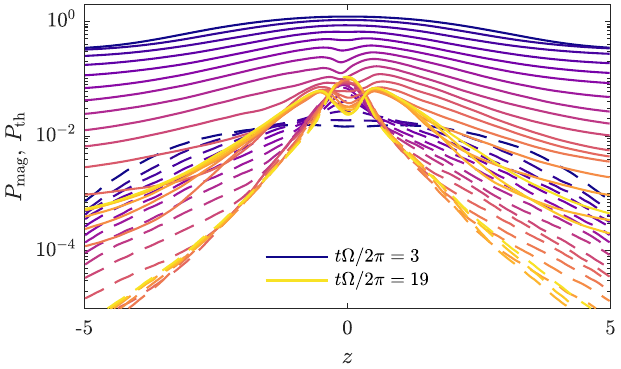}\vspace{-0.86cm}\\
\includegraphics[width=0.85\columnwidth]{\ffold 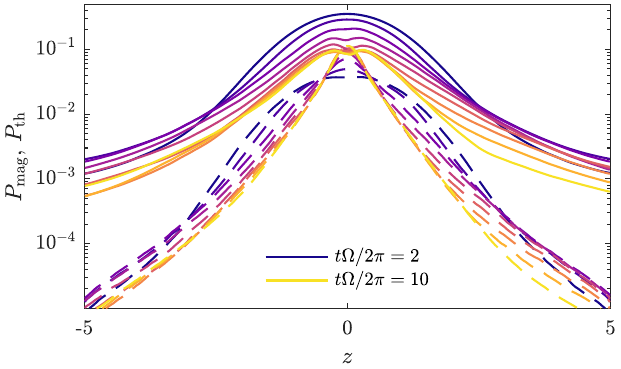}\vspace{-0.06cm}\\
\includegraphics[width=0.85\columnwidth]{\ffold 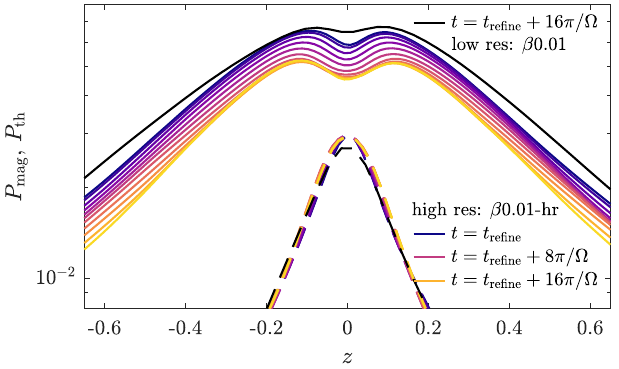}\vspace{-0.06cm}\\
\caption{ Magnetic and thermal pressures (as in \cref{fig: z profiles}) as a function of time, from dark purple to yellow in equal time increments of approximately 1 orbit, over 
the time range indicated on each panel. The top panel shows $\beta$0.1-H4, which starts with  $H_{\beta}=4$ and $\beta\approx 0.01$, 
so that it loses significant flux before reaching a state  similar to $\beta$R0.1 (cf. \cref{fig: z profiles numerics}); the middle panel shows $\beta$0.05, 
which  has less flux to lose over its initial phase, but behaves analogously, settling into a similar state (other than its smaller $H_{\rm th}$) once $\beta\simeq 1$ in the midplane. The lower panel shows $\beta0.01$-hr, which is refined from the steady state of $\beta0.01$ at $t_{\rm refine}\approx 72\pi/\Omega$ (see text). The refined version decreases in time, resettling into a somewhat higher-$\beta$ midplane, even though over the same time period
the coarser-resolution version $\beta0.01$ increases its midplane field somewhat (the black lines show $\beta$0.01 at the same time 
as the yellow lines for $\beta$0.01-hr).
This illustrates that the lower-$\beta$ midplanes seen in $\beta$0.01 and $\beta$0.001 (and other cases not shown)
are likely the result of  poor resolution of $H_{\rm th}$.\vspace{5pt}
} 
\label{fig: low beta approach}
\end{figure}

Figures \ref{fig: z profiles} \& \ref{fig: flux beta alpha} show that even when ZNVF simulations are initialized with $\beta \ll 1$ and reach the low-$\beta$ state, the midplane \revchng{usually reaches a gas-pressure dominated $\beta \sim 1$ steady state.  Here we explore how the system behaves as it loses  magnetic flux and evolves towards a $\beta\sim1$ midplane, also showing that the process is sensitive to resolving the thermal scale height}.

\Cref{fig: low beta approach} shows several numerical experiments that probe this physics, illustrating the time evolution 
of $P_{\rm th}$ and $P_{B}$. The top panel shows a simulation ($\beta$0.1-H4) with the same $H_{\rm th}$ and box size as $\beta$R0.1 or $\beta$0.1-mr,
but with a  more strongly magnetized initial condition with  $\beta\approx 0.01$. It thus starts with a  wider profile with $H_{\beta}\approx4$, 
effectively simulating the small central patch $|z|\lesssim 1.6$ of $\beta$0.01 ($L_{x}$ and $L_{y}$ would also be  similarly scaled).
After a short reorganization phase from the Guassian initial conditions, the system settles into a quasi-equilibrium state with $P_{B}$
flatter than $P_{\rm th}$,  similar to the central region of the generalized power-law profile \cref{eq: lorenztian equilibrium} \revchng{(see \cref{fig: analytic})}. This strong field then escapes through the boundary over ${\simeq}15$ orbits,
with the density becoming more peaked and $P_{B}$ developing its characteristic ``dip'' at $z=0$, before halting and reaching 
steady state at $\beta\sim1$ (cf. \cref{fig: z profiles numerics}). The same behaviour  is observed for $\beta$0.05 (middle panel) and other simulations (not shown), albeit with a shorter field-decrease 
phase   because they start from higher $\beta$ initially. In fact, the measured rate of field 
decrease over this phase, $2\pi \Omega^{-1} d\ln {B_{y}}/dt\approx  1/6$ , seems to be independent of boundary conditions, $\beta$, or resolution, 
as indicated by the thin black line in the lower panel of \cref{fig: flux beta alpha}. This might be expected for Parker-instability-driven turbulence given 
its $\beta$-independent growth rate $\gamma\sim\Omega$  (see \S\ref{subsec: parker}). Below (\S\ref{sec: dynamo}), we will argue  that the field is being continuously regenerated 
over this phase in a similar way  to the true saturated state, but that this regeneration is slower than its expulsion through the boundaries.

Two ZNVF simulations in \cref{tab:sims} stand out as exceptions to the above scenario: $\beta$0.01 and $\beta$0.001,
which both saturate with a $\beta\ll1$ midplane (see \cref{fig: z profiles}b; for  $\beta$0.001 especially the
gas is everywhere magnetically supported, with no $P_{B}$ dip at $z=0$). 
The thermal scale heights of these  simulations are poorly resolved, with only ${\simeq}9$ zones in the thermal  midplane ($|z|<H_{\rm th}$)
for $\beta$0.01, and only  ${\simeq}3$ zones for $\beta$0.001. A natural hypothesis is thus 
that the resolution halts the collapse of $\rho$ to the thermal scale height thereby artificially halting  the collapse at $\beta\ll1$.
We test this in the lower panel of \cref{fig: low beta approach} by taking a snapshot of $\beta$0.01 in the saturated state 
($t_{\rm refine}=72\pi/\Omega$) and  refining its resolution by a factor of two.\footnote{At $t_{\rm refine}$  the shear-periodic boundary conditions are truly periodic, allowing straightforward refinement over the $x$ boundaries. }
We compare the evolution of this refined simulation ($\beta$0.01-hr) to the unrefined version over the same time period.
While $P_{B}$ actually increases modestly over this period for  $\beta$0.01 (cf. purple and black lines), that of $\beta$0.01-hr decreases (cf. purple and yellow lines).
Although it saturates again with $\beta<1$ in $\beta$0.01-hr, 
this does support the general idea that $\beta\ll1$ midplanes are likely a result of insufficient resolution, at least in the shearing box and without net vertical flux. 
Nonetheless, it is interesting --- and perhaps physically important despite its unphysical origin ---  that halting the collapse of the density in the midplane also halts the loss 
of flux through the boundaries at $z\gg H_{\rm th}$. We will return to this point in the discussion (\S\ref{sec: conclusion}).  

\begin{figure}
\includegraphics[width=1.0\columnwidth]{\ffold 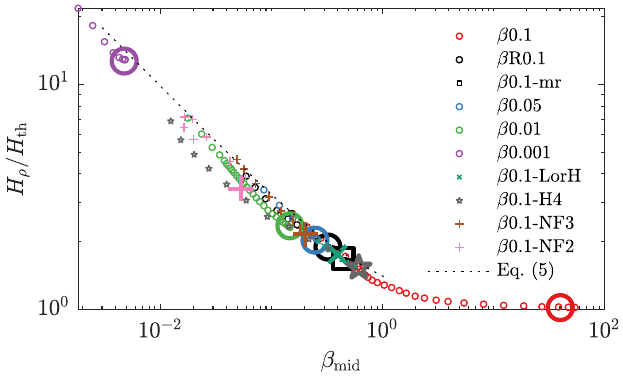}\vspace{-0.06cm}\\
\caption{The relation between the density scale height, $H_{\rho}$ and $\beta$ averaged over $|z|\leq H_{\rho}$ ($\beta_{\rm mid}$), 
for a number of simulations from \cref{tab:sims}. Small markers illustrate time slices separated by approximately 1 orbit ($t\Omega=6$), 
while the large markers indicate an average over the steady state. The dashed line shows the same 
quantities computed from the  equilibrium \eqref{eq: lorenztian equilibrium}, which is effectively $H_{\rho}\propto \beta^{-1/2}$ at $\beta\ll1$.\vspace{3pt}
} 
\label{fig: beta-a and dbydt}
\end{figure}

We probe the approach to the saturated state more generally in \cref{fig: beta-a and dbydt}, which 
plots the density scale height,  $H_{\rho}\equiv \int dz\,|z| \rho/\int dz\,\rho$,  and $\beta$ averaged over $|z|\leq H_{\rho}$ (termed $\beta_{\rm mid}$; note
this is slightly different to $\beta_{\rm mid}$ in \cref{fig: flux beta alpha}, which averages over $|z|\leq H_{\rm th}$).
The relationship between $H_{\rho}$ and $\beta_{\rm mid}$ is effectively probing the equilibrium --- lower $\beta$ generally causes a wider density 
profile unless $P_{B}$ becomes flat --- so the scatter plot of a simulation's evolution in $\beta_{\rm mid}$-$H_{\rho}$
space provides a crude measure of how flux is lost over the quasi-equilibrium phase (small markers) and where it settles at saturation (large 
markers). We see that all simulations collapse onto one path, as expected,  starting at lower $\beta$ (higher $H_{\rho}$) 
then moving to the right and down as they lose flux (the spacing of the points indicates the rate of decrease\footnote{$\beta$0.01
exhibits a very long phase of more slowly increasing $\beta$ (see also \cref{fig: flux beta alpha}), which 
apparently occurs once the collapse of $\rho$ is strongly affected by resolution, but not yet completely 
halted.}). $\beta$0.001 and $\beta$0.01 stop at lower $\beta$, as described above, as do cases 
with net vertical flux. The final $\beta_{\rm mid}$ of the other cases seems to increase 
modestly with coarser resolution (e.g., compare $\beta$0.1-mr or $\beta$0.1-LorH with $\beta$R0.1; see also App.~\ref{app: resolution dep}) and decrease modestly with $L_{z}/H_{\rm th}$ (e.g., compare $\beta$0.05 or $\beta0.01$-hr with various $\beta$0.1 cases). However, 
there also appears some randomness involved in the details of the final state (indeed, it also changes slowly in time);
 longer-time, finer resolution simulations are needed to better probe this collapse when $H_{\rm th}\ll L_{z}$ and $\beta\ll1$. In this 
context, static mesh refinement would be  
valuable in order to resolve  small midplane scales (${\ll}H_{\rm th}$), while avoiding the extremely costly timestep limitations that result from the strongly magnetized upper layers.

\begin{figure}
\includegraphics[width=1.0\columnwidth]{\ffold 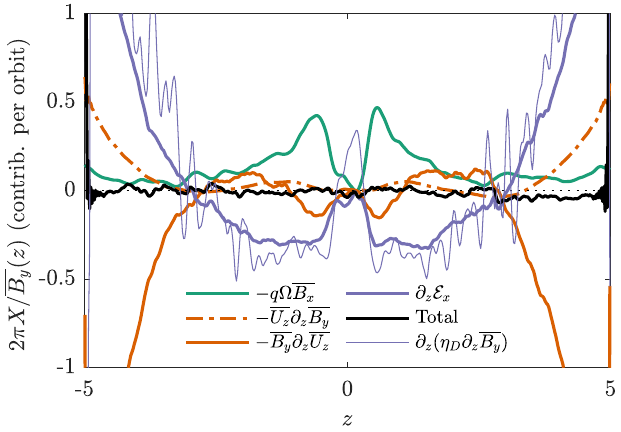}\vspace{-0.06cm}\\
\caption{Terms in the equation for $\partial_{t} \mean{B_{y}}$ in $\beta$R0.1, averaged over the
steady state in time (see \cref{eq: dynamo}). We see that near the midplane, $B_{y}$ flux is supplied by the shearing of $\mean{B_{x}}$, which is
balanced by  transport to higher $|z|$ from the turbulence ($\partial_{z}\mathcal{E}_{x}$) with small contributions from 
mean outflows on either side of the midplane. In contrast, at large $|z|$, $\mean{B_{y}}$ is sustained primarily 
by the upwards transport of flux by the turbulence  ($\partial_{z}\mathcal{E}_{x}$), which competes against 
strong vertical outflows through the boundary. This shows that, absent a dynamo effect to regenerate $\mean{B_{x}}$, 
the field would decay within several orbits.  The thin purple line compares the
measured $\partial_{z}\mathcal{E}_{x}$ with the dynamo-theory  result $\mathcal{E}_{x}\approx \eta_{\rm turb}\partial_{z}B_{y}$ with $\eta_{\rm turb}= \Omega^{-1}\mean{\delta \bm{u}^{2}}/3 $ (we 
smooth this with a 5-gridpoint moving average for clarity). \vspace{8pt}
 }
\label{fig: by dynamo}
\end{figure}

\vspace{0.5cm}
\section{Sustaining the low-$\beta$ state}\label{sec: dynamo}\vspace{0.3cm}

In this section we study how the low-$\beta$ state is maintained via a dynamo feedback that 
regenerates flux lost through the boundaries. We do not consider the high-$\beta$ cycles, since
this has been studied extensively in past work \citep[e.g.,][]{Lesur2008a,Squire2015c,Gressel2015}, focusing instead on how the system can maintain strong time-independent mean fields. 
(Note that the term ``dynamo'' is sometimes used to refer specifically to cyclic behavior in the accretion-disk literature, but 
here we use it more generally to describe the mechanism by which a large-scale field is generated.)
While we provide clear evidence that fields are regenerated and expelled on several-orbit time-scales --- viz., 
the system does not reach steady state because its activity shuts down, but because flux loss and growth are balanced --- 
the physics of this mechanism, and how this depends on parameters, remain poorly understood.

\subsection{Mean-field dynamo: background}

An $x,y$ average of the radial and azimuthal components of the induction equation \eqref{eq:MHD B} yields 
\begin{subequations}
\begin{align}
\partial_{t} \mean{B_{x}} =& - \partial_{z}(\mean{U_{z}}\,\mean{B_{x}}) + \mean{B_{z}}\partial_{z}\mean{U_{x}}- \partial_{z}\mathcal{E}_{y}, \label{eq: dynamo x}\\
\partial_{t} \mean{B_{y}} =& -q\Omega \mean{B_{x}} -\partial_{z}(\mean{U_{z}}\,\mean{B_{y}}) + \mean{B_{z}}\partial_{z}\mean{U_{y}}+ \partial_{z}\mathcal{E}_{x},\label{eq: dynamo y}
\end{align}\label{eq: dynamo}
\end{subequations} 
\!where $\bm{\mathcal{E}}\equiv \mean{\delta \bm{u}\times \delta \bm{B}}$ is the turbulent electromotive force (EMF) from the fluctuations. 
In \cref{eq: dynamo y}, the first term captures the stretching of the radial field to azimuthal by the mean shear flow, the second captures advection
and compression/rarefaction by the vertical outflow, the third captures stretching of a vertical field in a $z$-dependent shear (note that $\nabla\cdot\bm{B}=0$ implies
$\mean{B_{z}}=0$ unless there is a net vertical flux),  and the fourth term captures the effect of the fluctuations on the mean field. 
In the traditional mean-field dynamo paradigm \citep{Brandenburg2005,Rincon2019}, $\bm{\mathcal{E}}$ is Taylor expanded assuming 
the fluctuations are small scale compared to the mean, yielding \vspace{-3pt}
\begin{align}
{\mathcal{E}}_{i}\approx &\alpha_{D,ij}\mean{B_{j}}+ \tilde{\eta}_{D,ijk}\partial_{j} B_{k}+\dots \nonumber\\
\sim & \alpha_{D} \mean{B_{i} }- \eta_{D}(\nabla\times \mean{\bm{B}})_{i},\label{eq: alpha eta dynamo}
\end{align}
where the `transport coefficients' $\alpha_{D}$ and $\eta_{D}$ are determined by the dependence of the fluctuations on the mean fields. The second expression results from assuming 
homogeneous and isotropic fluctuations, which, while not valid in a stratified shear flow, is helpful for the simple analysis carried out here ($\alpha_{D}$ is the
diagonal part of $\alpha_{D,ij}$; $\eta_{D}$ is the fully antisymmetric part of $\tilde{\eta}_{ijk}$). 
The  so-called $\alpha$-$\Omega$ dynamo results  for non-zero $\alpha_{D}$, when the strong $\mean{B_{y}}$ generates $\mean{B_{x}}$ via the final term in \cref{eq: dynamo x}, which 
then regenerates $\mean{B_{y}}$ via $-q\Omega \mean{B_{x}}$.
The effect can be understood as a linear instability by ignoring the effect of $\mean{U_{z}}$ and  inserting \cref{eq: alpha eta dynamo} into \eqref{eq: dynamo}, then assuming $\mean{B_{i}}\propto e^{i\kappa z}$ and $|\alpha_{D} \kappa|\ll q\Omega$;
one finds solutions with frequency $\omega_{D} \approx -i\kappa^{2}\eta_{D} \pm \sqrt{i \kappa \alpha_{D} q \Omega}$, such that a branch of growing and oscillating solutions
exists at small $\kappa$, where the dynamo feedback (via $\alpha_{D}$) overcomes the turbulent diffusion (via $\eta_{D}$).
The crux, of course, is having fluctuations that cause a sufficiently large $\alpha_{D}$, which can result from the combination of
stratification and rotation creating a net helicity $\mean{\delta\bm{u}\cdot\nabla\times\delta\bm{u}}$ or current helicity $\mean{\delta\bm{B}\cdot\nabla\times\delta\bm{B}}$.
This formalism, or extensions of it, has been analysed in a number of papers in order to understand the cyclic behavior of the high-$\beta$ state (see, e.g., \citealt{Gressel2015,Mondal2023} and 
references therein). It is also possible to have  a similar large-scale dynamo with $\alpha_{D}=0$ as a result of off-diagonal components of $\tilde{\eta}_{D,ijk}$ \citep{Raedler2006,Squire2015b,Mondal2023}.

\subsection{Sustaining a strong azimuthal field against escape}

Mean-field theory is applied to the saturated low-$\beta$ state in 
 \cref{fig: by dynamo}. We illustrate the time average over the steady state of each term in \cref{eq: dynamo y}, 
 normalizing each by  $\mean{B_{y}}(z)/2\pi$, such that 
a value of $1$ implies that the particular term would grow or decay $\mean{B_{y}}$ by a factor $e^{\pm 1}$ over one orbit at each $z$. The sum
over all terms is effectively zero, as must be the case in steady state.

Our general conclusions are similar to \citet{Johansen2008} (figures 11-14), with the important 
difference that flux is continuously escaping the boundaries in our case. 
Fields are  replenished and expelled over several-orbit timescales, with strong
flux creation by shearing  for $H_{\rm th}\lesssim |z|\lesssim 3$ balanced by its diffusion to larger $|z|$ by turbulence ($\partial_{z}\mathcal{E}_{x}$).
For $|z|\gtrsim 3$, $\mean{B_{y}}$ is instead replenished by fluctuation-induced diffusion from smaller $z$ (i.e., turbulent transport; $\partial_{z}\mathcal{E}_{x}$), 
while being rapidly carried out of the box by the expanding  outflows ($\mean{B_{y}}\partial_{z}\mean{U_{z}}$; advection by $\mean{U_{z}}$
is only a minor effect because $\mean{U_{z}}$ is purely compressive). With the thin purple line, we compare the measured $\partial_{z}\mathcal{E}_{x}$ from $\mean{\delta \bm{u}\times \delta \bm{B}}$
to the expectated  $\eta_{D}$ from quasi-linear dynamo theory, $\eta_{D} = \mean{\delta \bm{u}^{2}}/3\Omega$, assuming the correlation time of the turbulence is ${\simeq}\Omega^{-1}$ 
(there is no contribution to $\eta_{D}$ from $\delta \bm{B}$ fluctuations in quasi-linear theory;  \citealp{Rincon2019}). 
We see surprisingly good agreement, with $\partial_{z}\mathcal{E}_{x}\approx \partial_{z}(\eta_{D} \partial_{z} \mean{B_{y}})$ across the full domain; this indicates that
the fluctuations are predominantly affecting $\mean{B_{y}}$ as a turbulent diffusion, as expected (the effect of a possible $\alpha_{D}$ on $\partial_{t}\mean{B_{y}}$ would be very small, being
proportional to $\mean{B_{x}}$).

\begin{figure}
\includegraphics[width=1.0\columnwidth]{\ffold 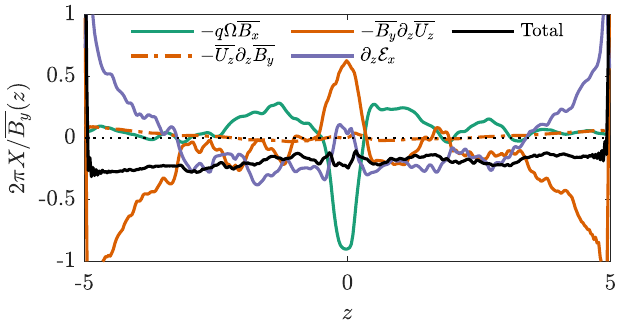}\vspace{-0.86cm}\\
\includegraphics[width=1.0\columnwidth]{\ffold 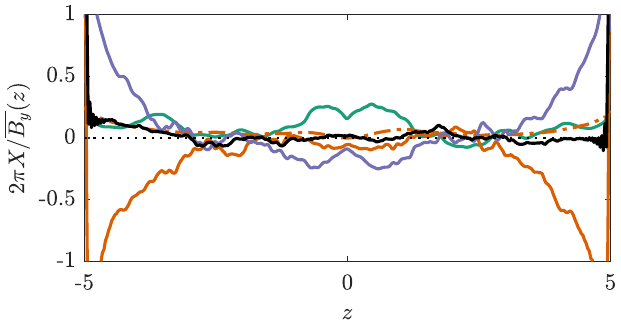}\vspace{-0.06cm}\\
\caption{As in \cref{fig: by dynamo}, but for the cases $\beta0.1$-H4 (top) during its collapse phase (averaged from $t\Omega/2\pi\approx6$ to 8; cf. \cref{fig: low beta approach})
and $\beta0.001$ (bottom).}
\label{fig: by dynamo all}
\end{figure}

\Cref{fig: by dynamo all} shows the same $\mean{B_{y}}$-dynamo analysis for two other cases of relevance. 
The first, $\beta$0.1-H4, is discussed in \S\ref{sec: low beta}; it starts from $\beta\ll1$ and loses all of its 
flux until reaching $\beta\sim 1$ in the midplane. We consider the middle of this flux-decrease phase, averaging from  $t\Omega/2\pi\approx6$ to 8 (fourth to sixth line from the top in 
\cref{fig: low beta approach}a), but the results are similar at all times during this phase. As expected,  the total $\mean{B_{y}}$ contribution from all terms is nearly constant and negative, indicating $\mean{B_{y}}$ is decreasing in time without changing shape. But, we also still see robust field creation for $|z|\gtrsim H_{\rm th}$, and the balance of terms is similar 
 to the true saturated state in \cref{fig: by dynamo}. In other words, during this flux-decrease phase, a similar dynamo continues to operate, but the growth 
via $q\Omega\mean{B_{x}}$ is overwhelmed by the turbulent diffusion and outflows. The
 properties for $|z|\lesssim H_{\rm th}$ are explained by our previous observations that the midplane density is increasing by inwards collapse  (thereby 
requiring a large positive contribution from $\mean{U_{z}}$) but the midplane $\mean{B_{y}}$ is not (thereby requiring a large negative 
contribution  from $\mean{B_{x}}$ to balance that from $\mean{U_{z}}$). How the system 
conspires to reverse the midplane $\mean{B_{x}}$ in order to achieve this is not at all obvious, but it is nonetheless effectively guaranteed if $\mean{B_{y}(z,t)}$ is
to maintain a similar spatial form as the  total flux decreases.

The second example in \cref{fig: by dynamo all} is $\beta$0.001, which maintains a $\beta\ll1$ midplane because the collapse of density 
is halted by the simulation's low number of grid cells per $H_{\rm th}$. While 
it is not surprising that the coarse grid stops the density's collapse to ${\simeq}H_{\rm th}$, the observation that this \emph{also} stops the escape
of flux is  interesting (see also \S\ref{sec: low beta}). Indeed, we see that the system's dynamo continues to operate, supporting the field against the turbulent diffusion in the midplane and rapid outflows  to the boundaries.
While the loss/growth rate is somewhat lower than in the resolved case ($\beta$R0.1; \cref{fig: by dynamo}), it still implies 
 complete regeneration of the flux over $4$ to $5$ orbits in regions away from the midplane that should, in principle, be ignorant of $H_{\rm th}$ and thus adequately resolved. Note also that the near-midplane turbulence in 
 this simulation is highly supersonic, as needed to cause a similar relative $\partial_{z}\mathcal{E}_{x}$ at $\beta\ll1$.
 
 A simple interpretation of these results is that, in principle, any $\beta\ll1$ state can be maintained, with 
 the dynamo generating a $\mean{B_{x}}$ near the midplane that in turn balances the continual loss of $\mean{B_{y}}$ through the 
 boundaries. However, without a reason for the density to stop at a particular scale height $H_{\rho}$, it 
 always collapses towards a thermal profile in the midplane, with the magnetic field (and thus the dynamo) being 
 forced to cooperate.  It stops once $P_{\rm th}$ provides significant pressure 
 support, corresponding to the  $\beta\sim 1$ midplane in the steady state of well-resolved simulations. 
 On the other hand, if other effects intervene --- the grid in our simulations, \revchng{but perhaps e.g., radiation, or physical midplane turbulence  
from another source in more realistic settings }--- the collapse is halted with $H_{\rho}\gg H_{\rm th}$, thereby also enabling a $\beta\ll1$ midplane.

\begin{figure}
\includegraphics[width=1.0\columnwidth]{\ffold 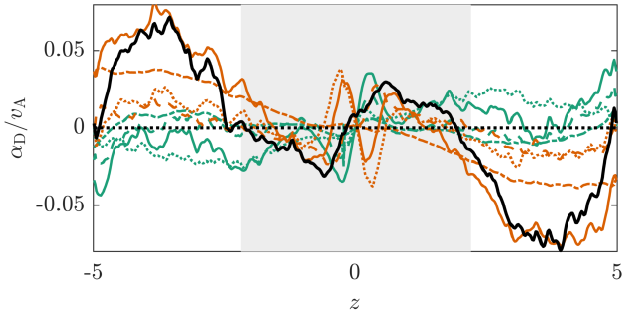}\vspace{-0.86cm}\\
\includegraphics[width=1.0\columnwidth]{\ffold 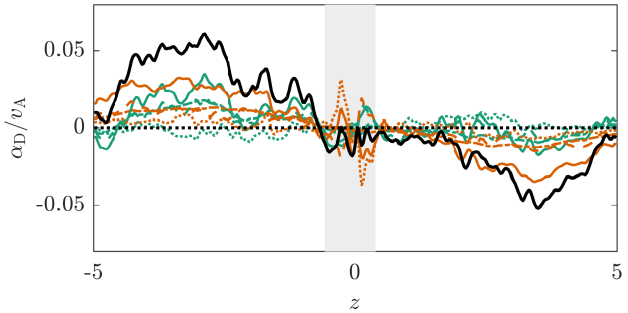}\vspace{-0.86cm}\\
\includegraphics[width=1.0\columnwidth]{\ffold 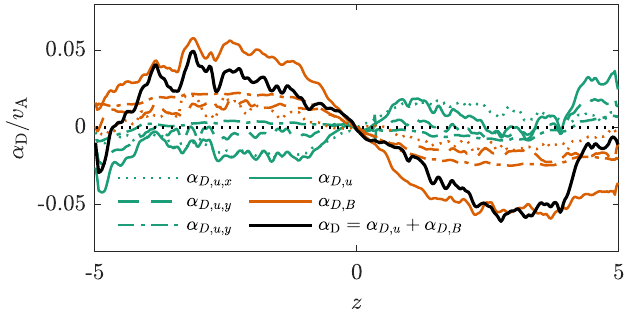}\vspace{-0.06cm}\\
\caption{Estimate of $\alpha_{D}$ from mean-field dynamo theory based on the helicity of the fluctuations (\cref{eq: alpha mean field}). We show the three 
simulations from  \cref{fig: by dynamo,fig: by dynamo all} ($\beta$R0.1, $\beta$0.1-H4 from $t\Omega/2\pi=4$ to $6$, and $\beta$0.001 from top to bottom). 
The shaded regions show where $\beta>0.03$, which 
seems empirically to predict the sign of $\alpha_{D}$. Although $\alpha_{D}$ is sufficiently large to explain the mean-field dynamo 
for the parameters we consider, it does not exhibit the required sign in $\beta>0.03$ regions, and moreover, this sign changes between cases that 
otherwise seem to exhibit similar dynamo mechanisms.
} 
\label{fig: alpha dynamo}
\end{figure}

\subsection{Generation of the radial field}\label{sub: radial field generation}

The results of \cref{eq: dynamo y,fig: by dynamo all} show only 
that $\mean{B_{y}}$ is driven primarily by $\mean{B_{x}}$, providing no information on $\partial_{t}\mean{B_{x}}$, which must also be continuously regenerated to balance its escape. 
Unfortunately a similar analysis for $\mean{B_{x}}$ is less useful: $-\partial_{z}\mathcal{E}_{y}$
can provide both turbulent diffusion through $\eta_{D}$, and growth of $\mean{B_{x}}$ through  $\mathcal{E}_{y}\sim \alpha_{D}\mean{B_{y}}$;
absent other contributions from mean flows, the two effects will generally cancel so that $\partial_{t}\mean{B_{x}}\approx 0$. This implies that $\partial_{z}\mathcal{E}_{y}\approx0$
does not yield useful information about the relative sizes of growth and diffusion terms. 
 While a number of techniques 
have been developed to overcome this and measure $\alpha_{D}$ and $\eta_{D}$ in simulations (e.g., the ``test-field'' method; \citealt{Schrinner2005}), all have limitations
and are rather complex, falling beyond the scope of this paper. Instead, motivated by the apparent success of quasi-linear dynamo 
theory for predicting $\partial_{z}\mathcal{E}_{x}$ (thin purple line in \cref{fig: by dynamo}), here we attempt  to test the
basic $\alpha$-$\Omega$ dynamo paradigm in which $\alpha_{D}$ is driven 
by the fluctuations' helicity, using the
simplest, standard quasi-linear result for helical velocity and magnetic fluctuations: \begin{equation}
\alpha_{D} \approx -\frac{1}{3\Omega}(\mean{\delta \bm{u}\cdot\nabla\times \delta \bm{u}}- \mean{\delta \bm{B}\cdot\nabla\times \delta \bm{B}}/\rho).\label{eq: alpha mean field}
\end{equation}
We again  assume  a
correlation time  $\Omega^{-1}$   \citep{Pouquet1976,Rincon2019}.
This $\alpha$ effect then enters the mean induction equation \eqref{eq: dynamo x} as $-\partial_{z}(\alpha_{D}\mean{B_{y}})$, 
so 
this  form of dynamo feedback would require $\alpha_{D}<0$ at $z>0$ since $\mean{B_{x}}<0$ is needed to drive $\mean{B_{y}}>0$   when $\partial_{z}\mean{B_{y}}<0$ (likewise, $\alpha_{D}>0$ at $z<0$).\footnote{In principle,
the expression $-\partial_{z}(\alpha_{D}\mean{B_{y}})$ could allow a dynamo driven by an $\alpha_{D}$ gradient instead, in which 
case $\partial_{z}\alpha_{D}>0$ would be needed. However, this is not the standard $\alpha$-$\Omega$ mechanism 
and many other terms beyond the simple scalar expansion for $\alpha_{D}$ in \cref{eq: alpha eta dynamo} may become important also.}
To estimate the value of $\alpha_{D}$ that would be required to drive the observed dynamo, we can use the approximate power-law 
equilibrium \eqref{eq: pure power law equilibria} with $\va/\Omega = z\sqrt{2/a}$ to write $\partial_{z}\mathcal{E}_{y}\sim \alpha_{D}\partial_{z}B_{y}\sim \sqrt{a/2}\Omega \mean{B_{y}} \alpha_{D}/\va$
(where $a/2 = |\partial\ln \mean{B_{y}}/\partial\ln z|$). Thus, to regenerate $B_{x}$ in one orbit requires $\alpha_{D}/\va \simeq(2\pi)^{-1} \sqrt{2/a}\,\mean{B_{x}}/\mean{B_{y}}$;
we have $|\mean{B_{x}}/\mean{B_{y}}|\lesssim 0.05 $ in ZNVF flux cases, suggesting $\alpha_{D}/\va\sim 0.01$ should be sufficient to sustain the dynamo.

We test these ideas in \cref{fig: alpha dynamo} for the simulations shown in \cref{fig: by dynamo,fig: by dynamo all}, all of which exhibit a continuously regenerated $\mean{B_{x}}$.
As well as \cref{eq: alpha mean field}, we plot the contribution from each direction of $\delta \bm{u}$ and $\delta \bm{B}$ fluctuations ($\alpha_{D,u,x}=-\mean{\delta u_{x}(\nabla\times \bm{u})_{x}}/3\Omega$, 
$\alpha_{D,B,y}=\mean{\delta B_{y}(\nabla\times \bm{B})_{y}}/3\Omega$ etc.)  to attempt to extract general 
features and trends.
Most importantly, although the magnitude of $\alpha_{D}$ is in principle sufficiently large to 
produce the observed dynamo feedback, we see no consistent sign of $\alpha_{D}$ between each of the 3 cases shown, including for $\alpha_{D,u}$ and $\alpha_{D,B}$ individually 
(this could be relevant, for example, if the magnetic contribution to $\alpha_{D}$ in \cref{eq: alpha mean field} was more or less efficient at driving a dynamo than the 
velocity contribution; see e.g., \citealt{Gressel2010,Rincon2019}). Instead, we find empirically that the sign of $\alpha_{D}$ seems to be well predicted by whether $\beta>0.03$ (grey-shaded 
regions; $\alpha_{D}>0$),  or $\beta<0.03$ ($\alpha_{D}<0$), a feature that seems robust across 
all of our simulations, including those with net vertical flux (not shown). This change  is driven primarily by the stronger magnetic fluctuations
 at low $\beta$, and may relate to a greater dominance of the Parker instability compared to the MRI, although this is hard to diagnose.
$\beta$R0.1 is of particular relevance to our overall conclusion: with $\alpha_{D}>0$ at $z>0$, it would not sustain the observed $\mean{B_{x}}$ if \cref{eq: alpha mean field} were correct, despite
apparently having a rather similar dynamo to $\beta$0.1-H4 and $\beta$0.001 (for which $\alpha_{D}<0$ at $z>0$).

We thus conclude that the dynamo mechanism regenerating $\bar B_x$ is likely not the standard $\alpha$ effect driven by flow or current helicity.  
This is consistent with previous simulations of buoyancy-driven large-scale dynamos, which have seen similar inconsistencies between the measured sign of helicity and theoretical expectations \citep[e.g., ][and references therein]{Tharakkal2023}. One
possible explanation is that there exist two mechanisms --- driven, e.g., by MRI or the Parker instability --- which can both 
provide the required dynamo feedback but in different ways (e.g., in low-$\beta$ Parker-dominated regions, $\alpha_{D}$ does have the required sign). This 
resembles ideas proposed in  \citet{Johansen2008} and is \revchng{also consistent with analyses of differences between 
midplane and coronal regions in the high-$\beta$ state \citep{Gressel2010,Held2024}. Similarly, the analysis of  \cite{Held2024}  
suggests that local  vertical fields play a key role in low-$\beta$ regions, generating radial fields via vertical gradients in 
the flow; this interplay would not show up in our analysis to our horizontal average but would be interesting to study further in the low-$\beta$ state. }
Another non-exclusive possibility is that the feedback is unrelated to the helicity, instead relying only on the shear flow via off-diagonal 
$\tilde{\eta}_{D,ijk}$ terms \citep{Raedler2006,Rogachevskii2008,Squire2016a}. This would more easily explain the apparent robustness of the field regeneration, 
but, other than at low Reynolds numbers \citep{Squire2015}, the properties of this type of dynamo remain poorly understood \citep{Kaepylae2020,Zhou2021,Mondal2023}.

\begin{figure}
\includegraphics[width=1.0\columnwidth]{\ffold 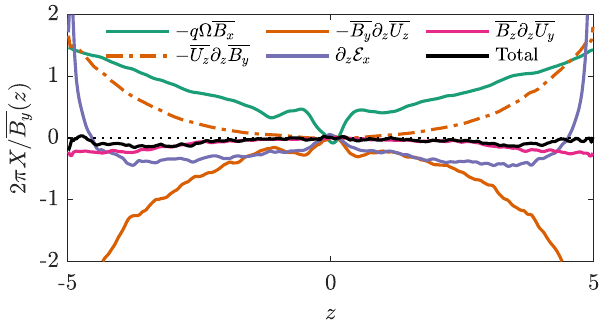}\vspace{-0.86cm}\\
\includegraphics[width=1.0\columnwidth]{\ffold 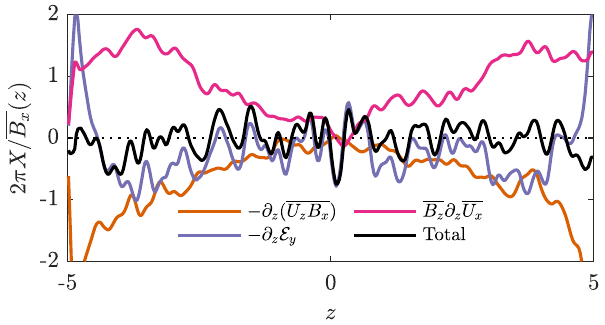}\vspace{-0.06cm}\\
\caption{Dynamo contributions for   $\beta$0.1-NVF2 with net vertical flux. The top panel shows the same analysis as  \cref{fig: by dynamo}, again showing  that $\mean{B_{y}}$
is regenerated against very strong outflows by  stretching of the radial field ($-q\Omega \mean{B_{x}}$), though with a rather different form to the ZNVF case and $\partial_{z}\mathcal{E}_{x}<0$
across effectively the full domain. The lower panel shows the same analysis for $\mean{B_{x}}$, with terms normalized to $\mean{B_{x}}/2\pi$ such that they again 
represent the relative contribution per orbit. We see that $\mean{B_{x}}$ is sustained by a self-generated vertical shear of $\mean{U_{x}}$ stretching the (constant) mean $B_{z}$ ($\mean{B}_{z}\partial_{z}\mean{U_{x}}$). 
This shows that the NVF case is sustained by a {laminar} dynamo; effectively a single large-scale MRI channel mode (see text).}
\label{fig: bz dynamo}
\end{figure}

\subsection{The dynamo loop with net vertical flux}

The dynamo mechanism with net vertical flux requires separate attention, both because of previous results and 
because it  operates  differently. In \cref{fig: bz dynamo} we plot (for $\beta_z = 100$), the dynamo contributions 
from \cref{eq: dynamo} for both $\mean{B_{y}}$ (as in \cref{fig: by dynamo}) and $\mean{B_{x}}$, each normalized as before
to show the relative contribution per orbit at each $z$. The $\mean{B_{y}}$ balance shows significantly stronger outflows than in \cref{fig: by dynamo}, balanced 
by a larger shear-generation term, $-q\Omega\mean{B_{x}}$  due to its strong $\mean{B_{x}}$ (this also causes the large mean magnetic field $\alpha_{\mean{B}}\approx \alpha$ contribution in \cref{fig: with bz}e).
Fluctuations, which have a similar amplitude to the ZNVF low-$\beta$ states, play little role in the balance (see \cref{fig: with bz}d). In the lower panel, we see that $\mean{B_{x}}$ is also being sourced in a laminar way, via stretching of the
mean $\beta_{z}=100$ vertical field by a self-induced vertically sheared radial flow  ($\mean{B_{z}}\partial_{z}\mean{U_{x}}$). 

This behavior, which is clearly different from that at ZNVF,  is simply the manifestation of a single, box-scale MRI channel mode developing across the  domain. Indeed the relative signs,  symmetries, and sizes of $\mean{\bm{B}}$ and $\mean{\bm{U}}$
in this $\beta_{z}=100$ state are   similar to those of a growing MRI mode. 
The solution resembles that discussed in the context of winds in  \citet{Lesur2013} but with the opposite symmetry: here and in \citet{Bai2013a}, $\mean{B_{y}}$ has a peak in the  midplane (like $\cos(z)$);
for the solutions in \citet{Lesur2013}, $\mean{B_{y}}$ passes through zero at $z=0$ (like $\sin(z)$).
By extension, effectively all of the angular
momentum stress comes from this channel mode, explaining the strong dominance of  $\alpha_{\mean{B}}$ compared to the turbulent contributions. 

An additional  important feature is that the symmetry of this $\mean{U_{x}}$ may not be possible to realize in global 
simulations. In particular, as pointed out by \citet{Bai2013a} in the context of the outflow/wind (see their figure 12), this 
$\partial_{z}\mean{U_{x}}>0$ requires that $\mean{U_{x}}$ and $\mean{B_{x}}/\mean{B_{z}}$  have opposite signs above and below the midplane, implying a radial inflow 
for $z>0$ and outflow for $z<0$ or vice versa (depending on the sign of $B_{z}$). The flow is also fast ($\mean{U_{x}}\gtrsim c_{s}$ at larger $z$; \citealp{Bai2013a}) thereby 
requiring a strongly asymmetrical corona that seems at odds with those expected and observed  in corresponding global simulations. Indeed,
rather than the single-signed strong toroidal field seen here and in past local studies, \citet{Zhu2018,Mishra2020} observe a sign flip in $\mean{B_{y}}$ (and $\mean{B_{x}}$) across 
the midplane, with approximately symmetric inflow both above and below the midplane. We note that this state still appears to have
the structure of a single large-scale channel mode, albeit one with the same symmetry as those studied by \citet{Lesur2013}.
This  change causes $\beta\gtrsim 1$ in the midplane of the global solution of \citet{Mishra2020} at $\beta_{z}=100$, although the accretion stress above and below remains mean-field dominated.
While further study of these solutions is clearly needed, 
the analysis  helps to elucidate the cause of the strong influence of net flux on both local and global simulations --- it destabilizes a box-scale MRI mode, 
which then dominates the generation of the mean fields and angular momentum transport, driving strong outflows.

\section{Discussion \& Conclusions}\label{sec: conclusion}

In this work, we have introduced and characterized a novel strongly magnetized state of self-sustaining accretion disk turbulence in the local  stratified shearing box.
The system maintains a strong mean azimuthal field in the midplane, with $\beta\lesssim 1$, modestly sub-Alfv\'enic fluctuations ($|\delta \bm{B}|\lesssim \mean{B_{y}}$), and large accretion stresses, $\alpha\simeq0.5$. 
A net vertical flux --- traditionally thought to be necessary to sustain such large $\alpha$ --- is not required, although 
the field structure and turbulence properties do superficially resemble  previous NVF results \citep[e.g.,][]{Bai2013a} even 
though the sustenance mechanism of  the ZNVF state is different.
So long as the vertical domain size is sufficiently large (in units of $H_{\rm th}$), the state is also robust to  choices of vertical boundary conditions and numerical options --- indeed, we show explicitly that the azimuthal 
flux is continuously escaping through the boundaries.  

We argue that the transition to this low-$\beta$ state, as opposed to the well-known high-$\beta$ cyclic dynamo discussed in many previous works  \cite[e.g.,][]{Davis2010,Simon2012},
is controlled by the initial magnetization of the midplane: if there is sufficient  azimuthal flux after the initial conditions rearrange to create a $\beta\lesssim 1$
midplane and $\beta\ll1$ corona, the system can maintain the low-$\beta$ state; if there is not, it inevitably loses all midplane flux, transitioning into 
high-$\beta$ dynamo cycles. The initial azimuthal (or radial) flux thus acts like a ``switch'' between 
two states of vastly different $\alpha$ in a system with otherwise identical parameters.  Empirically we find that an initial $\beta \simeq 0.1$ in the initial Gaussian toroidal field is the dividing line between the disk evolving to the low-$\beta$ or high-$\beta$ state.  Given the well-known convergence and microphysics dependence of ZNVF shearing box results in the high-$\beta$ state \citep{Fromang2007a,Ryan2017}, the transition between the two states found here, and indeed the properties and existence of the low-$\beta$ state, may also be sensitive to these physical and numerical parameters. This should clearly be studied more in future work, where 
 mesh refinement may be helpful in allowing the thermal midplane to be better resolved than has been possible here.  The importance of turbulent magnetic flux transport in determining the properties of the low-$\beta$ state also specifically motivates future simulations with explicit resistivity and varying magnetic Prandtl number.

Given the variety of rapid state transitions observed in X-ray transients, dwarf novae \cite[e.g.][]{Smak1984} and changing-look AGN \citep{Ricci2023}, 
the strong dependence of the efficiency of angular momentum transport on the ``initial'' magnetic field may be of particular interest. 
For instance, in dwarf novae, $\alpha$ is seen to vary in between $\alpha \sim 0.1$-$0.4$ in the hot state and $\alpha \sim 0.01$-$0.03$ 
in the cold state \citep{King2007,Kotko2012}, rather similar to the difference seen here between the low- and high-$\beta$ states, while
X-ray binaries may exhibit yet higher $\alpha\sim 0.2$-$1$ \citep{Tetarenko2018}.
While changes in the net vertical flux, the MRI,  and/or winds can  potentially 
explain this variability \citep[e.g.,][]{Gammie98,Hirose2014,Begelman2014,Scepi2018,Zhu2018}, \revchng{as well as mechanisms unrelated to changes in the magnetic field \citep[e.g.,][]{King1998,Scepi2024},}
the possibility hinted at by our results has the nice feature that 
extra magnetic flux in \emph{almost any configuration} (toroidal, radial, or \revchng{vertical}) could be sufficient to cause significant changes in the efficiency of angular momentum transport.    
For instance, in feeding such a disk, if a parcel of gas with higher-than average magnetic field
caused the toroidal field strength to cross the $\beta\lesssim1$ threshold, it would trigger the transition from the high- to low-$\beta$ state.   \revchng{As seen in \cite{Machida2006}, whose global simulations reach a $\beta\sim0.1$ state with similar properties to that studied here,} the same could occur if the disk undergoes a thermal collapse, amplifying the toroidal field by flux freezing.
This would result in significantly more rapid accretion, potentially then depleting  the disk of magnetic flux and density to cause an eventual transition back to the
high-$\beta$ state. \revchng{This scenario is not necessarily at odds with various previous proposals; the new insight here is that the magnetic-field strength adds bimodality to the 
local turbulence, with  large, sudden changes to $\alpha$ resulting from small changes in disk parameters.
However, further study with a more detailed treatment of radiation is certainly needed, for instance to understand whether such transitions, which occur over tens of orbital periods in our simulations, disagree with observed smoother transitions through intermediate states (e.g., \citealp{Skipper2016}).}

\subsection{The cause of the state transition}

In the course of analysing the low-$\beta$ state and its transition to the high-$\beta$ state, we have run a wide variety of simulations with different initial conditions 
and other options, leading to various more detailed conclusions. These are listed in \S\ref{sec: key points}, to which we refer
the reader for a more in depth summary of our computational results. Forgoing a summary here, we instead propose a scenario for how the system attains bimodality, giving rise to the distinct  low- and high-$\beta$ states. This physics remains far from certain --- indeed our
attempts at diagnosing the field-generation mechanism remained partially unsuccessful (see \S\ref{sub: radial field generation}) --- but 
may provide a helpful framework for further development. 
The scenario has three main elements:
\vspace{-0.2cm}\paragraph{(1) Azimuthal flux must be continuously generated} The disk midplane and its atmosphere are highly unstable, exhibiting strong trans- and super-sonic 
turbulence. This rapidly transports flux upwards, implying that in order to sustain itself, the system must continuously generate flux through a (non-cyclic) large-scale dynamo. 
The flux is regenerated and removed on several-orbit timescales, as shown in \cref{fig: by dynamo,fig: by dynamo all}, implying the dynamo must be efficient and robust. Given that the mean shear  generates strong azimuthal 
flux from radial flux, the turbulent fluctuations must generate radial flux from azimuthal flux to ``close the loop.''
\vspace{-0.2cm}\paragraph{(2) Parker-instability-driven turbulence generates radial from azimuthal flux}  
Both the Parker instability and MRI seem to play important roles in maintaining the low-$\beta$ state, as evidenced by \revchng{the similarity of the midplane turbulence to the high-$\beta$ state (\cref{fig: alpha})},
our simple turbulence  model  (\cref{fig: turbulence}), and previous works \citep{Johansen2008,Held2024}.
However,  evidence suggests that fluctuations driven by the Parker instability generically and naturally 
generate azimuthal from radial flux, with MRI playing a subsidiary role.
First,  even when the system  is not in steady state with $\beta$ increasing in time, the dynamo still 
operates (flux is just being lost faster than it is created; \cref{fig: by dynamo all}); in this regime 
$\beta\ll1$ so the MRI is suppressed (likewise $\beta\ll1$ resolution-limited states sustain a dynamo). Second, 
in the midplane  ($|z|\lesssim H_{\rm th}$) the net radial field is absent or of the wrong sign to support the dynamo (\cref{fig: butterfly,fig: by dynamo}); 
this is presumably where MRI drives turbulence, since the Parker instability is stable.
Together these suggest that Parker-driven fluctuations drive the dynamo itself, while MRI is likely needed
to drive midplane turbulence that can diffuse $\mean{B_{y}}$ to maintain the strongly magnetized midplane.
\vspace{-0.25cm}\paragraph{(3) As $\beta$ grows, the Parker instability shuts off near the midplane} This is a simple consequence of the Parker instability criterion, $z\, \partial \ln B_{y}/\partial z<0$ (see \S\ref{subsec: parker}),
and the magnetized equilibrium. For  $\beta_{\rm mid}\gtrsim 1$, gravitational support is provided by thermal  rather than magnetic pressure, implying that $P_{B}$ flattens
(or reverses its gradient) some distance above the midplane. Thus, as $\beta_{\rm mid}$ rises above unity, the Parker-unstable region moves further from the midplane to larger $|z|/H_{\rm th}$.

\vspace{0.1cm}Together with points (1) and (2) --- that the Parker  instability is needed for sustaining the azimuthal flux, which rapidly escapes 
if it is not continuously regenerated ---
point (3) implies that the Parker-induced feedback  becomes less efficient at regenerating the mean field for $\beta_{\rm mid}\gtrsim1$. This would 
then further raise $\beta$ by its inability to regenerate more flux, running away until 
the flux is completely expelled from the midplane. Thus results the high-$\beta$ state, which presumably sustains turbulence 
via a different midplane mechanism (unrelated to the Parker instability) exhibiting characteristic ${\sim }10$-orbit reversal of $\mean{\bm{B}}$, $\delta \bm{B}\gg \mean{\bm{B}}$, and $P_{B}\ll P_{\rm th}$. 
This scenario thus qualitatively explains why the system switches states based on the initial flux, with no ``intermediate'' state available (see \cref{fig: flux beta alpha}) --- as $\beta_{\rm mid}$ rises above unity, the feedback mechanism becomes continuously weaker. 
If we further assume that the primary role of the MRI is to generate midplane turbulence, the scenario remains consistent with the behavior at $\beta \ll 1$. Specifically: (i) a steady state with $\beta \ll 1$ cannot be maintained despite a similar dynamo, because the MRI fails to sustain supersonic turbulence at $z = 0$ (where Parker instability shuts off), which would  be required to support a density scale height $\gg H_{\text{th}}$; and (ii) in contrast, a steady state with $\beta \ll 1$ can persist if resolution is insufficient --- when the midplane is poorly resolved, physical (turbulent) diffusion is replaced by numerical diffusion, preventing the collapse of the density.


\subsection{Subsidiary results}
In addition to the main results of this work, we have noted subsidiary results relating to the high-$\beta$ dynamo cycles and
turbulence with net vertical flux:
\vspace{-0.2cm}\paragraph{Dynamo cycles} Periodic reversals of $\mean{B_{y}}$, which have been traditionally associated with the MRI, appear to persist in the absence of rotation, 
albeit with some morphological differences \revchng{(see App.~\ref{app: shear and rotation}, \cref{fig: no rotation})}. This does not seem to disagree with previous works, which 
mostly have not explored the effect of setting $\Omega=0$, but may be of interest for understanding the mechanism via which this turbulence self sustains (idealized
dynamo-focused simulations exhibit cyclic behavior without rotation, so this is perhaps not unexpected; e.g.,  \citealt{Brandenburg2008,Squire2015b}). Similar cycles are also observed with 
a net vertical flux in the absence of rotation \revchng{(\cref{fig: with bz no rot})}.
\vspace{-0.2cm}\paragraph{MRI channel modes with net vertical flux}  With strong NVF ($\beta_{z}\lesssim100$), the large angular momentum transport is almost entirely driven by the
mean magnetic-field stress ($\alpha_{\mean{B}}=-\mean{B_{x}}\,\mean{B_{y}}$;  \citealp{Bai2013a}), with $\mean{B_{x}}$ and $\mean{B_{y}}$ maintained  by the mean flows ($\partial_{z}\mean{U_{x}}$ stretches $\mean{B_{z}}$ into $\mean{B_{x}}$; the Keplerian flow stretches $\mean{B_{x}}$ into $\mean{B_{y}}$). This structure is effectively a single MRI 
channel mode stretching across the full vertical extent of the domain, although its appearance is rather similar to the low-$\beta$ ZNVF state.
As also noted  by \citet{Bai2013a} in the context of the winds produced by their simulations, the symmetry of the solution likely implies it cannot occur in more realistic, global settings, 
since it requires strong (supersonic) oppositely directed radial flows on either side of the  midplane to sustain the mean fields and large $\alpha$. Indeed, unlike the shearing box, which 
exhibits a single-signed $\mean{B_{y}}$ at any particular
time, similar ($\beta_{z}\approx 100$) global NVF simulations develop strong ($\beta\lesssim 1$) mean fields
that reverse their polarity across the midplane (\citealp{Mishra2020}; see also \citealt{Zhu2018}).
We note in passing that this global structure  in \citet{Zhu2018,Mishra2020} is also essentially a single
channel mode, but with the opposite symmetry about $z$: effectively $\mean{B_{y}}\propto \sin(z)$ and $\mean{U_{x}}\propto \cos(z)$ in the global domain, rather than $\mean{B_{y}}\propto \cos(z)$ and $\mean{U_{x}}\propto -\sin(z)$ in the local one. This change allows the global system to maintain a symmetric flow structure across the midplane.

These aspects, which have only been touched upon in this work insofar as they related to our main results, deserve further study in future work. \newpage

\subsection{Application to global models \revchng{and $\beta\ll1$ fields}}\label{subsec: global apps}

Some of the most important questions raised by this work relate to its application 
to global simulation results, e.g., those of \citet{Gaburov2012}, \hpaper, and \citet{Guo2024}.   \citet{Gaburov2012} studied the disruption of a magnetized gas cloud by a black hole and found disks with midplane $\beta \sim 0.1$ supported almost entirely by magnetic pressure.   The global multi-physics solutions of \hpaper, which form self-consistently from 
cosmological initial conditions via capture of gas from a cloud complex, evolve to
an extremely strongly magnetized disk, with $\beta$ reaching ${\sim}10^{-4}$ in the midplane and large accretion rates $\alpha\gg1$.  \citet{Guo2024} studied the fueling of M87* and found that cooling of the hot intracluster gas produces a disk dominated by strong toroidal magnetic fields with $\beta \sim 10^{-2}$.   \revchng{Other global simulations have likely also yielded similar 
states, for instance \citet{Machida2006}, who observed a $\beta\sim0.1$ magnetically supported solution emerge from a cooling disk that contracted vertically, or \citet{Sadowski2016}, who reported a thermally stable strongly magnetized   disk 
arising from idealized initial conditions with a net radial field.}

Our results capture a number of general features of these solutions, 
 such as the modestly sub-Alfv\'enic fluctuations and dominant, \revchng{single-polarity toroidal field. 
 While local simulations with NVF also create strong toroidal fields and some outwardly similar features, 
 the different morphology observed in  global simulations with NVF \citep[e.g.,][]{Zhu2018,Mishra2020}, suggests that the results of  \hpaper, \citet{Guo2024}, and 
 \citet{Gaburov2012} do not result from ``patches'' of vertical field driving the MRI (moreover, these simulations do not appear to involve a  vertical field that is at all coherent; see, e.g., \hpaper\ figure 6).} 
The vertical  profile shape of the magnetic field and 	density found here are also very similar to those in \hpaper\ and \citet{Guo2024}.  

\revchng{However,  one key feature in which our solutions differ markedly is the midplane $\beta$: while some of these global disks sustain  $\beta \ll 1$ in the midplane, our local simulations always
collapse to $\beta\sim1$ given sufficient resolution}.  This difference in midplane $\beta$ is particularly important in understanding AGN, since low $\beta$ leads to much lower gas densities, which helps the disk avoid fragmentation due to gravitational instability (a long-standing problem in fueling luminous AGN; \citealt{Goodman2003}).

Given that our local simulations also 
saturate at $\beta\ll1$ when $H_{\rm th}$ is poorly resolved (the density collapses to the grid resolution; see \cref{fig: z profiles}), one possible cause of the difference is simply that the global cases are affected by resolution, artificially lowering $\beta$.   However, it is also plausible that physical processes not captured in the shearing box simulations presented here could play a significant role, leading to differences. \revchng{Clearly, our simulations omit global effects, and we discuss below how mechanisms such as global instabilities or radial advection of flux may be needed to sustain low $\beta$. Similarly,  \hpaper\ in particular incorporates a much wider set of physical processes --- e.g., multiphase thermodynamics, accretion from a dynamic interstellar medium (which induces warps), radiation, and self gravity --- which may also have relevance.}

Motivated by the limitations of the shearing box in capturing some potentially important physics, in a companion paper (Guo et al., in prep) we use \textsc{Athenak} simulations to study a simple model of the formation and evolution of a global disk initially dominated by a toroidal magnetic field, which forms from the inflow of a rotating gas cloud seeded with azimuthal magnetic flux.   The global results in Guo et al.~(in prep) are similar in many respects to the low-$\beta$ state presented here
\revchng{and seem to exhibit a similar resolution dependence, showing a $\beta \ll 1$ midplane only when the thermal scale-height of the disk is not resolved (see App.~\ref{app: resolution dep}). This comparison of idealized local and global results supports the idea that the low-$\beta$ state is robust but that a midplane with $\beta \ll 1$ is not a generic outcome}.


The idealized global simulations in Guo et al.~(in prep) and the local simulations presented here still differ dramatically in physics complexity from the simulations of \hpaper.\footnote{This is less true for \citealt{Gaburov2012} who studied a more idealized problem; indeed the idealized global model of Guo et al.~(in prep) was chosen to be similar to \citet{Gaburov2012}'s calculation.}  It thus remains an open possibility that in some circumstances global simulations can realize a (resolved) $\beta \ll 1$ low-$\beta$ state, even though the local simulations and simplified global models do not.   To speculate on the conditions under which $\beta\ll1$ may be sustained, we refer back to \cref{fig: by dynamo all}, which 
shows: (i) even as the system is losing net flux, it is regenerated near the midplane (just lost more slowly than it is produced) and (ii) 
that field is regenerated well above the midplane and sustains $\beta_{\rm mid}\ll1$,  so long as the 
midplane density collapse is halted (in this case, by the grid resolution). These observations hint that the \emph{cause}
of the loss of flux (leading to $\beta_{\rm mid}\sim1$) could be the density collapse: in the absence of some particular  scale height $H_{\rho}$ for the
system to ``choose,'' it collapses to (nearly) $H_{\rm th}$. If so, 
then one might expect that any other physical process that halts the density collapse with $H_{\rho}\gg H_{\rm th}$ might also --- as a side effect --- 
halt the loss of flux. Another (non-exclusive) possibility is the opposite: that halting the loss of flux could stop the collapse of the density. 

As relates to the former possibility, there are a number of ways in which the density collapse could be halted in the global case. 
\revchng{Other than radiation, which can provide another non-thermal pressure that supports the midplane against collapse \citep[e.g.,][]{Jiang2020}}, the most 
obvious possibility is supersonic  turbulence in the midplane (it must be supersonic so 
that the turbulent pressure $P_{\rm turb}$ exceeds $P_{\rm th}$, thus leading to a larger scale-height and lower midplane density). As discussed extensively in \hpaper\ and \S\ref{subsec: MRI}, 
there exist other global instabilities 
that can
have faster growth rates than the standard MRI at $\beta\ll1$ \citep{Pessah2005,Das2018}.
These are driven by global currents and thus depend on the radial gradient of the (super-thermal) toroidal field, 
reducing to the local MRI when ${d\ln B_{\phi}}/{d\ln R}=-1$ such that  $\bm{J}\propto \nabla\times (\mean{B_{\phi}}\hat{\bm{\phi}})\approx r^{-1} \partial_{r}(r \mean{B_{\phi}})\hat{\bm{z}}\approx0$
(see \S\ref{subsec: MRI} and \citealt{Begelman2023}).
Such instabilities could plausibly produce supersonic turbulence in global simulations, and it is possible that they would be present in \hpaper\ but not in Guo et al's more idealized global simulations,
given the different conditions in each.   In support of this idea, we note that the 
$B_{\phi}$ gradient in \hpaper's simulation is close to, but not exactly, $\hat{B}=-1$ (see their figure 3), as would be expected if the instability was driving the system close to marginality (see also their figure 16).   If correct, this implies the existence of yet another switch-like transition between accretion states, with the system 
choosing between (i) ultra-low $\beta$, with active global instabilities and the Parker dynamo maintaining $\beta\ll1$, (ii) low $\beta$, as studied here in the shearing box, with $\beta_{\rm mid}\sim 1$, and (iii) 
high-$\beta$ cycles. The choice between these would be determined primarily by the total azimuthal and/or radial magnetic flux available, as well as ${d\ln B_{\phi}}/{d R}$.


In addition to the above scenario, the maintenance of $\beta\ll1 $ in a global context could be explained by various other 
\revchng{effects, from radiation pressure and multi-phase gas structure to an additional source of magnetic flux.
The latter could stop }
the density's collapse by maintaining $\beta\ll1$ (i.e., the reverse causation to that discussed above)
if it were sufficient to overwhelm vertical field losses. Such a source could arise as a simple consequence of radial flux advection, as discussed extensively in \hpaper, adding additional 
growth and loss terms into the $\partial_{t}\mean{B_{y}}$ equation \eqref{eq: dynamo}, which depend on the profile and relative
compression of $B_{\phi}$ and $B_{R}$ as they are transported in the disk.  
Indeed, because of the large accretion rates, the accretion timescales in the \hpaper\ simulation can be extremely short, of order several orbital times,
and thus potentially shorter than the timescale over which the flux is lost vertically in our local simulations.
While such radial advection seems insufficient to maintain $\beta \ll 1$ at the midplane in Guo et al.~(in prep), these simulations explored just one class of idealized setup; other global solutions may advect flux inwards more efficiently.


Overall, clearly more work on this subject is needed. While global setups will undeniably be the most relevant for study of these issues, aspects of the \citet{Pessah2005}
instabilities may be possible to capture in local domains.   The importance of thermodynamics, more complex initial conditions, and other physics present in various global studies should also be explored to see if they influence the magnetization in the low-$\beta$ state.   Whatever the outcome, there clearly exists a rich array of behaviors in strongly magnetized disks, in particular
various \revchng{magnetic} ``switches'' --- state transitions that depend on the magnetic-field strength and geometry. This has interesting applications for our understanding of quasars, X-ray binaries, CVs, and other high-energy accreting systems.

\section*{Acknowledgments}

We thank Omer Blaes, Charles Gammie, Minghao Guo, Matt Kunz,  Eve Ostriker,  Jim Stone,   Yashvardhan Tomar, and especially Yuri Levin for enlightening discussion, as well as the anonymous referee, whose useful suggestions improved the manuscript.    JS acknowledges the support of the Royal Society Te Ap\=arangi, through Marsden-Fund grant MFP-UOO2221  and  Rutherford Discovery Fellowship RDF-U001804. This work was also supported in part by Simons Investigator grants from the Simons Foundation (PH, EQ) and by NSF AST grant 2107872.  This research was part of the Frontera computing project
 at the Texas Advanced Computing Center, which is made possible by National Science Foundation award OAC-1818253. Further computational support was provided by the New Zealand
 eScience Infrastructure (NeSI) high performance computing facilities, funded jointly by NeSI’s
 collaborator institutions and through the NZ MBIE.  This research benefited from interactions at the Kavli Institute for Theoretical Physics, supported by NSF PHY-2309135.

\bibliographystyle{aasjournal}


\begin{appendix}


\begin{table*}[htb]
\tiny
\begin{center}
\scalebox{1}{
 \begin{tabular}{c c c c c c c c c c c}  Name  & $\beta_{y0} $ & $\beta_{x0} $ &  $\beta_{z} $ &  $H_{\rm th}=\sqrt{2}\tfrac{c_{s}}{\Omega} $ & $N_{x}\times N_{y}\times N_{z}$ & $(t_{\rm av},t_{\rm fin})\tfrac{\Omega}{2\pi}$ & BCs &  Notes & Outcome & $\langle \alpha\rangle$ \\ [0.5ex] 
\hline\hline
$\beta$R0.1-long & 0.1 & 100  & $\infty$ & 0.43 & $336^{3}$  & (10,30) & Power-law & $L_{y}=40,\,L_{x}=10$  & Low-$\beta$  &  0.61 \\
\hline
$\beta$R0.1-short & 0.1 & 100  & $\infty$ & 0.43 & $336\times168\times 336$   & (25,32) & Power-law & $L_{y}=10,\,L_{x}=10$ & High-$\beta$  &  <0.05  \\
\hline
$\beta$R0.1-narrow & 0.1 & 100  & $\infty$ & 0.43 & $168\times336^{2}$  & (10,32) & Power-law & $L_{y}=20,\,L_{x}=5$ & Low-$\beta$  &  0.55  \\
\hline
$\beta$R0.1-short-narr & 0.1 & 100 & $\infty$ & 0.43 & $168^{2}\times 336$  & (30,60) & Power-law & $L_{y}=10,\,L_{x}=5$ & High-$\beta$  &  0.01  \\
\hline
$\beta$R0.05-short & 0.05 & 100  & $\infty$ & 0.31 & $336\times168\times 336$  & (10,34) & Power-law & $L_{y}=10,\,L_{x}=10$ & Low-$\beta$  &  0.44  \\
\hline\hline
$\beta$0.1-224 & 0.1 & $\infty$  & $\infty$ & 0.43 & $224^{3}$  & (10,40) & Power-law &   & Low-$\beta$  &  0.37 \\
\hline
$\beta$0.1-168 & 0.1 & $\infty$  & $\infty$ & 0.43 & $168^{3}$  & (10,40) & Power-law &   & Low-$\beta$  &  0.31 \\
\hline
$\beta$0.1-112 & 0.1 & $\infty$  & $\infty$ & 0.43 & $112^{3}$  & (10,40) & Power-law &   & Mixed  &  0.24 \\
\hline
$\beta$0.1-84 & 0.1 & $\infty$  & $\infty$ & 0.43 & $84^{3}$  & (20,40) & Power-law &   & Mixed (see \cref{fig: resolution dependence})  &  0.18 \\
\hline
$\beta$0.1-56 & 0.1 & $\infty$  & $\infty$ & 0.43 & $56^{3}$  & (20,40) & Power-law & Non-stationary & Low-$\beta$  &  0.48 \\
\hline
$\beta$0.1-36 & 0.1 & $\infty$  & $\infty$ & 0.43 & $32^{3}$  & (20,40) & Power-law &  & Low-$\beta$  &  1.8 \\
\hline\hline
$\beta$0.1-mr-lowflr & 0.1 & $\infty$  & $\infty$ & 0.43 & $336^{3}$  & (63.7,71.5) & Outflow & Restart of  $\beta$0.1-mr & Low-$\beta$  &  0.50 \\
\hline
$\beta$0.1-lr-lowflr & 0.1 & $\infty$  & $\infty$ & 0.43 & $224^{3}$  & (15,40) & Power-law &  $\rho_{\rm flr}=3\times 10^{-5}$ & Low-$\beta$  &  0.35 \\
\hline
$\beta$0.1-lr-lowerflr & 0.1 & $\infty$  & $\infty$ & 0.43 & $224^{3}$  & (15,26) & Power-law &  $\rho_{\rm flr}= 10^{-5}$ & Low-$\beta$  &  0.37 \\
\hline\hline
$\beta$R0.1-noshear & 0.1 & 100 & $\infty$ & 0.43 & $336^{3}$ & (10,48) & Power-law & no shear & High-$\beta$  &  -0.014 \\
\hline
$\beta$R0.1-norot & 0.1 & 100 & $\infty$ & 0.43 & $336^{3}$ & (10,48) & Power-law & no rotation & High-$\beta$  &  0.014  \\
\hline
$\beta$0.01-noshear & 0.01 & $\infty$ & $\infty$ & 0.14 & $336^{3}$ & (20,34) & Outflow & no shear & High-$\beta$  &  0.17 \\
\hline
$\beta$0.01-norot & 0.01 & $\infty$ & $\infty$ & 0.14 & $336^{3}$ & (20,42) & Outflow & no rotation & High-$\beta$  &  0.14  \\
\hline
$\beta$R0.1-NVF2-norot & 0.1 & $100$ & $100$ & 0.43 & $336^{3}$ & (5,27) & Power-law & no rotation & High-$\beta$  &  0.17 \\
\hline\hline
\end{tabular}
}
\end{center}\caption{A list of the  simulations discussed through the Appendices. These are used to test the dependence on numerical parameters, \revchng{as well 
as the effect of selectively removing rotation or shear.} Naming conventions and  
columns are the same as \cref{tab:sims}.}
\label{tab:sims app}
\end{table*}

\section{Dependence on box size}\label{app: box size dep}

\begin{figure}
\vspace{0.4cm}
\includegraphics[width=1.0\columnwidth]{\ffold 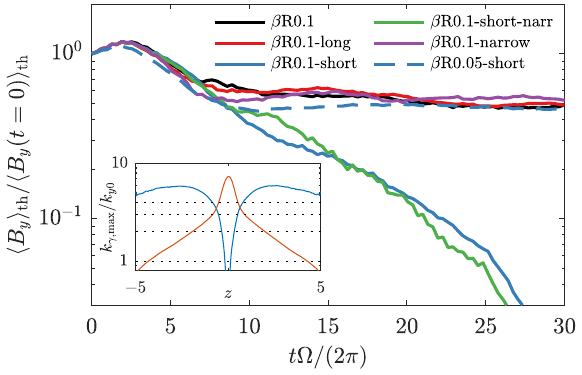}\\
\caption{Testing the effect of changing the horizontal box dimensions on the transition to the low-$\beta$ state, as diagnosed by the time evolution 
of $B_{y,{\rm mid}}/B_{y0}$ in the main panel (see \cref{fig: flux beta alpha}; plots of $\beta_{\rm mid}$ or $\alpha$ show similar features). 
The system appears robust to the choice of $L_{x}$, as indicated by $\beta$R0.1-narrow reaching the low-$\beta$ state, as well as to longer 
box lengths in the azimuthal direction ($\beta$R0.1-long). But, decreasing $L_{y}$ by a factor of 2 in $\beta$R0.1-short and $\beta$R0.1-short-narr (both with $L_{y}=L_{z})$ causes
an otherwise identical system to lose its flux and transition to the high-$\beta$ state. At lower $\beta$ and smaller $H_{\rm th}$ it can maintain 
the low-$\beta$ state even at $L_{y}=L_{z}$ ($\beta$R0.05-short). 
The inset shows the azimuthal wavenumber $k_{\gamma,{\rm max}}$ of the 
fastest growing Parker mode (blue; \cref{eq: parker wavenumber}) and MRI mode (red; $k_{y}\approx\Omega/v_{{\rm A}y}$) as
a function of $z$ for the profile of the $\beta$R0.1 run,  normalized by $k_{y0}=2\pi/L_{y}$. The dotted lines show the first four wavenumbers that fit in the box. }
\label{fig: box size dependence}
\end{figure}

In \cref{fig: box size dependence} we diagnose the dependence of our 
results on horizontal domain size by systematically changing $L_{y}$ and $L_{x}$ by factors
of 2 in the various simulations listed in \cref{tab:sims app}. The initial conditions and parameters are as for  $\beta$R0.1 (though with $N_{z}=336$ and Power-Law boundary conditions), which reach the low-$\beta$ state with the  default 
size $L_{x}=L_{y}/2=L_{z}$. Changing $L_{x}$ does 
not appear to have an effect, although it does lead to larger relative fluctuations in the mean field (not shown), 
while increasing $L_{y}$ leads to no noticeable differences in the low-$\beta$ state itself or how it is approached. However, 
decreasing $L_{y}$ by a factor of 2 causes the system to lose its 
flux and enter the high-$\beta$ state (the cyclic phases are not shown but do emerge at late times), while this same (smaller) $L_{y}$,  retains its flux in the low-$\beta$ state 
if the initial $\beta$ and $H_{\rm th}$ are also decreased (simulation $\beta$0.05-short). This is
consistent with the idea discussed in the main text that $\beta\approx 0.1$ initial conditions lie close to the boundary where the system can reach either 
the low- or high-$\beta$ state.

These results  suggest that  long-wavelength modes in $y$ are needed to sustain the low-$\beta$ state. The clear 
candidates are the MRI (\S\ref{subsec: MRI}) and/or the  Parker instability (\S\ref{subsec: parker}), which both have 
peak growth rates at finite $k_{y}$. To assess this physics, in the inset of \cref{fig: box size dependence} we plot $k_{\gamma,{\rm max}}$, the $k_{y}$
of the fastest-growing mode, for the quasi-interchange Parker instability (blue; from \cref{eq: parker wavenumber}) and the MRI (red; $k_{\gamma,{\rm max}}=\Omega/v_{{\rm A}y}$),
using the $x$-$y$ averaged fields of $\beta$R0.1. We see that the MRI and Parker instabilities appear well resolved at small and large $z$, respectively, 
although at $z\simeq 1$, just outside of the central density hump, it is closer to marginal (especially if $k_{y0}=2\pi/L_{y}$ were increased by a factor 
of two, as in $\beta$R0.1-short). 
It is also worth noting that the Parker-instability analysis in \S\ref{subsec: parker} neglects shear 
and rotation, and these effects --- especially the shear-induced  time-dependence of the radial wavenumber --- 
could change the system significantly \citep{Foglizzo1994,Foglizzo1995}. 

Given our hypothesis that 
MRI is necessary for maintaining turbulence in the midplane, while the Parker instability drives stronger turbulence and regenerates the field above, our tentative
conclusion is that scales several times larger than that of the fastest-growing Parker quasi-interchange mode are needed to sustain 
the low-$\beta$ state. This is qualitatively consistent with the arc-merging picture seen in 2D Parker-instability simulations in \citet{Johansen2008} 
as well as the fact that the Parker scale \eqref{eq: parker wavenumber} depends on $H_{\rm th}$ while that of the MRI does not (thus, the MRI alone would not explain the results of $\beta$0.05-short). However, 
we cannot rule out other possible explanations at the present time and a more detailed study is needed.

\section{Dependence on resolution}\label{app: resolution dep}
\begin{figure}
~\includegraphics[width=1.0\columnwidth]{\ffold 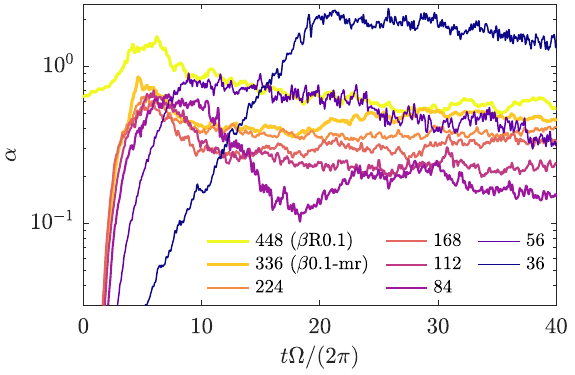}\vspace{-0.06cm}\\
\includegraphics[width=1.0\columnwidth]{\ffold 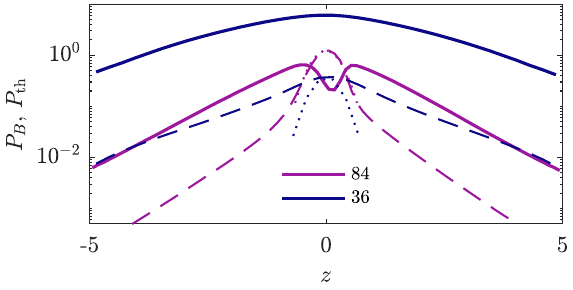}\vspace{-0.06cm}\\
\caption{Dependence of the low-$\beta$ state on numerical resolution. The top panel shows $\alpha(t)$
for a set of nearly identical simulations with $H_{\rm th}=0.43$ ($\beta_{y0}=0.1$) but with different $N_{z}$ as labeled in the legend (all cases have $N_{x}=N_{y}=N_{z}$). 
The bottom panel shows the time-averaged mean magnetic and thermal pressure profiles (solid and dashed lines, respectively), for the case with minimum transport ($N_{z}=84$) and the coarsest-resolution case, with high transport ($N_{z}=32$; we normalize such that $\int dz\,P_{\rm th}=1$).
Finer resolution cases approach those shown in the main text (\cref{fig: z profiles,fig: z profiles numerics}).
We see interesting non-monotonic behavior: if the thermal midplane is resolvable (for $N_{z}\gtrsim 56$ here), the level of turbulence, and thus $\alpha$, decreases
with coarsening resolution, which also enhances the collapse to the thermal profile in the midplane; if the midplane is not 
resolvable ($N_{z}\lesssim 56$) the flux is maintained, giving a $\beta\ll1$ midplane with significantly larger transport.}
\label{fig: resolution dependence}
\end{figure}

We explore the dependence of the results on numerical resolution via a set of simulations with 
successively {fewer grid points} and the same parameters as $\beta$0.1-mr (see \cref{tab:sims app}).
This reveals interesting non-monotonic behavior,  illustrated in the top panel of \cref{fig: resolution dependence}; the transport level decreases with {coarsening} resolution up to some point (around $N_{z}=56$ here), then suddenly increases again to $\alpha \gtrsim 1$.
This behavior can be explained by 
appealing to the phenomenology of the $\beta\ll1$ state described in \S\ref{sec: low beta}.
If the resolution is sufficiently \revchng{fine} that the thermal density profile can be resolved,
$\alpha$ decreases with \revchng{coarsening} resolution because the midplane turbulence amplitude decreases\footnote{Note that this is the reverse of the behavior seen at high resolution in the high-$\beta$ state, where the turbulence amplitude drops with finer resolution \citep{Ryan2017}. Presumably this difference arises because the low-$\beta$ state involves larger-scale fluctuations and has a stronger
mean field (see \cref{fig: snapshot}).} (not shown; 
the turbulence amplitude in the corona is not a strong function of resolution).  
As the turbulence level drops, the density in the midplane collapses to the thermal profile across 
a wider range in $z$, and thus starts to resemble the high-$\beta$ state, reaching $\beta >1$ 
in the midplane in poorly resolved cases. This is illustrated  in the lower panel of \cref{fig: resolution dependence} for 
 $\beta$0.1-84 (cf. \cref{fig: z profiles}).
However, at resolutions below ${\simeq} 1.5$ zones per $H_{\rm th}$ for this setup ($\beta$0.1-56 has 1.7 zones per $H_{\rm th}$),
the density collapse is halted by the grid itself, maintaining a $\beta\ll1$ midplane as a side effect. The midplane 
turbulence also maintains a larger amplitude, perhaps via overshoot from the Parker-unstable atmosphere, thus causing 
much larger transport (the $\beta$0.1-32 simulation also has stronger outflows and has lost around half its density by $t\Omega/2\pi=40$).

This behavior is broadly consistent with that discussed in the main text, although there do also appear to be 
important effects related to $L_{z}/H_{\rm th}$ and/or the initial conditions. 
In simulation $\beta$0.001, which has ${\simeq}1.1$  
zones per $H_{\rm th}$ like $\beta$0.1-32,  but with a much smaller $H_{\rm th}/L_{z}$, the density collapse is also supported by the resolution and it maintains a high-transport $\beta\ll1$ midplane.
However, $\beta$0.01 and $\beta$0.01-hr, which  have resolutions per $H_{\rm th}$ similar to $\beta$0.1-112 and  $\beta$0.1-224 respectively,
maintain $\beta<1$ in the midplane, albeit with a nearly thermal density profile across a wider range of $z/H_{\rm th}$  and  a flatter $P_{\rm B}$ profile (see \cref{fig: z profiles,fig: low beta approach}c). Similarly $\beta$0.1-tall has a transport level somewhat less than half that of $\beta$0.1-224, which has the same resolution 
per $H_{\rm th}$ but a smaller vertical extent (the different boundary conditions may also be important in this difference, since we see from \cref{tab:sims} that Outflow conditions 
\begin{figure}[h]
~~~~~\includegraphics[width=0.98\columnwidth]{\ffold 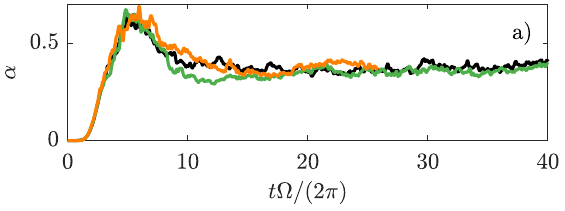}\vspace{-0.06cm}\\
\includegraphics[width=1.0\columnwidth]{\ffold 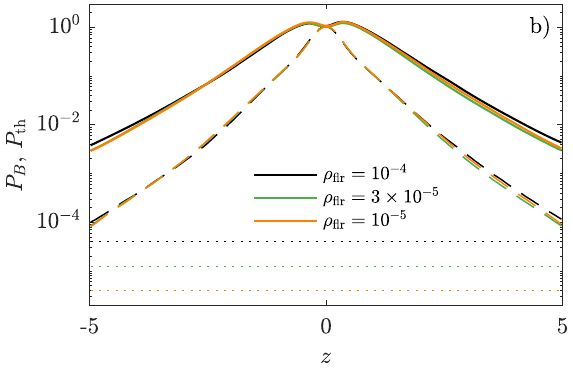}\vspace{-0.06cm}\\
\includegraphics[width=0.49\columnwidth]{\ffold 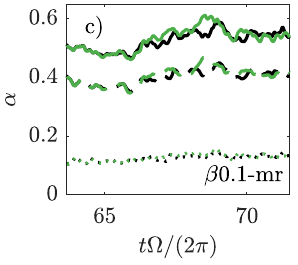}\includegraphics[width=0.51\columnwidth]{\ffold 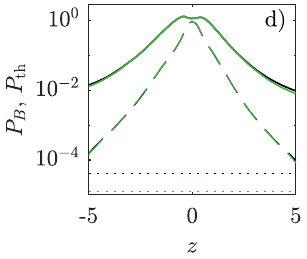}\vspace{-0.06cm}\\
\caption{Panels a-b show the effect of changing the density floor below the default value of $\rho_{\rm flr}=10^{-4}$ by 
a factor of ${\simeq}3$ ($\beta$0.1-lr-lowflr; $\rho_{\rm flr}=3\times10^{-5}$) and $10$ ($\beta$0.1-lr-lowerflr; $\rho_{\rm flr}=10^{-5}$).
Panel (a) shows $\alpha(t)$, while (b) shows the time-averaged density and pressure profiles, as in \cref{fig: resolution dependence}. The
horizontal dotted lines illustrate the density floor in each case.
Panels (c)-(d) show the similar test of restarting $\beta$0.1-mr at $t\Omega=400$ with a lower floor $\rho_{\rm flr}=3\times10^{-5}$.}
\label{fig: density floor}
\end{figure}
lead to somewhat lower $\langle \alpha\rangle$). While some of these aspects remain incompletely understood and worthy of further study, a full investigation is beyond the scope of this work.

\section{The effect of the density floor}\label{app: density floor dep}

Most simulations in the main text impose a floor on the density at $\rho_{\rm flr}=10^{-4}$ of the initial central density, which 
is important for stabilizing the numerical scheme and allowing somewhat larger timesteps. 
While undesirable, this is needed because the large $\va$ and highly supersonic turbulence 
at large $z$ 
 cause frequent  code crashes unless a very conservative CFL number is used (${\lesssim}0.1$), 
 significantly increasing the cost of the simulations (which are already far more expensive than those in the high-$\beta$ regime).
While  the low-$\beta$ simulations studied in the main text do not show any obvious effects and mostly have a minimum mean density that remains a factor of $\simeq2$-$3$
above the chosen floor, it is important to check that the maintenance of the low-$\beta$ state does not depend on $\rho_{\rm flr}$.

We do this through two sets of simulations that change $\rho_{\rm flr}$ in otherwise identical setups. In the first set ($\beta$0.1-224, $\beta$0.1-lr-lowflr and $\beta$0.1-lr-lowerflr), we 
decrease $\rho_{\rm flr}$ by factors of ${\simeq}3$ and $10$, starting from the standard Gaussian initial conditions. The time evolution of the transport and time-averaged profiles are shown in \cref{fig: density floor}a-b, demonstrating no statistically significant differences in the steady state, other than a slightly  (factor ${\simeq}1.1$-$1.2$) lower
density towards the domain's edge, which also decreases $\mean{B_{y}}$ due to force balance. The initial evolution is somewhat different because the density floor impacts the initial Gaussian 
density profile near the edge of the domain; this in itself is not cause for concern, since we are interested in the initial evolution only insofar 
as it determines the transition into the low- or high-$\beta$ quasi-steady states.

The other test is to restart a simulation in a well-formed low-$\beta$ state   ($\beta$0.1-mr at $t\Omega=400$) with 
$\rho_{\rm flr}=3\times10^{-5}$, the results of which are shown in  \cref{fig: density floor}c-d. Although the detailed evolution 
starts to diverge at later times, a consequence of the system being chaotic, differences in the time averaged quantities are small even in the lowest-density regions at the edge.

We note that these lower-floor simulations are at least ${\simeq}3$-$4$ times more expensive computationally than those with $\rho_{\rm flr}=10^{-4}$, making them impractical for production simulations. Other methods should be explored 
in future work, particularly mesh refinement in the midplane or Lagrangian numerical schemes, which will alleviate these issues as a result of their larger grid spacing in low-$\rho$ regions.

\begin{figure}
\includegraphics[width=1.0\columnwidth]{\ffold 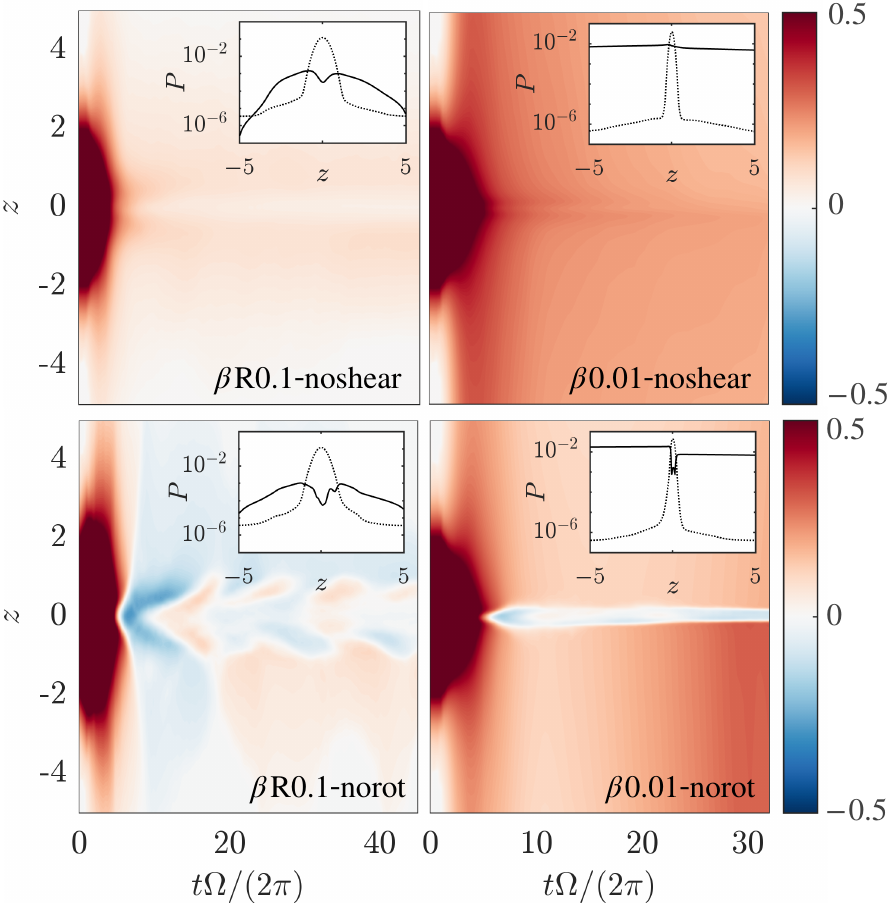}
\caption{Testing the physical ingredients needed to sustain the low-$\beta$ state, illustrating space-time `butterfly' plots for various simulations. The top two panels remove the mean shear flow but keep Coriolis effects by setting 
$q=0$ ($\beta$R0.1-noshear and $\beta$0.01-noshear);
the bottom panels remove the effect of rotation but retain the mean shear and the vertical gravity ($\beta$R0.1-norot and $\beta$0.01-norot).
In each case, the inset shows the $P_{B}=\mean{B_{y}}^{2}/8\pi$ (solid) and $P_{\rm th} = c_{s}^{2}\mean{\rho}$ (dotted) time-averaged 
over the steady state. All fail to maintain midplane magnetic flux.\vspace{3pt}
}
\label{fig: no rotation}
\end{figure}
\begin{figure}
\begin{center}
\includegraphics[width=0.8\columnwidth]{\ffold 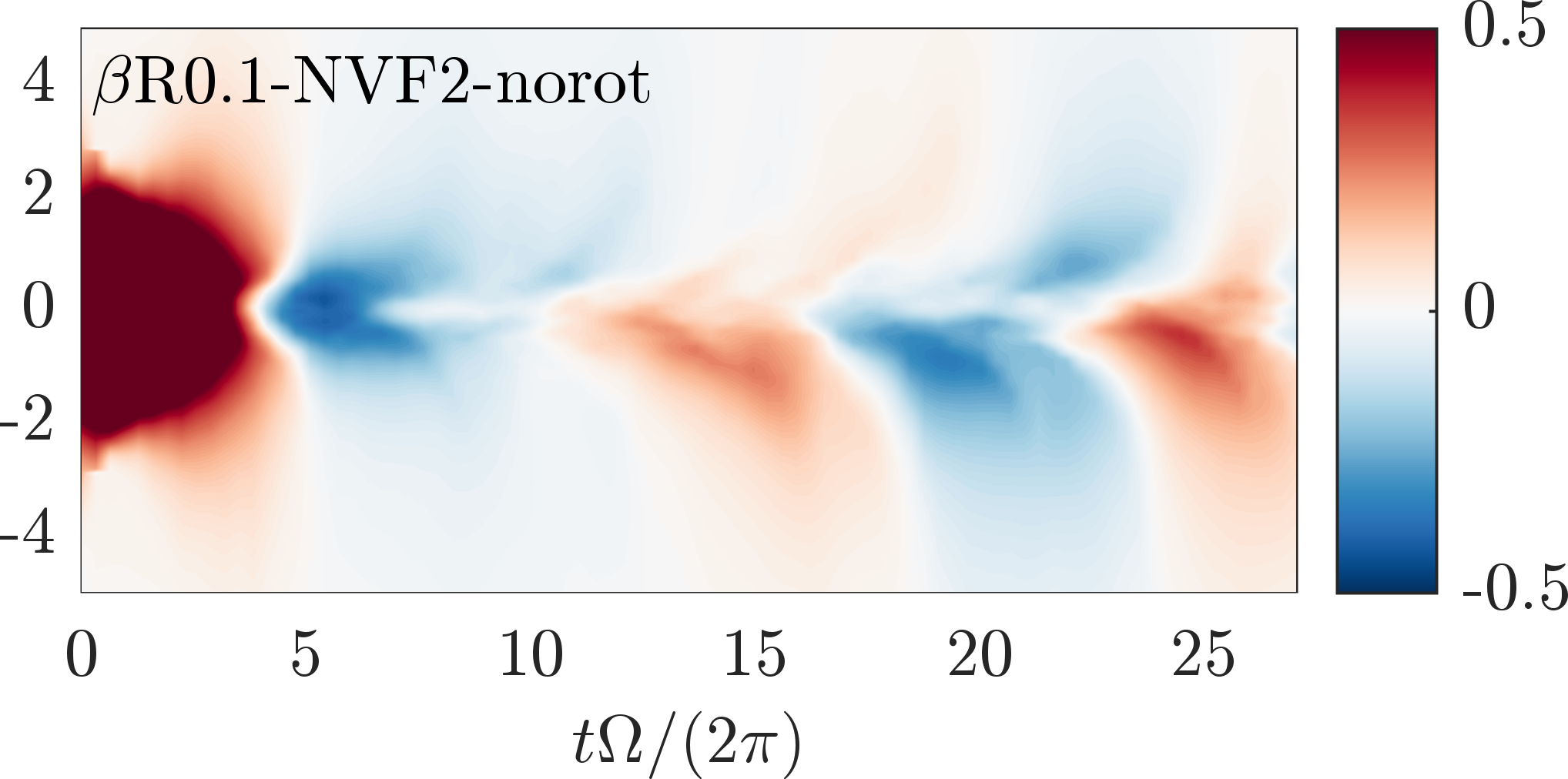}
\caption{\revchng{Space-time butterfly plot showing $\mean{B_{y}} $ for simulation $\beta$R0.1-NVF2-norot, 
which has a $\beta_{z}=100$ vertical flux, but with rotational effects removed. The system does not sustain the strong $\beta\lesssim1 $ azimuthal field
that occurs with rotation (cf.~\cref{fig: with bz})}.}
\label{fig: with bz no rot}
\end{center}
\end{figure}

\section{\revchng{The necessity of shear and rotation}}\label{app: shear and rotation}

\revchng{To help diagnose the mechanism(s) by which the low-$\beta$ state is maintained, in \cref{fig: no rotation} we rerun two cases 
that  transition into the low-$\beta$ state in the standard shearing box ($\beta$R0.1 and $\beta$0.01), but selectively 
remove either shear or rotation. The purpose of these modifications is not to explore physically relevant scenarios, but rather as numerical 
experiments to probe the important physics. Removing the shear simply involves setting $q=0$ in \crefrange{eq:MHD rho}{eq:MHD B},
while removing rotation is achieved by dropping the first two terms on the second line of \cref{eq:MHD u}, or equivalently, by taking the limit $\Omega\rightarrow 0$ with  $q\Omega=S={\rm const.}$ Both cases  retain the vertical stratification via $-\rho \Omega^2 z \hat{\bm{z}}$.}

The top two panels of \cref{fig: no rotation} show the effect of removing the mean shear flow; these clearly do not maintain interesting dynamics. The flux continuously escapes in $\beta$R0.1-noshear, while  $P_{B}$ becomes flat in $\beta$0.01 due to its Outflow boundary conditions, and both develop perfect thermal profiles in the midplane. The detailed structure at larger $z$ is  affected by the density floor and so is untrustworthy, but the obvious difference with \cref{fig: butterfly} nonetheless clearly demonstrates the importance of the shear. 

 As disussed in \S\ref{subsec: MRI},
removing rotation kills the MRI, as well as changing the turbulence properties and Parker instability. 
The lower panels of  \cref{fig: no rotation} show $\beta$R0.1-norot and $\beta$0.01-norot in which the rotation is artificially removed. Neither of these maintain the low-$\beta$ state, developing
a  thermally dominated midplane. Interestingly, 
dynamo cycles develop in $\beta$R0.1-norot, despite the lack of the MRI. Assuming these are of similar origin 
to those in the high-$\beta$ state (given they have a similar period and structure)
it is clear that the nomenclature `MRI dynamo cycles' is not accurate. Given this system's small $H_{\rm th}$ and the density floor, 
this would be worth a dedicated study in future work. In the lower-$\beta$ case ($\beta$0.01-norot) we do not see this feature, 
presumably due to the poorly resolved midplane and outflow boundary conditions that lead to $P_{B}\approx {\rm const.}$ at $z\gtrsim 2H_{\rm th}$.
Notwithstanding, given the clear difference with $\beta$0.01, we can conclude that rotation is crucial for maintaining the low-$\beta$ state.

\revchng{In \cref{fig: with bz no rot}, we remove the rotation for a case with a strong vertical field $\beta_{z}=100$.   This causes the system 
to lose its azimuthal flux, generating cyclic behavior in $\mean{B_y}$ and a thermal midplane. This is  similar to the high-$\beta$ ZNVF system (cf. \cref{fig: no rotation}) although with somewhat stronger mean-field cycles.  The behavior is  consistent with our interpretation in \S\ref{sub: NVF} that the NVF low-$\beta$
state is effectively a single large-scale MRI mode, which requires rotation to sustain itself. It also supports the idea that the dynamo cycles are not a consequence of the MRI, which does not exist in the absence of rotation. }\vspace{10pt}

\end{appendix}

\end{document}